\documentclass[twocolumn,prd,aps,longbibliography,superscriptaddress,preprintnumbers,tightenlines,
showpacs,nofootinbib,eqsecnum,amsfonts,amsmath]{revtex4}
\pdfoutput=1

\usepackage{color}
\usepackage{calc}
\usepackage{graphicx}
\usepackage{tensor}
\usepackage{bm}
\usepackage{url}
\usepackage{times}
\usepackage{multirow}
\usepackage[varg]{txfonts}

\usepackage{float}
\usepackage{dcolumn}
\usepackage[nolist,nohyperlinks]{acronym}
\usepackage{xspace}
\usepackage[english]{babel}
\usepackage[abs]{overpic}
\usepackage{pict2e}
\usepackage[caption=false]{subfig} % prevent loading of caption package: http://tex.stackexchange.com/questions/22388/subfigures-with-revtex
\allowdisplaybreaks[1]
\usepackage[utf8]{inputenc}
\usepackage{braket}
\usepackage{stackengine}
\usepackage{gensymb}

\usepackage{doi}%<----------

\usepackage{boldline}
\usepackage{longtable}
\DeclareMathAlphabet{\mathcalstd}{OMS}{cmsy}{m}{n}
\DeclareMathAlphabet{\mathpzc}{OT1}{pzc}{m}{it}

\newcommand{\UIB}{Departament de F\'isica, Universitat de les Illes Balears, IAC3 -- IEEC, Crta. Valldemossa km 7.5, E-07122 Palma, Spain}

\newcommand{\UoB}{School of Physics and Astronomy and Institute for Gravitational Wave Astronomy, University of Birmingham, Edgbaston, Birmingham, B15 9TT, United Kingdom}

\begin{document}

\preprint{LIGO-P1900279}

%%%%%%%%%%%%%%%%%%%%%%%%%%%%%%%%%%%%%%%%%%%%%%%%%%%% Title page %%%%%%%%%%%%%%%%%%%%%%%%%%%%%%%%%%%%%%%%%%%%%%%%%%%%

\title{A first survey of spinning eccentric black hole mergers: Numerical relativity simulations, hybrid waveforms, and parameter estimation}
%Numerical Relativity waveform catalog, hybridization and parameter estimation of spinning eccentric black-hole binaries}

\author{Antoni Ramos-Buades}
\affiliation{\UIB}
\author{Sascha Husa}
\affiliation{\UIB}
\author{Geraint Pratten}
\affiliation{\UIB}
\affiliation{\UoB}
\author{Héctor Estellés}
\affiliation{\UIB}
\author{Cecilio García-Quirós}
\affiliation{\UIB}
\author{Maite Mateu-Lucena}
\affiliation{\UIB}
\author{Marta Colleoni}
\affiliation{\UIB}
\author{Rafel Jaume}
\affiliation{\UIB}

%%%%%%%%%%%%%%%%%%%%%%%%%%%%%%%%%%%%%%%%%%%%%%%%%%% Abstract %%%%%%%%%%%%%%%%%%%%%%%%%%%%%%%%%%%%%%%%%%%%%%%%%%%%%%
\begin{abstract}
 We analyze a new numerical relativity dataset of spinning but nonprecessing binary black holes on eccentric orbits, with eccentricities from approximately $0.1$ to $0.5$,  with dimensionless spins up to $0.75$ included at mass ratios $q=m_1/m_2 = (1, 2)$, and further nonspinning binaries at mass ratios $q = (1.5, 3, 4)$. 
A comparison of the final mass and spin of these simulations with noneccentric data extends previous results in the literature on circularization of eccentric binaries to the spinning case.
For the $(l,m)=(2,2)$ spin-weighted spherical harmonic mode we construct eccentric hybrid waveforms that connect the numerical relativity data to a post-Newtonian description for the inspiral, and we discuss the limitations in the current knowledge about post-Newtonian theory which complicate the generation of eccentric hybrid waveforms. 
We also perform a Bayesian parameter estimation study, quantifying the parameter biases introduced when using three different quasicircular waveform models to estimate the parameters of highly eccentric binary systems. We find that the used aligned-spin quasicircular model including higher order modes produces lower bias  in certain parameters than the nonprecessing quasicircular model without higher order modes and the quasicircular precessing model.
\end{abstract}


\today

\maketitle

% ======================
%  ACRONYMS
% ======================
\acrodef{PN}{post-Newtonian}
\acrodef{EOB}{effective-one-body}
\acrodef{NR}{numerical relativity}
\acrodef{GW}{gravitational-wave}
\acrodef{BBH}{binary black hole}
\acrodef{BH}{black hole}
\acrodef{BNS}{binary neutron star}
\acrodef{NSBH}{neutron star-black hole}
\acrodef{SNR}{signal-to-noise ratio}
\acrodef{aLIGO}{Advanced LIGO}
\acrodef{AdV}{Advanced Virgo}

\newcommand{\PN}[0]{\ac{PN}\xspace}
\newcommand{\EOB}[0]{\ac{EOB}\xspace}
\newcommand{\NR}[0]{\ac{NR}\xspace}
\newcommand{\BBH}[0]{\ac{BBH}\xspace}
\newcommand{\BH}[0]{\ac{BH}\xspace}
\newcommand{\BNS}[0]{\ac{BNS}\xspace}
\newcommand{\NSBH}[0]{\ac{NSBH}\xspace}
\newcommand{\GW}[0]{\ac{GW}\xspace}
\newcommand{\SNR}[0]{\ac{SNR}\xspace}
\newcommand{\aLIGO}[0]{\ac{aLIGO}\xspace}
\newcommand{\AdV}[0]{\ac{AdV}\xspace}

%%%%%%%%%%%%%%%%%%%%%%%%%%%%%%%%%%%%%%%%%%%%%%%%%%%%%%%
\section{Introduction}\label{sec:introduction}

The detections of gravitational wave signals \cite{PhysRevLett.116.061102, PhysRevLett.116.241103,Abbott:2017vtc,Abbott:2017gyy,Abbott:2017oio,TheLIGOScientific:2017qsa,LIGOScientific:2018mvr, Romero-Shaw:2019itr} have been found  to be consistent with models of the waveform emitted from the merger of compact objects under the assumption of quasicircularity of the binary's orbit prior to the merger. The assumption of quasicircularity motivated by the efficient circularization 
of binaries as a consequence of the emission of gravitational waves \cite{PhysRev.131.435,PhysRev.136.B1224} simplifies significantly the complexity of the signal and has accelerated the development
%allowed the emergence of a pletora 
 of inpiral-merger-ringdown (IMR) waveform models: several mature IMR models for quasicircular  coalescences, i.e.~neglecting eccentricity, are now publicly available \cite{Husa2015,PhysRevD.93.044007,phenomp,seobnrv4,london2018,PhysRevD.98.084028,Blackman:2017dfb,PhysRevResearch.1.033015, PhysRevD.99.064045,Khan:2018fmp,Khan:2019kot,phenX,phenXHM,phenT}, and are being used to search and infer the parameters of observed binary black hole systems \cite{LIGOScientific:2018mvr}.

Recently, population synthesis studies \cite{Belczynski:2016obo,Park:2017zgj,PhysRevD.97.103014,Samsing:2013kua} have shown that active galactic nuclei and globular clusters can host a population of moderate and highly eccentric binaries emitted in the frequency band of ground-based detectors. Therefore, the increase in sensitivity of the detectors will increase the likelihood of detecting binary systems with non-negligible eccentricities. The modeling of the gravitational waveforms from eccentric black hole binaries is complicated by the addition of a new timescale to the binary problem, the periastron precession \cite{1985AIHS...43..107D}. This new timescale induces oscillations in the waveforms due to the asymmetric emission of gravitational radiation between the apastron and periastron passages. 

The orbits of eccentric black hole binaries are typically described using the quasi-Keplerian (QK) parametrization \cite{PhysRevD.70.064028}, which is currently known up to 3 post-Newtonain (PN) order \cite{PhysRevD.70.104011}. This parametrization has proved to be a key element in developing inspiral PN waveforms \cite{PhysRevD.93.064031,Moore:2018kvz,Moore:2019xkm,Loutrel:2018ydu,Moore_2019}. The generation of IMR eccentric models relies on the connection of an eccentric PN inspiral with a circular merger \cite{PhysRevD.97.024031,PhysRevD.98.044015}. Alternatively, one can substitute the PN waveform with one produced within the effective one body (EOB) formalism describing an eccentric inspiral \cite{PhysRevD.96.044028,PhysRevD.96.104048,Chiaramello:2020ehz}. Some eccentric IMR waveform models show good agreement with numerical waveforms up to $e \sim 0.2$ for nonspinning configurations \cite{PhysRevD.97.024031}. Recent work has shown possible extensions of the eccentric PN and EOB inspiral waveforms to include spin effects \cite{PhysRevD.98.104043,Chiaramello:2020ehz}.
A key step in the generation of IMR waveform models is the production of hybrid waveforms~\cite{PhysRevD.82.064016,Ohme_2012,PhysRevD.84.064029,MacDonald_2011,Ajith_2012,PhysRevD.84.064013,Bustillo:2015ova} between PN/EOB inspiral and numerical relativity (NR) waveforms. The hybridization procedure consists of smoothly attaching a PN/EOB inspiral waveform to a NR one in order to get the full description of the gravitational radiation of the binary system. The generation of datasets of hybrid waveforms has been used in the quasicircular case to calibrate and validate the accuracy of IMR waveform models \cite{phenX,phenXHM,PhysRevD.98.084028,Blackman:2017dfb,PhysRevResearch.1.033015}.  

In this paper we present the input data and some key tools required  for the development of an IMR eccentric waveform 
model calibrated to eccentric hybrid PN-NR waveforms.
In Sec. \ref{sec:NRcatalog} we first present our NR catalog of nonspinning and spinning eccentric binaries, computed with the private \texttt{BAM} code \cite{PhysRevD.77.024027} and the open-source EinsteinToolkit (\texttt{ET}) \cite{Loffler:2011ay,maria_babiuc_hamilton_2019_3522086}. This includes a discussion of our procedure to specify the initial parameters of the eccentric simulations in Sec.  \ref{sec:NRcatalogA}, a study of
the remnant quantities in Sec.~\ref{sec:NRcatalogB}, and  a new method for measuring the eccentricity of NR waveforms with arbitrarily high eccentricity  in Sec.~\ref{sec:NRcatalogC}. We find that the final spin and mass are consistent within the error estimates with the quasicircular case, which extends the study in \cite{PhysRevD.77.081502} to the eccentric spinning case.
We hybridize the dominant gravitational waveform $(l=2,m=2)$ mode between numerical relativity and post-Newtonian waveforms in Sec. \ref{sec:PNEcc}. This will provide the input data for future work on constructing waveform models that contain the inspiral, merger, and ringdown,  and it allows us to perform injections into detector noise which contain a long inspiral phase.
%,
%We also point out the limitations of the current post-Newtonian theory in the description of eccentric systems.
%
In Sec. \ref{sec:PE} we use such injections of hybrid waveforms, as well as those of pure numerical relativity waveforms, to study the parameter biases introduced when using quasicircular waveform models to estimate the parameters of highly eccentric spinning systems.
% into gaussian detector noise
Unless explicitly noted, we are working in geometric units $G=c=1$. To simplify expressions we will also set the total mass of the system $M=1$ in Secs. \ref{sec:NRcatalog} and \ref{sec:PNEcc}. We define the mass ratio $q=m_1/m_2$ with the choice $m_1 > m_2$, so that $q>1$. We also introduce the symmetric mass ratio $\eta=q/(1+q)^2$, and we shall denote the black hole's dimensionless spin vectors by $\vec{\chi}_i= \vec{S}_i/m^2_i$ for $i=1,2$.

 %%%%%%%%%%%%%%%%%%%%%%%%%%%%%%%%%%%%%%%%%%%%%%%%%%%%%%%%
\section{Numerical Relativity dataset} \label{sec:NRcatalog}
 %%%%%%%%%%%%%%%%%%%%%%%%%%%%%%%%%%%%%%%%%%%%%%%%%%%%%%%%

\subsection{Overview} \label{sec:NRcatalog0}

\begin{figure}[ht!]
\centering
\includegraphics[scale=0.35]{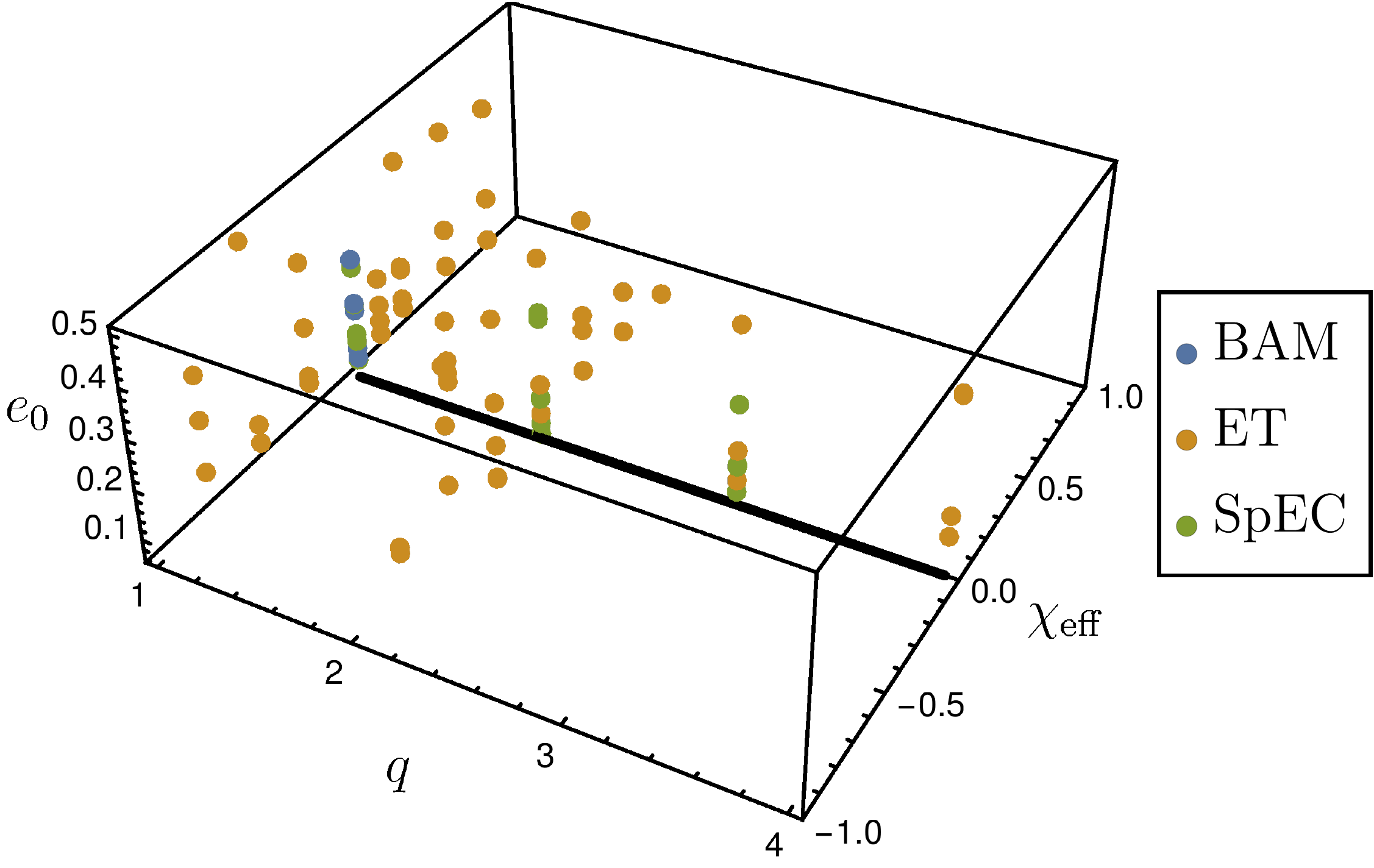}
\caption{Initial eccentricity $e_0$, mass ratio $q$ and effective spin parameter $\chi_{\text{eff}}= (m_1 \chi_{1,z}+m_2 \chi_{2,z})/(m_1+m_2)$ for the numerical relativity simulations generated with the \texttt{BAM}, EinsteinToolkit and \texttt{SpEc} \cite{SpEC} codes. The thick black line represents the cases with $\chi_{\text{eff}}=0$.}
\label{fig:NRsims}
\end{figure}

We present a catalog of 60 eccentric NR simulations performed with the nonpublic \texttt{BAM} code \cite{PhysRevD.77.024027} and the open-source \texttt{ET} code \cite{Loffler:2011ay,maria_babiuc_hamilton_2019_3522086} with the multipatch \texttt{LLAMA} infrastructure \cite{Pollney:2009yz}. The numerical setup of both codes is the same as in \cite{PhysRevD.99.023003}. Most of the simulations are run with the \texttt{ET} code using the \texttt{LLAMA} module due to its ability to extract the waves at larger extraction radii. 
The different simulations and their initial conditions are described in Table \ref{tab:tabNR3}. In Fig. \ref{fig:NRsims} we show our choices of mass ratio $q$, initial eccentricity $e_0$, and effective spin parameter, $\chi_{\text{eff}}= (m_1 \chi_{1,z}+m_2 \chi_{2,z})/(m_1+m_2)$. We have also added  20 public eccentric SXS simulations presented in \cite{PhysRevD.98.044015}.

\subsection{Initial parameters of eccentric NR simulations} \label{sec:NRcatalogA}
%%%%%%%%%%%%%%%%%%%%%%%%%%%%%%%%%%%%%%%%%

We use conformally flat Bowen-York initial data \cite{PhysRevD.21.2047} in the center-of-mass frame, where the free parameters are the spins and masses of the two black holes, the separation, and the momentum of one of the two black holes (the momentum of the second black hole is then equal in magnitude but opposite in direction). We first choose the masses and spins as displayed in Fig.~\ref{fig:NRsims}.
To be able to construct hybrid waveforms, the minimal separation, i.e. the separation at periastron, has to be large enough that
the PN approximation is still roughly valid. We then use a simple PN approximation as discussed below to compute the apastron separation required to achieve a chosen value of the eccentricity, and a further PN approximation to compute the appropriate value of the momentum corresponding to this value of the eccentricity. Owing to the simplicity, i.e., the low order, of the PN approximations used, neither the periastron separation nor the measured eccentricity will exactly coincide with the specified values.  In this study we choose our initial choice for the approximate periastron separation to be $r_{\text{min}} \sim 9M$, with slightly different values to account for mass ratio and spin effects which can significantly increase the computational cost of the simulations. We start our simulations at the apastron, where the PN approximation employed to specify the initial momentum and the agreement with the PN data that we use for hybridization, will be more accurate than during other points of the orbit. 

Larger choices of eccentricity for the same configuration of masses and spins thus lead to a larger merger time and number of orbits, as one can see in Table \ref{tab:tabNR3}. For instance, focusing on simulations with identification numbers (IDs) 34, 35, and 36, one observes an increase in the merger time when increasing the initial eccentricity. This increase in merger time also implies an increase in the computational cost of the simulation.

Using the QK parametrization at Newtonian order, one can relate the initial minimum and maximum separations by

\begin{equation}
r_{\text{min}}= r_{\text{max}} \frac{1-e}{1+e}.
\label{eq:eq2}
\end{equation}
%%%
As stated above, for our simulations we choose $r_{\text{min}} \sim 9M$ such that the PN approximation is still roughly valid. Then  for $e_0 =  0.1,0.2,0.5$ Eq. \eqref{eq:eq2} implies that  $r_{\text{max}} = 11 M$, $13.5 M$, $27 M $, respectively. These values of  $r_{\text{max}}$ are rough estimates 
based on a Newtonian order calculation; in practice, we slightly modify those values of initial separations to account for the increase of computational cost depending on the mass ratio and the spins of the simulations as observed in Table \ref{tab:tabNR3}. For instance, in the case of negative spin components the merger time is significantly reduced \cite{PhysRevD.97.084002}; thus, we increase $r_{\text{max}}$  for $e=0.1,0.2$ cases to produce longer NR waveforms which  are  easier to hybridize afterward.

To produce initial data for a desired eccentricity we then make use of Eq. (3.25) of \cite{PhysRevD.99.023003} to perturb the initial tangential momentum of the black holes by a factor $\lambda_t$ from its quasicircular value. The expression for $\lambda_t$ in terms of the eccentricity at 1PN order is 
\begin{equation}
\lambda_t (r,e_0, \eta, \text{sign})= 1+ \frac{e_0 }{2}\times \text{sign}\times \left[ 1-\frac{ 1}{r}(\eta +2) \right],
\label{eq:eq1}
\end{equation}
where $\eta$ is the symmetric mass ratio, $r$ is the orbital separation and $\text{sign}=\pm 1$ depends on the initial phase of the eccentricity estimator \cite{PhysRevD.99.023003}.  We refer the reader to Sec. III D of \cite{PhysRevD.99.023003} for an explicit derivation of Eq. \eqref{eq:eq1}. Taking Eq. \eqref{eq:eq1}, we compute the correction factor applied to the momentum as the mean between the inverse of the expression with the plus sign plus the expression with the minus sign, 

\begin{equation}
\begin{split}
\bar{\lambda}^0_t (r,e_0, \eta) & = \frac{1}{2} \left[ \lambda_t(r,e_0, \eta,+1)^{-1}+\lambda_t(r,e_0, \eta,-1) \right] \\
& = \frac{8 r^2-e_0^2 (\eta -r+2)^2}{4 r (e_0 (-\eta +r-2)+2 r)}.
\end{split}
\label{eq:eq3}
\end{equation}
We use the combination of factors in Eq. \eqref{eq:eq3} because we have experimentally tested it to see that it works more accurately than just specifying a value of $\lambda_t(r,e,\text{sign})$ with a given sign. In Table \ref{tab:tabNR3} one can compare the value of the desired initial eccentricity, $e_0$, specified in Eq. \eqref{eq:eq3}, and the actually measured initial eccentricity, $e_\omega $, from the orbital motion of the simulation. Both quantities are also displayed in Fig. \ref{fig:eccMeasEccDes}, where we have differentiated  among nonspinning and positive and negative spin simulations. The results point out that the use of Eq. \eqref{eq:eq3} produces differences of less than $10 \%$ between $e_\omega$ and $e_0$ in nonspinning cases at low eccentricities of the order of $0.1$. However, when spins are present or the eccentricities are higher, the inaccuracy of the formula becomes manifest, with differences of the order of $20\% -30 \%$, this is due to the fact that Eq. \eqref{eq:eq1} was derived assuming a nonspinning binary in the low eccentric limit. Additionally, one can check in Table \ref{tab:tabNR3} and Fig. \ref{fig:eccMeasEccDes} to see that the differences between $e_\omega$ and $e_0$ are smaller for the cases with positive spins than in cases with negative spins because in Eq. \eqref{eq:eq1} the radiation reaction effects, which are more significant for negative spins, are also not taken into account.

\subsection{Final state of spinning eccentric systems} \label{sec:NRcatalogB}
%%%%%%%%%%%%%%%%%%%%%%%%%%%%%%%%%%%%%%

We compare the final state of the eccentric NR simulations with the predicted final mass and final spin of the quasicircular (QC) NR fits \cite{PhysRevD.95.064024} as an indicator of circularization of the coalescence process as the binary merges. This is basically an extension of \cite{PhysRevD.77.081502} to the eccentric spinning case with more moderate values of the eccentricity but longer NR evolutions.

The final mass and final spin of the simulations are computed using the apparent horizon (AH) of the remnant black hole and are shown in Table  \ref{tab:tabNR3}. 
The magnitude $S$ of the angular momentum of the final black hole can be computed from the integral
\begin{equation}
S = \frac{1}{8 \pi} \oint_{\mbox{AH}} K_{ij} n^i \phi^j dA,
\label{eq:eq5}
\end{equation}
see the discussion in \cite{PhysRevD.75.064030, PhysRevD.67.024018}. Here for the \texttt{BAM} code \cite{PhysRevD.77.024027} the vector $\phi^j$ is a coordinate-based approximation to the (approximate) axial Killing vector of the black hole horizon as in \cite{PhysRevD.75.064030}, and for the EinsteinToolkit code  the  \texttt{QuasiLocalMeasures} module is used, which constructs
an approximate Killing vector with rotational symmetry around the spin axis as in \cite{PhysRevD.54.4899, Thornburg:2003sf}.
The vector $n^i$ is a spacelike unit normal to the horizon surface and $K_{ij}$ is the extrinsic curvature.
The final mass can be computed from the Christodoulou formula in terms of the black hole (BH) angular momentum and 
AH area $A$ as
\begin{equation}
M_f = \sqrt{M^2_{irr}+ \frac{S^2}{4 M^2_{irr} }}, \quad M_{irr}=\sqrt{\frac{A}{16\pi}}. 
\label{eq:eq4}
\end{equation}
where $M_{irr}$ is the irreducible mass. The dimensionless final spin can then be computed as $\chi_f = S/M_f^2$.

\begin{figure}[hbt!]
\centering
\begin{minipage}[hbt!]{\columnwidth}
\captionsetup{justification=centering}
\includegraphics[scale=0.42]{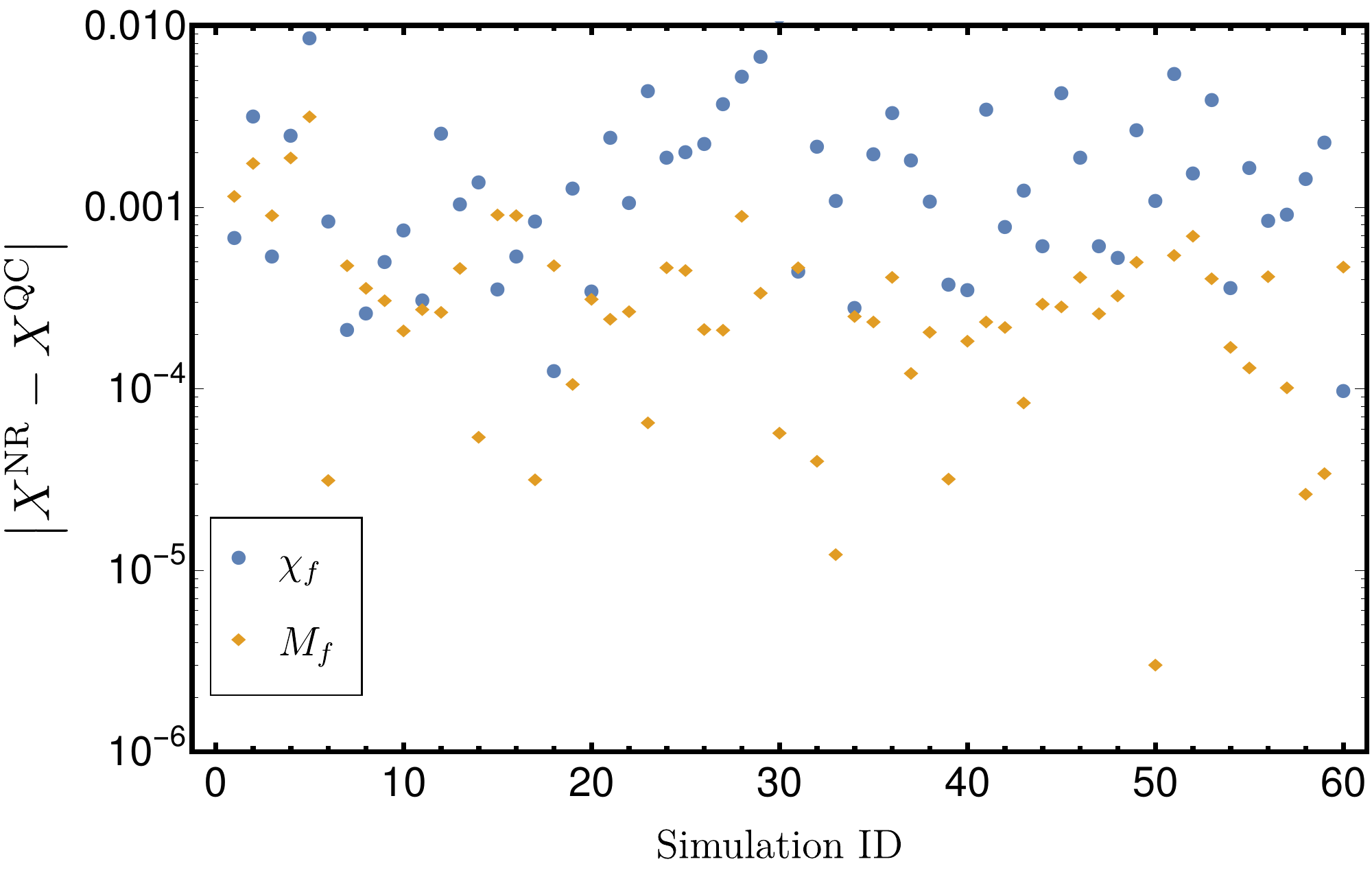}
\end{minipage}

\begin{minipage}[hbt!]{\linewidth}
\captionsetup{justification=centering}
\includegraphics[scale=0.42]{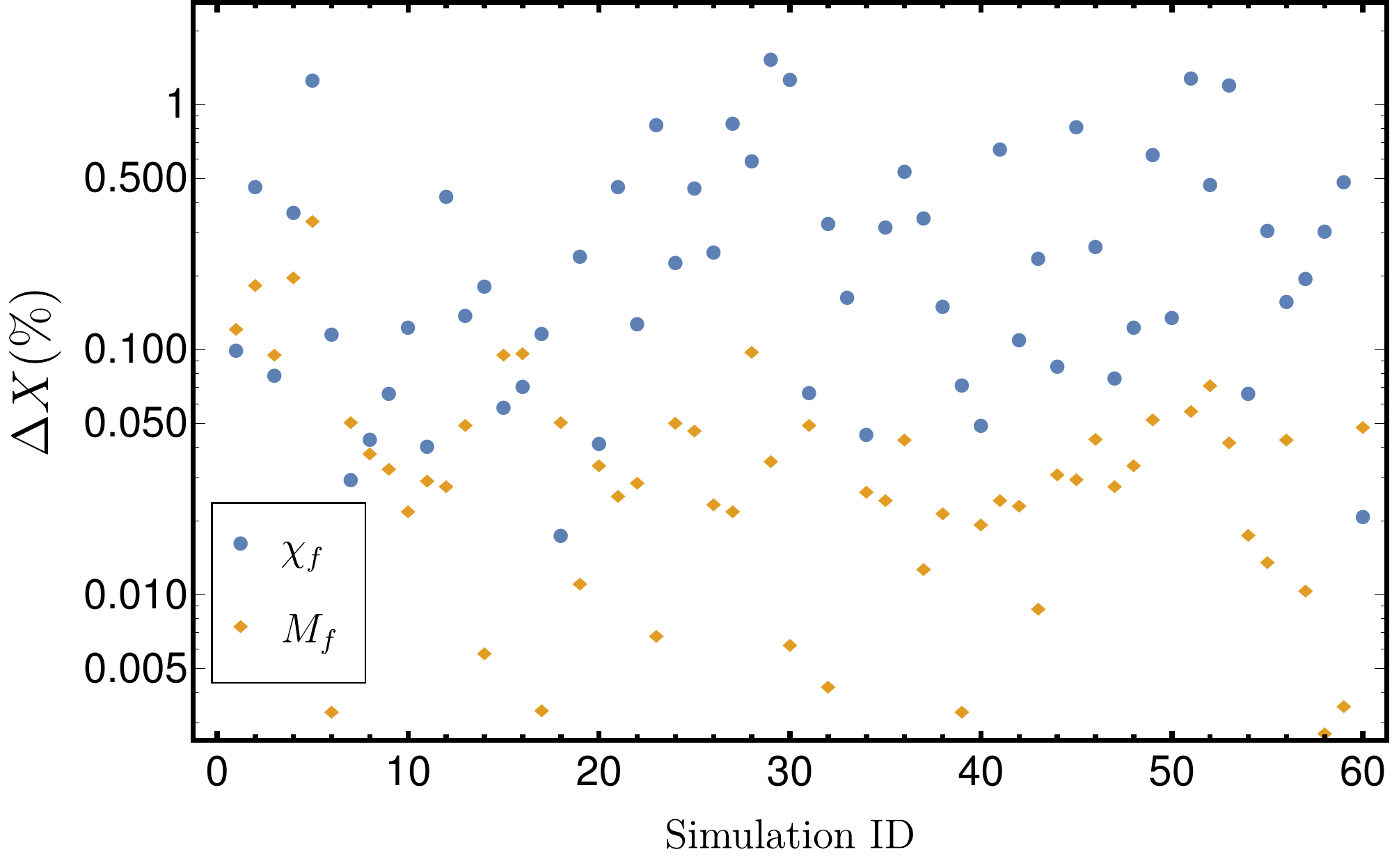}
\end{minipage}
\caption{ (Top panel) The absolute difference between the final mass and spin of the simulations and the QC NR fits as a function of the ID of the simulations in Table \ref{tab:tabNR3}. (Bottom panel) The absolute relative error for the phase and amplitude, $\Delta X = |X^{NR}/X^{QC}-1|\times 100$ for $X=M_f,\chi_f$, relative error of the final mass and final spin of the simulations against the QC NR fits as a function of the ID of the simulations in Table \ref{tab:tabNR3}.}
\label{fig:FinalStateErrors}
\end{figure}

In Fig. \ref{fig:FinalStateErrors} we have computed the absolute and relative errors between the eccentric simulations and the quasicircular NR final mass and final spin fitting formulas \cite{PhysRevD.95.064024},
\begin{equation}
\Delta X = \left| \frac{X^{NR}}{X^{QC}}-1 \right| \times 100, \quad X=M_f \text{ or }\chi_f.
\label{eq:eq6}
\end{equation}
The results in Fig. \ref{fig:FinalStateErrors} show that the differences in the final spin are generally higher than in the final mass. However, the differences with respect to the quasicircular fitting values are as high as $\sim 1 \%$ which is entirely consistent with numerical errors and gauge artifacts on the apparent horizon surfaces and inaccuracies in the fits. Hence, we can conclude that within the current knowledge of systematic errors (compare with \cite{PhysRevD.95.064024}), the final state of the eccentric simulations, up to the values of eccentricity studied here, is consistent with the quasicircular values. Identifying small physical deviations between the quasicircular and eccentric final states will require numerical simulations with improved error estimates.

\subsection{Measuring the eccentricity of highly eccentric systems} \label{sec:NRcatalogC}
%%%%%%%%%%%%%%%%%%%%%%%%%%%%%%%%%%%%%%%%%%%%%

This subsection aims to extend the discussion on the measurement of the eccentricity in NR presented in \cite{PhysRevD.99.023003} to highly eccentric systems. An eccentricity parameter is chosen to describe the noncircularity of orbits, such that for bound orbits its value ranges between 0 and 1, corresponding to circular and extremely elliptical configurations, respectively. Such an eccentricity can  be defined naturally only in Newtonian gravity, whereas in general relativity the eccentricity is a gauge dependent quantity. To measure the eccentricity in NR data one defines quantities known as eccentricity estimators, which estimate the eccentricity from the relative oscillations of a certain combination of dynamical quantities such as the orbital separation or orbital frequency, or wave quantities like the amplitude or frequency of the $(l,m)=(2,2)$ mode. All of these different estimators are usually defined such that they agree in the Newtonian limit and in the low eccentricity limit.

In \cite{PhysRevD.99.023003}, where we studied the reduction of residual eccentricity in initial datasets, we chose our eccentricity estimator based on the orbital frequency as
\begin{equation}
e_\omega(t)= \frac{\omega (t)-\omega(e=0)}{2 \omega(e=0)} \equiv  \frac{\omega (t)-\omega^{\text{fit}}(t)}{2 \omega^{\text{fit}}(t)},
\label{eq:eq8}
\end{equation}
 where $\omega (t)$ is the orbital frequency of the simulation and $\omega (e=0)$ is the orbital frequency in the quasicircular limit. We note that when dealing with numerical simulations, the quasicircular frequency in Eq. \eqref{eq:eq8} is typically replaced by a fit, $ \omega^{\text{fit}}(t)$, of the nonoscillatory part of the frequency~\cite{PhysRevD.99.023003}. This eccentricity estimator  is  used largely to measure the residual eccentricity of NR simulations of quasicircular black hole binaries. We remark that while in Eq. \eqref{eq:eq8} we decide to use the orbital frequency calculated from the BH motion, one can also  use the gravitational wave frequency extracted from the waves as discussed below.  Furthermore, gauge effects can impact the eccentricity measurement from the orbital frequency of NR codes as extensively discussed in the small eccentricity limit in \cite{Purrer:2012wy,PhysRevD.82.124016}. Here we follow the practice used in the literature \cite{Purrer:2012wy,PhysRevD.99.023003,PhysRevD.82.124016,PhysRevD.83.104034} to avoid contamination of the eccentricity measurement through the gauge quantities, like the choice of the value of the $\eta$-parameter in the gamma driver condition \cite{Alcubierre:2002kk}, which can  lead to residual oscillations in the orbital frequency complicating the determination of the eccentricity.   
 
In \cite{PhysRevD.99.023003} we argued that the procedure shown there, based on Eq. \eqref{eq:eq8}, to measure the eccentricity is limited to values as high as $e\sim 0.1$ due to the lack of an accurate ansatz to fit the higher order contributions beyond the sinusoidal contribution. While the lack of an ansatz for high eccentricities is a clear limitation, the use of Eq. \eqref{eq:eq8} biases the eccentricity measurement due to its reliance on a noneccentric fit of the orbital frequency and due to the fact that Eq. \eqref{eq:eq8} for high eccentricities does not reduce to the common definition of eccentricity in the Newtonian limit.

Therefore, we decide to change to another estimator \cite{PhysRevD.66.101501}, constructed also from the orbital frequency,
\begin{equation}
e_{\omega}(t)= \frac{\omega^{1/2}_p - \omega^{1/2}_a}{\omega^{1/2}_p + \omega^{1/2}_a},
\label{eq:eq9}
\end{equation}
where $\omega_a,\omega_p$ are the orbital frequencies at apastron and periastron, respectively. The eccentricity estimator in Eq.  \eqref{eq:eq9} does not depend on any noneccentric fit of the orbital frequency. Furthermore, as shown in Appendix \ref{sec:AppendixB} the eccentricity estimator from Eq. \eqref{eq:eq8} in the Newtonian limit at high eccentricities does not reduce to the eccentricity parameter and it is not normalized, while the eccentricity estimator from Eq.  \eqref{eq:eq9} fulfills all these conditions.

\begin{figure}[!]
\centering
\captionsetup{justification=centering}
\includegraphics[scale=0.45]{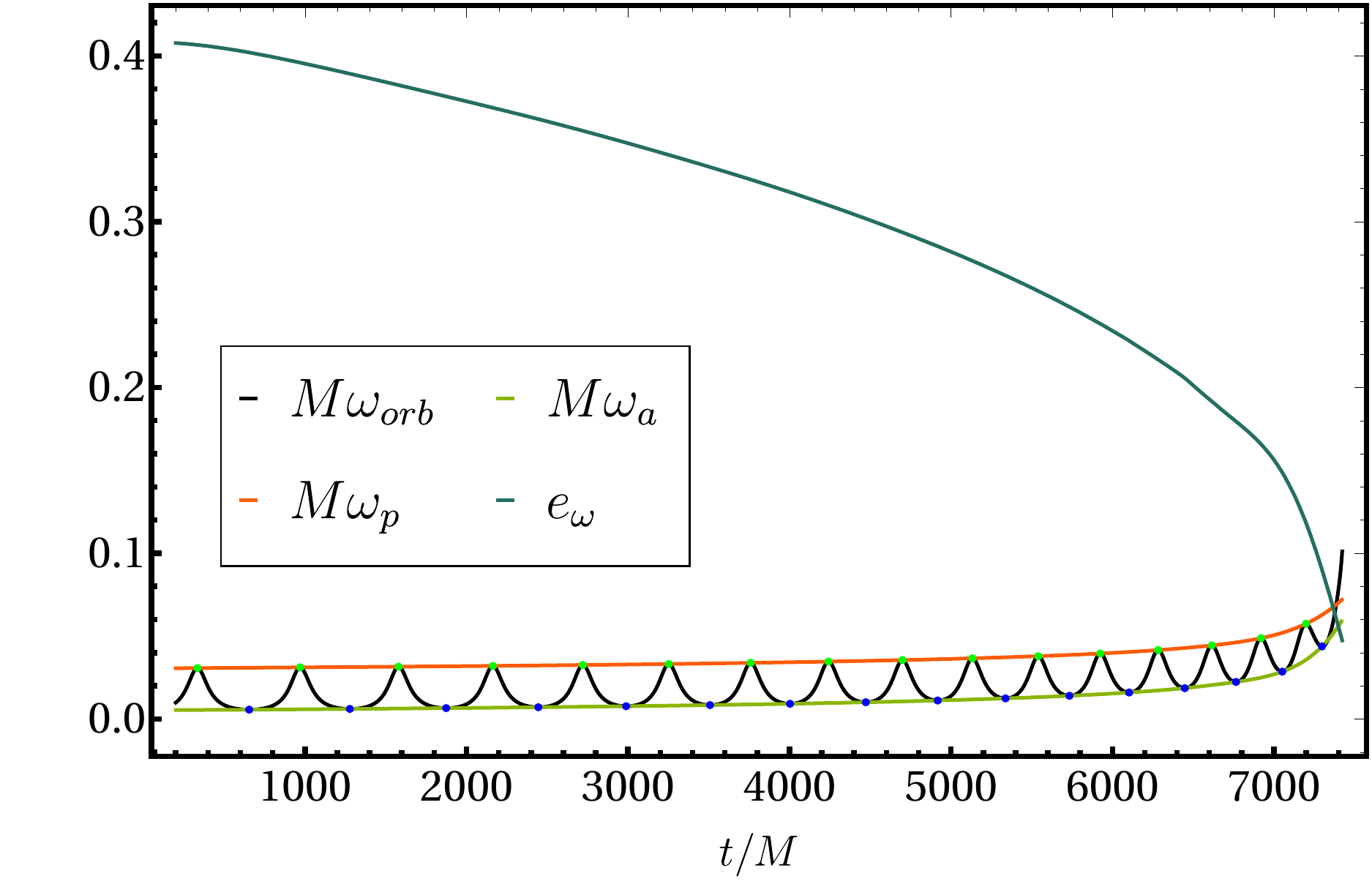}
\caption{ Time evolution of the orbital frequency $M \omega_{orb}$, the orbital frequency at apastron $M \omega_a$, the orbital frequency at periastron $M \omega_p$ and the eccentricity estimator $e_\omega$ defined in Eq. \eqref{eq:eq9}.}
\label{fig:EccMinMaxbo}
\end{figure}

 We measure the eccentricity from the maxima and minima of the orbital frequency corresponding to the periastron and apastron passages, respectively. Additionally, we produce an interpolated function from the maxima $\omega_p$  and the minima $\omega_a$ and substitute them into Eq. \eqref{eq:eq9}, so that one can estimate the evolution of the eccentricity from those points. The interpolation is calculated using the Hermite method implemented in the function \texttt{Interpolation} in \textit{Mathematica} \cite{Mathematica}.  The new procedure to measure the eccentricity is shown in Fig. \ref{fig:EccMinMaxbo}, where the time evolution of the orbital frequency, the interpolated functions of the maxima and minima of the orbital frequency and the eccentricity are shown for the configuration with ID 60 from Table \ref{tab:tabNR3}. As expected the eccentricity is a monotonically decaying function, whose value at $t=200M$, after the burst of junk radiation, is $e_\omega=0.415 \pm 0.005$. The error in the eccentricity, $\delta e_\omega$, is computed using error propagation: from Eq. \eqref{eq:eq9} we obtain
\begin{equation}
\delta e_{\omega}= \frac{  \delta \omega}{\left( \omega_a^{1/2} + \omega_p^{1/2} \right)^2} \left[ \frac{\omega_a^{1/2}}{\omega_p^{1/2}} + \frac{\omega_p^{1/2}}{\omega_a^{1/2}} \right],
\label{eq:eq91}
\end{equation}
where we have assumed $\delta \omega_a =\delta \omega_p= \delta \omega$. Motivated by the results of the error in the convergence analysis of the orbital frequency in \cite{NRUIB} we have taken as a conservative estimate   $\delta  \omega =0.0001$.  The error estimate of Eq. \eqref{eq:eq91} is the statistical error associated with the eccentricity measurement taking into account the error of the orbital frequency from different resolutions of the NR simulations. We remark that this error does not take into account systematics coming from the use of a different eccentricity estimator, nor contributions from the interpolation error when the number of minima and maxima is small due to the short length of the simulations. Because of the difficulties in quantifying the systematics associated with the choice of eccentricity estimator estimator and the fact that the interpolation error is a subdominant effect for most of the simulations here, we restrict for simplicity our eccentricity error calculation to Eq. \eqref{eq:eq91}.

The main drawback of this method is that when the simulations are so short that there is only one minimum and one maximum it becomes inefficient and inaccurate. Furthermore, one could choose the frequency of the $(l,m)=(2,2)$ mode and compute the orbital frequency as $\omega_{orb} \approx \omega_{22}/2$, employing the same method discussed in this section. Nevertheless, as pointed out in \cite{Purrer:2012wy} the usage of the orbital frequency from the $(2,2)$ mode requires additional postprocessing of the data due to the presence of high frequency noise when taking a time derivative of the phase of the $(2,2)$ mode.  As a conclusion, if one has long enough highly eccentric simulations, the method introduced in this section allows one to  measure the eccentricity as a monotonically decaying function for the whole inspiral, which is a key tool to be used to construct a time domain eccentric waveform model.

\section{Hybridization of eccentric waveforms} \label{sec:PNEcc}
In the eccentric case the hybrization of the PN-NR waveforms is a challenging problem. The higher the eccentricity the stronger the interaction between the binary components at each periastron passage, which can break the post-Newtonian, weak-field and low velocity, approximation and generate a secular dephasing between both waveforms. Moreover, the lack of a general description in PN theory of eccentric black hole binary systems poses the main difficulty. 
 Therefore, we briefly review the status of the PN theory for eccentric systems in Sec. \ref{sec:ReviewPN}. In Sec. \ref{sec:Exhyb} we show an example of our procedure to hybridize eccentric PN-NR waveforms.

\subsection{Review of eccentric post-Newtonian theory} \label{sec:ReviewPN}

As far as the authors know by the time of writing this communication, the orbital averaged gravitational wave energy flux for eccentric binaries is known up to 3PN order \cite{PhysRevD.80.124018} using the 3PN QK parametrization \cite{PhysRevD.70.104011}. Our strategy consists in evolving the $3.5$PN Hamilton's equations of motion in Arnowitt, Deser, and Misner transverse-traceless (ADMTT) gauge\cite{PhysRev.116.1322,Deser:1960zzc,PhysRev.117.1595} for a point particle binary,
\begin{equation}
\frac{d \bm{X}}{dt}=\frac{\partial \mathcal{H}}{\partial \bm{P}}, \quad \frac{d \bm{P}}{dt}=-\frac{\partial \mathcal{H}}{\partial \bm{X}}+\bm{F}, \quad \frac{d \bm{S}_i}{dt}=\frac{\partial \mathcal{H}}{\partial \bm{S}_i}\times \bm{S}_i, \quad i=1,2,
\label{eq:eq10}
\end{equation}
with $\bm{X}$,$\bm{P}$, and $\bm{S}_i$ being the position, momentum and spin vectors in the center-of-mass frame, $\mathcal{H}$ the Hamiltonian described in Sec. II of \cite{PhysRevD.99.023003} and $\bm{F}$ the radiation reaction force described in \cite{Buonanno:2005xu} enhanced with the eccentric contribution to the energy flux from \cite{PhysRevD.80.124018}. The eccentric term in the flux is expressed in the QK parametrization and depends only on the orbital frequency $\omega$, which is computed while evolving the system, and the eccentricity $e_t$, for which we use its 3PN expression in terms of the orbital energy and the angular momentum of the system, which are variables computed at each time step.
 
The solution of the PN point particle equations, Eq. \eqref{eq:eq10}, can be used to compute the gravitational radiation emitted by the system. Here, the lack of general PN expressions for the waveforms of point particles evolving on quasielliptical orbits sets a strong limitation. The instantaneous terms of the waveform multipoles are  known up to 3PN order for general non-spinning systems with arbitrary eccentricity \cite{PhysRevD.91.084040}. Recently, the complete description of the 3PN nonspinning multipoles was computed including tail, tail-of-tail, and memory terms within the QK parametrization for low eccentricities \cite{Boetzel:2019nfw, PhysRevD.100.084043}. At this point using the 3PN instantaneous terms  only \cite{PhysRevD.91.084040} introduces more errors than the quadrupole order due to the missing tail and tail-of-tail terms that enter at 1.5PN, 2.5PN and 3PN orders, respectively. Additionally, the translation of the generic solution we obtain from solving Eq. \eqref{eq:eq10} to the QK form of the waveform modes in \cite{Boetzel:2019nfw, PhysRevD.100.084043} is more involved due to the fact that they split the dynamical variables into adiabatic and postadiabatic contributions. Therefore, we will restrict here to the quadrupole formula to generate the $(l,m)=(2,2)$ mode and leave for future work the generation of full 3PN waveforms, which will additionally allow us to construct multimode eccentric hybrids.

\subsection{Hybridization example} \label{sec:Exhyb}
%%%%%%%%%%%%%%%%%%%%%%%%%%$

The hybridization of PN and NR waveforms consists of determining the time shift and phase offset which minimizes the difference between both waveforms in a certain time window. This hybridization procedure is well established in the quasicircular case \cite{Bustillo:2015ova,PhysRevD.82.064016,MacDonald:2011ne,PhysRevD.87.024009,PhysRevD.84.064013}.  The time shift is usually computed by  minimizing  a suitable quantity 
that measures disagreement of the two waveforms, such as an overlap integral \cite{PhysRevD.87.024009,PhysRevD.84.064013}, or the deviation between phase or frequency of the $(2,2)$ mode \cite{Bustillo:2015ova}. However, in the eccentric case the calculation of the time shift requires alignment of the peaks due to eccentricity of both waveforms in the hybridization window.  This alignment is complicated to obtain with the phase because the peaks corresponding to each periastron passage are not very pronounced and they are difficult to estimate. One could use the frequency of the $(2,2)$ mode. However, it is a quantity obtained from a time derivative of the phase, which for NR waveforms tends to be noisy.  As a consequence, for simplicity we use the amplitude of the $(2,2)$ mode to determine the time shift of the waveform because it is a clean quantity with clearly defined peaks. We remark that aligning the oscillations of the amplitude of PN and NR waveforms in a certain hybridization window is equivalent to minimizing their difference as the maximum agreement between both quantities is obtained when they are aligned at the eccentric peaks.

As an example, we take the NR simulation with ID 60 of Table \ref{tab:tabNR3}, which is a mass ratio $q=4$ nonspinning configuration with an initial eccentricity of $e^0_\omega=0.415 \pm 0.005$ and initial orbital separation at apastron $D_0=27.5M$. We take the initial conditions of the NR simulation defined by the initial position vector, momenta (velocities in the case of \texttt{SpEc} waveforms), and dimensionless spin vectors: $\{\bm{X},\bm{P}/\bm{V},\bm{S}_{1},\bm{S}_{2}\}_{t=0}$. The fact that PN and NR coordinates for the initial data agree up to 1.5PN order \cite{Tichy:2002ec,Yunes:2006iw,PhysRevD.74.104011} makes this identification a good approximation. However, we have checked to see that the differences between the PN and NR initial conditions can produce discrepancies between the NR and PN waveforms of the order of $10 \%$.

\begin{figure}[hbt!]
\centering
\begin{minipage}[hbt!]{\columnwidth}
\captionsetup{justification=centering}
\includegraphics[scale=0.45]{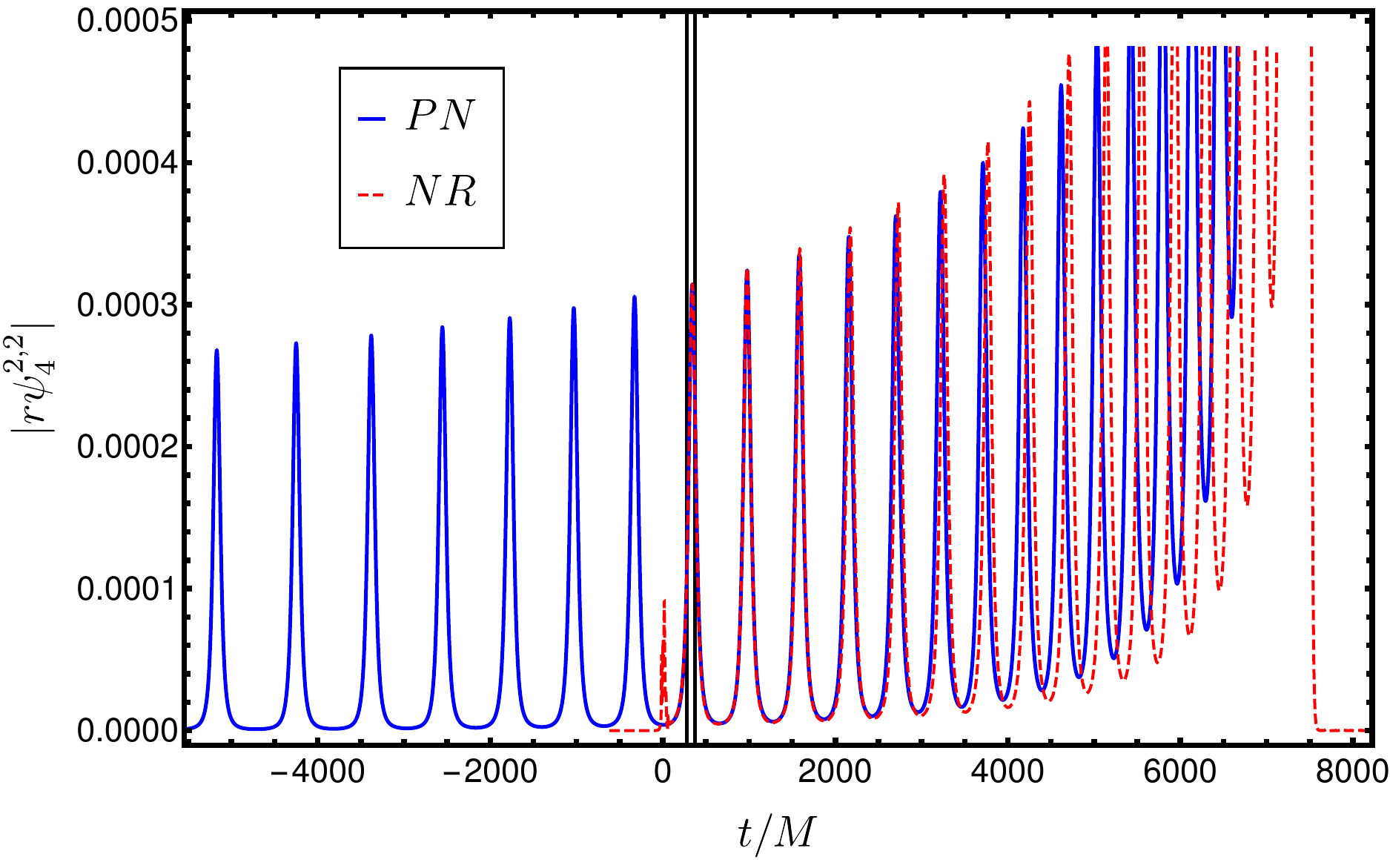}
\end{minipage}

\begin{minipage}[hbt!]{\linewidth}
\captionsetup{justification=centering}
\includegraphics[scale=0.45]{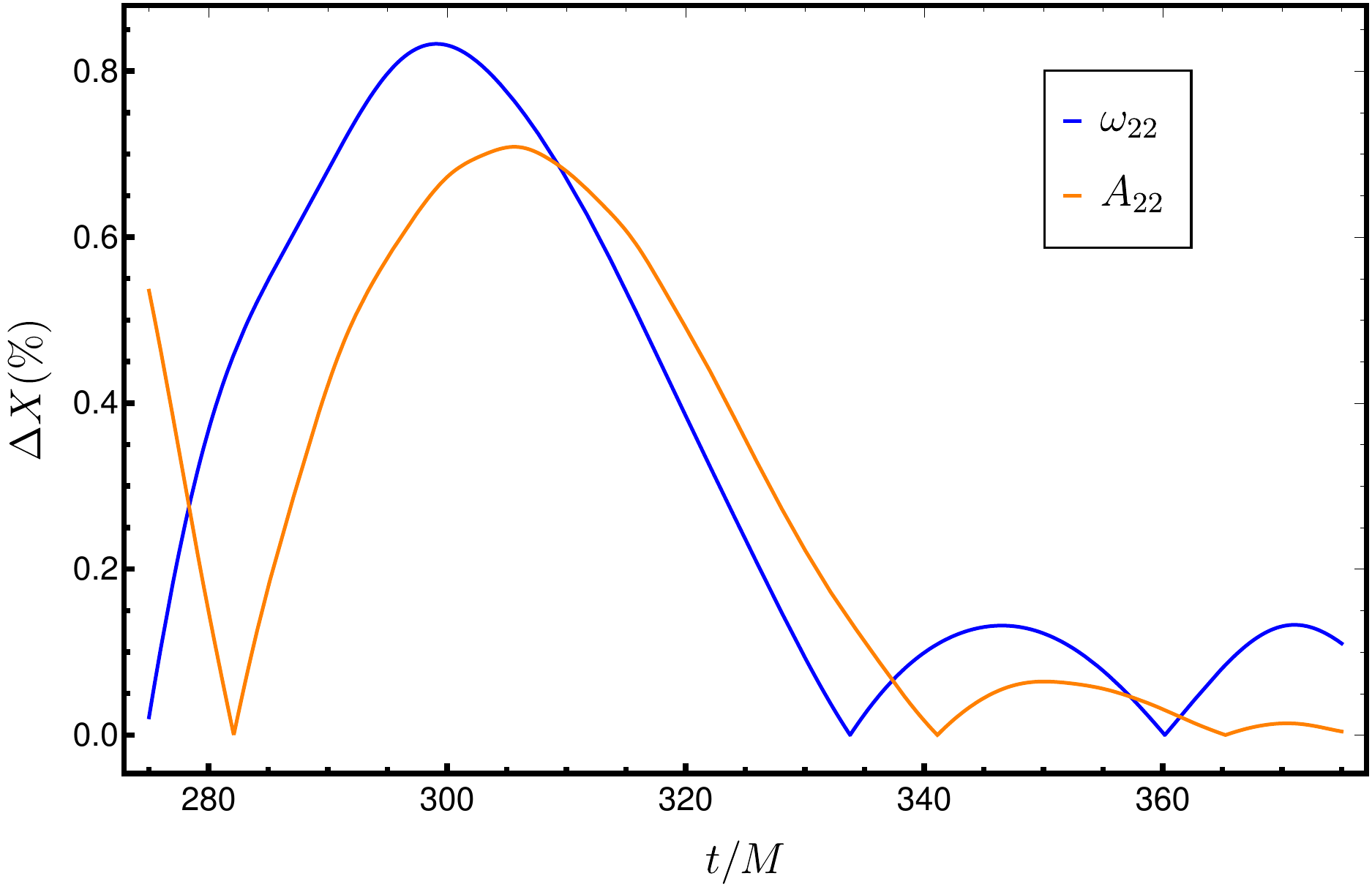}
\end{minipage}
\caption{ (Top panel) The time domain amplitude of the $| r \psi^{2,2}_{4} |$ mode. The blue thick and the red dashed curves represent the PN and NR waveforms, respectively, and the vertical black lines highlight the hybridization window. (Bottom panel) The absolute value of the relative error for the gravitational wave frequency and amplitude, $\Delta X = |X^{hyb}/X^{NR}-1| \times 100$ for $X=\omega_{22},A_{22}$, of the hybrid against the NR waveform in the matching region is displayed.}
\label{fig:HybridErrors}
\end{figure}

To leverage these differences we decide to modify the initial condition vector of the PN evolution by modifying the initial separation by a $\delta r$ such that the difference in the amplitude of the Newman-Penrose scalar, $\psi_4$, for the $(2,2)$ mode between PN and NR is minimal. In our example we obtained $\delta r = 0.08$. The outcome of such a calculation can be observed in the top panel of Fig. \ref{fig:HybridErrors}, where the time domain amplitudes of the PN and NR waveforms are shown. We do not show the full time domain range of the hybrid waveform in the top panel of  Fig. \ref{fig:HybridErrors} to better display the matching PN/NR region. The procedure is also applied to eccentric aligned-spin configurations. We find that initial highly eccentric configurations require larger $\delta r$ than low eccentric ones, and that the hybridization errors for high negative spins, where radiation reaction plays a dominant role, are 1 order of magnitude higher than for nonspinning or low spins due to the lack of expressions for PN spinning eccentric waveforms.

The procedure to construct the hybrid waveform is similar to the one presented in \cite{Bustillo:2015ova}.  We first choose the matching region to be after the junk radiation burst; in our particular case we take $t/M \in (275,375)$, which corresponds to less than one gravitational wave cycle as shown in the top panel of Fig.       \ref{fig:HybridErrors}.  To understand the choice of this short hybridization window for eccentric waveforms, we first explain the criteria for hybridizing quasicircular ones, following \cite{QCHybrids}. Quasicircular waveforms are  hybridized over several cycles as the low frequency approximant, typically EOB, is very accurate and resembles faithfully the NR behavior during the late inspiral. Furthermore, hybridization over several cycles is required to accurately compute the time alignment between waveforms by averaging out residual oscillations due to eccentricity and high frequency numerical noise coming from NR. 
In the eccentric case, the time alignment is much easier to compute as the peaks in the gravitational wave (GW) frequency ease such an alignment, so there is no need to use several cycles. Moreover, the inaccuracy of the current low frequency eccentric approximants also sets a clear limitation to faithfully reproduce the NR waveforms along several cycles. Hence, we have chosen a small hybridization window to ensure small errors in the GW amplitude and frequency between PN and NR. Choices of hybridization window including several cycles make that error increase to $10\%$ or larger depending on the case, due to the inaccuracy of the PN approximant. We have also checked to see that the election of different peaks for hybridization in the GW amplitude (in the inspiral regime)  does not significantly  change the errors, maintaining them below $1 \%$  as quoted in the lower panel of Fig. 4.

After choosing the hybridization region, we have to compute the time shift  $\tau$  and phase offset $\varphi_0$, which reduce the difference between the PN and NR waveforms in the matching window,
\begin{equation}
h^{PN}(t)= e^{i \varphi_0} h^{NR}(t+\tau).
\label{eq:eq101}
\end{equation}
To align the waveforms in time we choose $\tau$ such that it minimizes the amplitude difference along the matching window. For the phase offset  we decide to align the phases at the beginning of the window, $\varphi_0=\phi^{NR}(t_0 - \tau)-\phi^{PN}(t_0)$, where $t_0$ is the initial time of the window. Once $\tau$ and $\varphi_0$ are calculated the hybrid waveform is constructed as a piecewise function
\begin{equation}
h^{hyb}(t)=  
\begin{cases}% alignment adjustment
     e^{i \varphi_0} h^{PN}(t+\tau)  \hspace{2.95cm}   \text{ if } t< t_1 \\
     w^{-}(t)e^{i \varphi_0} h^{PN}(t+\tau)+  w^{+}(t) h^{NR}(t) \quad \text{if } t_1 < t < t_2  \\
 h^{NR}(t)    \hspace{4.12cm} \text{if }   t  > t_2
\end{cases}
\label{eq:eq11}
\end{equation}
where $t_1=275M$ and $t_2=375M$. The functions $w^{\pm}(t)$ denote the blending functions defined in the interval $\left[t_1,t_2 \right]$ that monotonically go from 0 to 1 and 1 to 0, respectively,
\begin{equation}
w^+(t)_{[t_1,t_2]}= \frac{t-t_1}{t_2-t_1}, \quad w^-(t)_{[t_1,t_2]}= 1- w^+(t).
\label{eq:eq12}
\end{equation}
 The result of the application of such a hybridization procedure can be observed in the bottom panel of Fig. \ref{fig:HybridErrors}, where the absolute value of the relative errors between the hybrid and NR amplitudes and frequencies are shown. The quantity $\Delta X$ is defined as $\Delta X= |X^{hyb}/X^{NR}-1|\times 100$ for $X=A_{22},\omega_{22}$. The errors in the gravitational wave frequency and amplitude are both below  $1 \%$, with those for the amplitude being slightly smaller due to the choice of the amplitude as the quantity with which to minimize the agreement between PN and NR waveforms. 

\begin{figure}[!]
\centering
\captionsetup{justification=centering}
\includegraphics[scale=0.45]{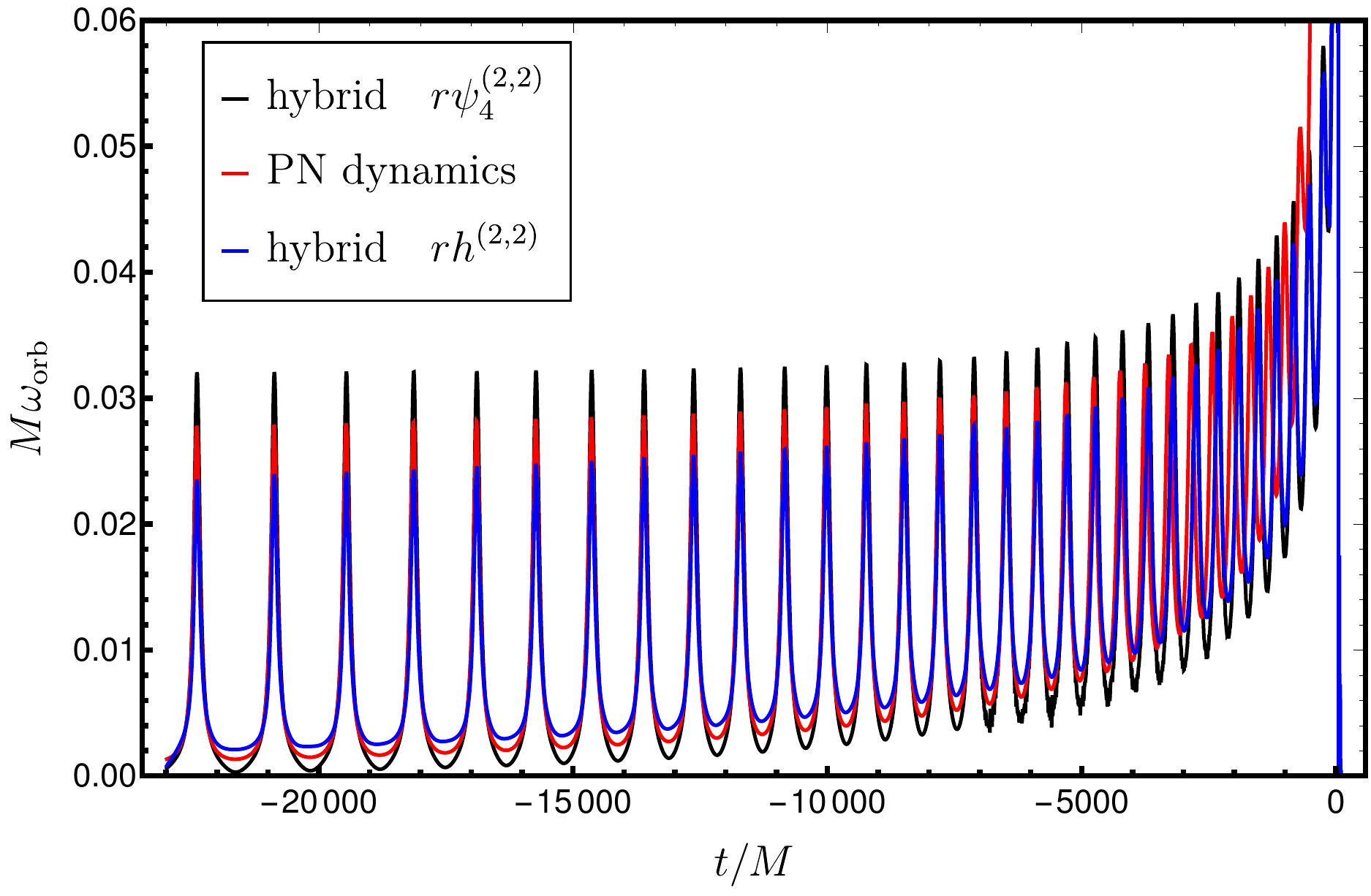}
\caption{ Time evolution of the orbital frequency $M \omega_{orb}$ computed from the phase of $r \psi^{2,2}_4$, the orbital frequency computed from the dynamics, $\omega=|\frac{\bm{v}\times \bm{r}}{r^2}|$, and from the phase of the strain $h^{2,2}$.}
\label{fig:OmegaComparisons}
\end{figure}

Finally, note that the PN waveform used to produce the hybrid is evolved backward in time from $D_0/M=27.5$ to $D_f/M=60$.  This makes the initial eccentricity increase with respect to the NR waveform. 
 Next, we explicitly show the systematics affecting the measurement of the initial eccentricity of the hybrid. We display in Fig. \ref{fig:OmegaComparisons} the time evolution of the orbital frequency for the same hybrid waveform of Fig. \ref{fig:HybridErrors},  $\omega_{orb} \approx \dot{\phi}_{22}/2$, computed from the phase of the $(2,2)$ mode of the Newman-Penrose scalar and  the strain computed using the fixed-frequency integration algorithm \cite{Reisswig:2010di}. We also compute the orbital frequency from the PN dynamics as
\begin{equation}
\omega=\left| \frac{\bm{v}\times \bm{r}}{r^2}\right|
\label{eq:eq100}
\end{equation}
where $r=|\bm{r}|$, and $\bm{v},\bm{r}$ are the velocity and the position vectors in the center-of-mass frame. The curves from Fig. \ref{fig:OmegaComparisons} indicate that the orbital frequencies computed from $\psi_4$ and $h$ overestimate and underestimate, respectively, the values of eccentricity with respect to the ones from the dynamics. This is confirmed from the values for the initial eccentricity one obtains from the orbital frequency of the strain, $\psi_4$ and the dynamics, $e_0^h=0.55\pm 0.01$, $e_0^{\psi_4}=0.84\pm 0.03$, and  $e_0^{\text{dyn}}=0.65\pm 0.01$, respectively. These three values of eccentricity are measured at the same initial time,  $t=600M$. These results lead to the conclusion that the eccentricity measured from the frequency of the $(2,2)$ mode  is higher for $\psi_4$ than for $h$. This can be understood using the fact that $h \approx \int \int \psi_4 dt' dt $; therefore, $h$ is a smoother function than $\psi_4$. As shown in Fig. \ref{fig:OmegaComparisons}, this is not a particular result of our procedure to measure the eccentricity but a general fact which can be reproduced by any method to measure the eccentricity based on the oscillations of the frequency of the $(2,2)$ mode.  We have  decided to show the orbital frequency from the PN dynamics, as it contains more cycles and eases the visualization of the effect, but the same effect can be obtained with the orbital frequency from the BH motion of a NR simulation. Moreover, we remark that these differences were also noted in  \cite{Purrer:2012wy}, where Puerrer et al. explicitly computed the factor between the eccentricity estimator calculated from the gravitational wave frequency of $h$ and $\psi_4$ in the low eccentric limit. Thus, one expects to see these discrepancies even augmented as the eccentricity increases, as is the case for the waveforms studied in this article.  
We also note that we choose not to integrate backward too long in the past of the binary due to the inaccuracy of the eccentric PN fluxes, which makes the solutions inaccurate for extremely high eccentricities and the inaccuracy of the PN expressions for the waveform, which also become more and more inaccurate for high eccentricities.

\section{Parameter estimation with eccentric signals} \label{sec:PE}
%%%%%%%%%%%%%%%%%%%%%%%%%%%%%%%%%%

In this section we employ the waveforms introduced in Secs. \ref{sec:NRcatalog} and \ref{sec:PNEcc} for data analysis studies. First, we analyze the impact of the eccentricity when computing overlaps against quasicircular models. Second, we perform parameter estimation studies injecting eccentric NR and hybrid waveforms into detector noise and compute parameter biases   using three different IMR  quasicircular models available in the LIGO libraries, \texttt{LALSUITE} \cite{lalsuite}.

\subsection{Match calculation}
A generic black hole binary evolving in a quasielliptical orbit is described by 17 parameters.  The intrinsic parameters are the individual masses of the binary $m_1$, $m_2$, the six components of the two spin vectors $\vec{S}_1$ and $\vec{S}_2$, the orbital eccentricity $e$ and the argument of the periapsis $\Omega$. The extrinsic parameters describing the sky position of the binary with respect to the detector are the distance from the detector to the source $r$, the coalescence time $t_c$, the inclination $\iota$, the azimuthal angle $\varphi$, the right ascension ($\phi$), the declination ($\theta$), and the polarization angle  ($\psi$). These parameters together describe the strain induced in a detector from a passing gravitational wave \cite{PhysRevD.47.2198};
\begin{equation}
\begin{split}
h(t,\zeta,\Theta)& =  \left[ F_+ (\theta,\phi,\psi)h_+(t-t_c;\iota,\varphi, \zeta)  \right. \\
& \left.  +  F_\times (\theta,\phi,\psi)h_\times (t-t_c;\iota,\varphi, \zeta)  \right].
\end{split}
\label{eq:eq13}
\end{equation}
 Where $\Theta=\{ t_c, r, \theta, \phi,\iota, \varphi, \psi  \}$ is the set of extrinsic parameters, $\zeta=\{m_1, m_2, \vec{S}_1, \vec{S}_2, e, \Omega \}$ are the intrinsic parameters and $F_+$, $F_\times$ are the antenna patterns functions defined in \cite{PhysRevD.47.2198} . The detector response is written in terms of the waveform polarizations $(h_+,h_\times)$ which combine to define the complex waveform strain
\begin{equation}
h(t)=h_+ - i h_\times = \sum_{l=2}^{\infty}\sum_{m=-l}^{l} Y^{-2}_{lm}(\iota, \varphi ) h_{lm} (t-t_c;\zeta),
\label{eq:eq14}
\end{equation}
 where $Y^{-2}_{lm}(\iota, \varphi )$ are spin-weighted -2 spherical harmonics, and where $h_{lm}$ refers to the $(l,m)$ waveform mode. 
The comparison between two waveforms is usually quantified by an overlap integral, which is a noise-weighted inner product between  signals \cite{Jaranowski2012}, and which can be maximized over subsets or all of the parameters of the signal. Given a real-valued detector response, the inner product between the signal $h_{resp}^S(t)$ and the model $h_{resp}^M(t)$ is defined as
\begin{equation}
	\braket{h^S_{resp}|h^M_{resp}}= 4 \text{Re} \int^{+ \infty}_{0} \frac{\tilde{h}^S_{resp}(f) \tilde{h}^{M*}_{resp}(f)}{S_n(|f|)}df,
\label{eq:eq15}
\end{equation}
where $\tilde{h}$ denotes the Fourier transform of $h$, $h^*$ denotes the complex conjugate of $h$, and $S_n(|f|)$ is the one sided noise power spectral density (PSD) of the detector.
 
 The normalized match optimized over a relative time shift and the initial orbital phase can be written as %and the polarization angle can be written as
\begin{equation}
\mathcal{M}(\iota_S,\varphi_{0_S}) = \max_{t_c, \varphi_{0_S}} \left[ 	\frac{\braket{h^S_{resp}|h^M_{resp}}}{\sqrt{\braket{h^S_{resp}|h^S_{resp}}\braket{h^M_{resp}|h^M_{resp}}}}\right] .
\label{eq:eq16}
\end{equation}
The match is close to 1 when the model is able to faithfully reproduce the signal, while values of the match close to 0 indicate large disagreement between the two waveforms.
 In Eq. \eqref{eq:eq16} the match is computed for given values of the angles $(\iota_S,\varphi_{0_S})$ of the signal and maximizing over phase and time shifts. We will take only the $h_{22}$ mode of the eccentric hybrids and a quasicircular (QC) waveform model and compute single mode mismatches maximized over a time shift $t_0$ and a phase offset $\phi_0$  as
\begin{equation}
\mathcal{M}\mathcal{M} = \max_{t_0, \phi_{0}} \left[ 	\frac{\braket{h^{\text{hyb}}_{22}|h^{\text{QC}}_{22}}}{\sqrt{\braket{h^{\text{hyb}}_{22}|h^{\text{hyb}}_{22}}\braket{h^{\text{QC}}_{22}|h^{\text{QC}}_{22}}}}\right] .
\label{eq:eq116}
\end{equation}
To simplify the comparisons, we introduce the mismatch,	$1- \mathcal{M}\mathcal{M}$. Values of the mismatch close to zero indicate good agreement between the signal and the model, while the higher the mismatch the larger the difference between both waveforms, indicating that the model is not able to accurately represent the signal.

Having set the notation for the calculation of the mismatch, we compute the mismatch between the eccentric $(2,2)$ mode hybrids computed in Sec. \ref{sec:PNEcc} and the quasicircular model PhenomX \cite{phenX, phenXHM}, which is an upgrade of the aligned-spin PhenomD model \cite{PhysRevD.93.044006,PhysRevD.93.044007}, with calibration to a larger NR dataset and also to extreme-mass ratio waveforms. We employ the Advanced LIGO' `zero detuned high power' PSD \cite{LIGOPSD} to compute the overlap in Eq. \eqref{eq:eq15}. The integral of Eq. \eqref{eq:eq15} is evaluated within a frequency range of $20$-$2000$ Hz.  The nonmonotonic behavior of the GW frequency of eccentric systems complicates the determination of the frequency range of a signal in the detector band. The ideal case would be one in which the initial apastron and periastron frequencies are below $20$Hz. This would mean that the whole waveform starts before the cutoff frequencies of the detectors and one observes the complete eccentric inspiral of the binary. Another possibility is that both frequencies are above $20$ Hz, then the signal is very short and much of the inspiral waveform is lost. Finally, it is also possible that during some part of the waveform the periastron frequencies are above $20$ Hz and apastron frequencies are below $20$ Hz. The latter is typically the case for our hybrid waveforms.
 
\begin{figure}[hbt!]
\centering
\captionsetup{justification=centering}
\includegraphics[scale=0.5]{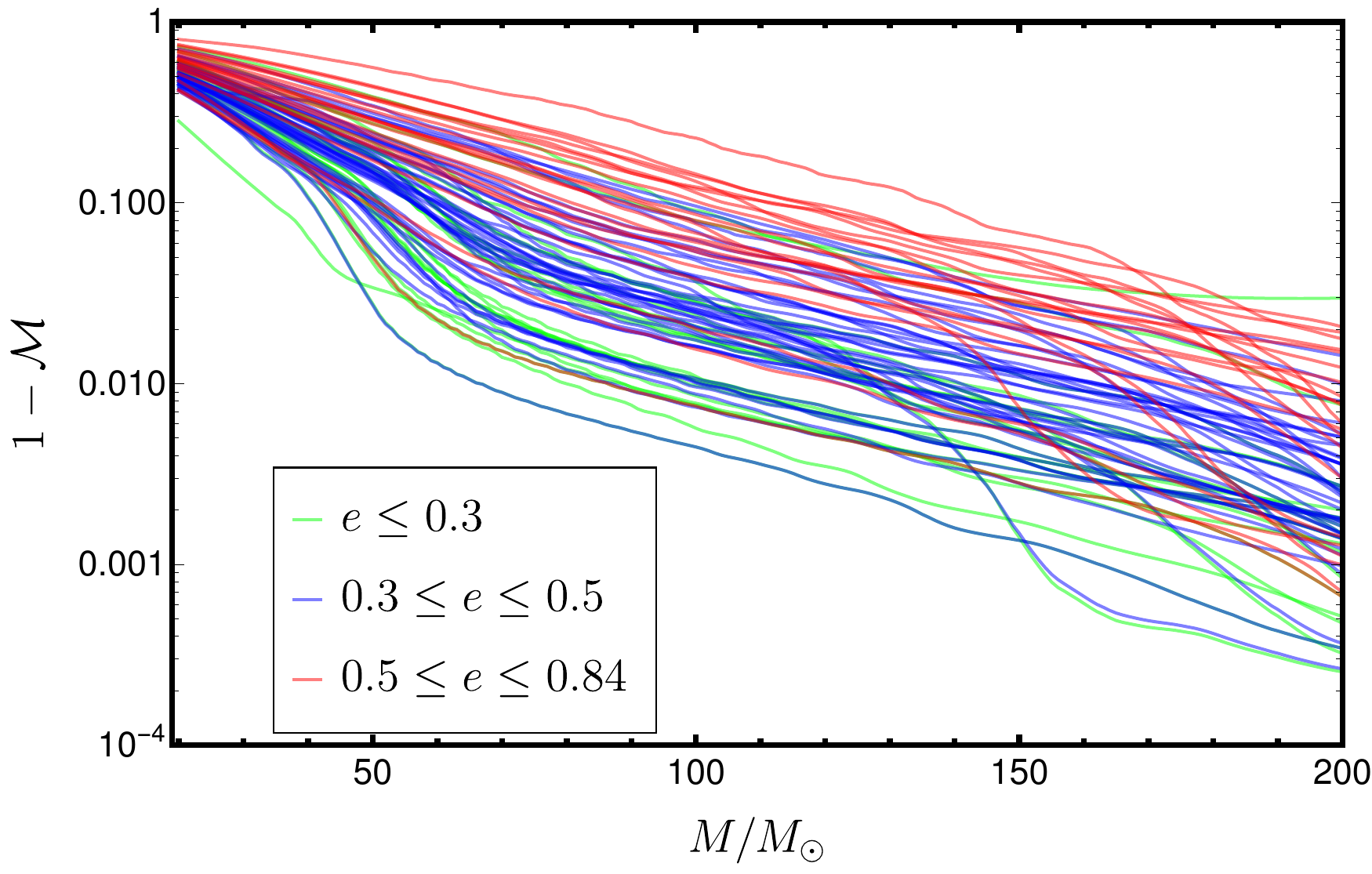}
\caption{ Mismatches for the $(l,m)=(2,2)$ mode between the eccentric hybrid waveforms corresponding to the cases presented in Table \ref{tab:tabNR3} and the quasicircular PhenomX waveform model as a function of the total mass of the system. The green, blue, black, and red lines correspond to eccentric PN-NR hybrid waveforms with initial eccentricities $e \leq 0.3$, $0.3 \leq e \leq 0.5$, and $0.5 \leq e \leq 0.84$, respectively. }
\label{fig:mmMatch22}
\end{figure}

In Fig. \ref{fig:mmMatch22} we show the single mode mismatches between the eccentric hybrids and PhenomX for a range of total mass for a system where $M_T \in\{20,200\} M_\odot$. As expected, for larger total masses of the system most of the waveform in the frequency band of the detector is in the merger and ringdown parts and the mismatches are even below the $3 \%$ threshold. This is consistent with the results obtained in Sec. \ref{sec:NRcatalogB}, which show the agreement in the final state between the eccentric simulations and the quasicircular fits. However, the lower the total mass the higher the mismatch, this is due to the fact that at low frequencies there is more inspiral part of the waveform in the frequency band, and therefore the inability of the quasicircular model to resemble the eccentric inspiral becomes notorious. One can also appreciate in Fig. \ref{fig:mmMatch22} that, generally, the higher the initial eccentricity the higher the mismatch for the whole mass range.

\subsection{Eccentric injections into detector noise}
In this section we show some applications of the eccentric waveform dataset to parameter estimation. We inject eccentric hybrids into Gaussian noise realization recolored to match the spectral density of the Advanced LIGO and Virgo detectors at design sensitivity. All simulations with the same injected signal are performed with the same noise realization.  We do not address the challenge of detecting eccentric signals and instead assume that the signal has been detected by standard CBC search pipelines \cite{PhysRevD.81.024007}. We perform parameter estimation using the python-based \texttt{BILBY} code \cite{Ashton:2018jfp}. Of the numerous stochastic samplers implemented in \texttt{BILBY}, we employ the nested sampler \texttt{CPNEST} \cite{JohnVeitchCpnest} and use waveform approximants implemented in \texttt{LALSUITE} \cite{lalsuite} as the model templates. The eccentric hybrids make use of the numerical relativity injection infrastructure \cite{Galley:2016mvy,Schmidt:2017btt}.

State-of-the-art non-spinning eccentric IMR models \cite{PhysRevD.97.024031,PhysRevD.98.044015,PhysRevD.96.044028,PhysRevD.96.104048} have not yet been implemented in \texttt{LALSUITE}. The only eccentric waveform models in \texttt{LALSUITE} are inspiral nonspinning frequency domain approximants \cite{PhysRevLett.112.101101,PhysRevD.90.084016,PhysRevD.93.064031}. We have decided not to use such inspiral waveform models to avoid bias induced by the sharp cutoff at the end of the waveform \cite{Mandel:2014tca}. For a study of the eccentricity measurement using such inspiral approximants, see \cite{PhysRevD.98.083028,Moore:2019vjj}. We restrict to IMR quasicircular approximants and perform parameter estimation analysis on the injected eccentric signals by sampling in the 15 parameters of a quasicircular black hole binary.

\begin{table}[h!]
\begin{center}
%\resizebox{10.cm}{!}{
 \def\arraystretch{1.3 }
\begin{tabular}{c c c c c c c c c c c c c c c c }
\hline  
\hline
ID &  & Simulation &  &    q  & &${\chi}_{1,z}$ & & $\chi_{2,z}$ &    &$D/M$ & &  $e_\omega \pm \delta e_\omega$  \\
%\hline
\hline
61&   & \texttt{SXS:BBH:1355}&  &     1& & $0$ & & $0$ & &   12.97      & &   $0.090 \pm 0.003$ \\
%\hline
62&   & \texttt{SXS:BBH:1359}&  &    1&  &$0$ &  &$0$  & &    15.73 	  & &   $0.146 \pm 0.003$ \\
%\hline
63&   & \texttt{SXS:BBH:1361}&  &   1&  &$0$ & & $0$   & &    16.69 	  & &   $0.209 \pm 0.003$ \\
%\hline
\hline
\hline
\end{tabular}
% }
\end{center}
\caption{Summary of the injected  NR simulations. The first column denotes the identifier of the simulation, and the second column indicates the name of the simulation as presented in \cite{PhysRevD.98.044015}. The next columns show the mass ratio, the z component of the dimensionless spin vectors, the initial orbital separation and the initial orbital eccentricity measured using the procedure detailed in Sec. \ref{sec:NRcatalogC}.}
\label{tab:tabNRPE}
\end{table}

We inject three  NR equal mass nonspinning simulations described in Table \ref{tab:tabNRPE} into a network of gravitational wave detectors composed of the LIGO-Hanford, LIGO-Livingston \cite{TheLIGOScientific:2014jea}, and Virgo interferometers \cite{TheVirgo:2014hva}, each operating at design sensitivity. We set a reference frequency of $f_{\text{ref}}=20 \text{ Hz}$, where the waveforms start. Some injected parameters are displayed in Table \ref{tab:tabNRbias}, while the declination is $\delta=-1.21$ rad, the right ascension $\alpha=1.37$ rad, and the coalescence phase $\phi=0$ rad. From these simulations, the $\{(l,m\}=\{(2, \pm 2), (3,\pm 2), (4, \pm 4), (5, \pm 4), (6, \pm 6)\}$ modes are used. We do not inject odd m modes because they are zero by symmetry. For the injected signal we choose the luminosity distance $D_L=430$ Mpc, which is similar to the first detection of a gravitational wave signal, GW150914 \cite{LIGOScientific:2018mvr}, which produces a high network signal-to-noise ratio (SNR) as shown in Tables \ref{tab:tabNRbias} and \ref{tab:tabHybridbias}.

We employ a uniform-in-volume prior on the luminosity distance, $p(D_L|H)\propto D_L^2 $,  between 50 and 1500 Mpc.  The inclination and polarization angles  both  have uniform priors between $(0,\pi)$. We use the standard priors for the extrinsic variables, as in Table I of \cite{Ashton:2018jfp}. Instead of sampling in the component masses, we sample in   mass ratio q and   chirp mass $\mathcal{M}_c$, with ranges $(0.05,1)$ and $(15, 60) M_\odot$, respectively. The spin priors are set differently according to the approximant. If the approximant is nonprecessing, we set the option of \texttt{aligned\_spin=True} in the \texttt{BBHPriorDict} function of \texttt{BILBY}, which samples in the dimensionless spin z components between -0.8 and +0.8. For precessing approximants, we sample in the tilt angles $(\theta_1,\theta_2)$, the angle between the spin vectors $\phi_{12}$, the angle between \textbf{J} and \textbf{L} $\phi_{JL}$, and the dimensionless spin magnitudes $(a_1,a_2)$. The priors for  $a_1,a_2,\theta_1,\theta_2,\phi_{JL}$, and $\phi_{12}$ are the same as in Table I of \cite{Ashton:2018jfp}.
We also define a uniform prior for the coalescence time of 2s centered at the injection time.

We take three quasicircular models as approximants: 1) IMRPhenomD \cite{PhysRevD.93.044006,PhysRevD.93.044007}, a nonprecessing model with only the $(2,\pm 2)$ modes, 2)  IMRPhenomHM \cite{london2018},  nonprecessing model including higher order modes, and 3) IMRPhenomPv2 \cite{phenomp},  an effective precessing model.

\begin{figure*}[!]

\noindent\begin{minipage}{\textwidth}

%A
\noindent\begin{minipage}[h]{.45\textwidth}
\includegraphics[scale=0.42]{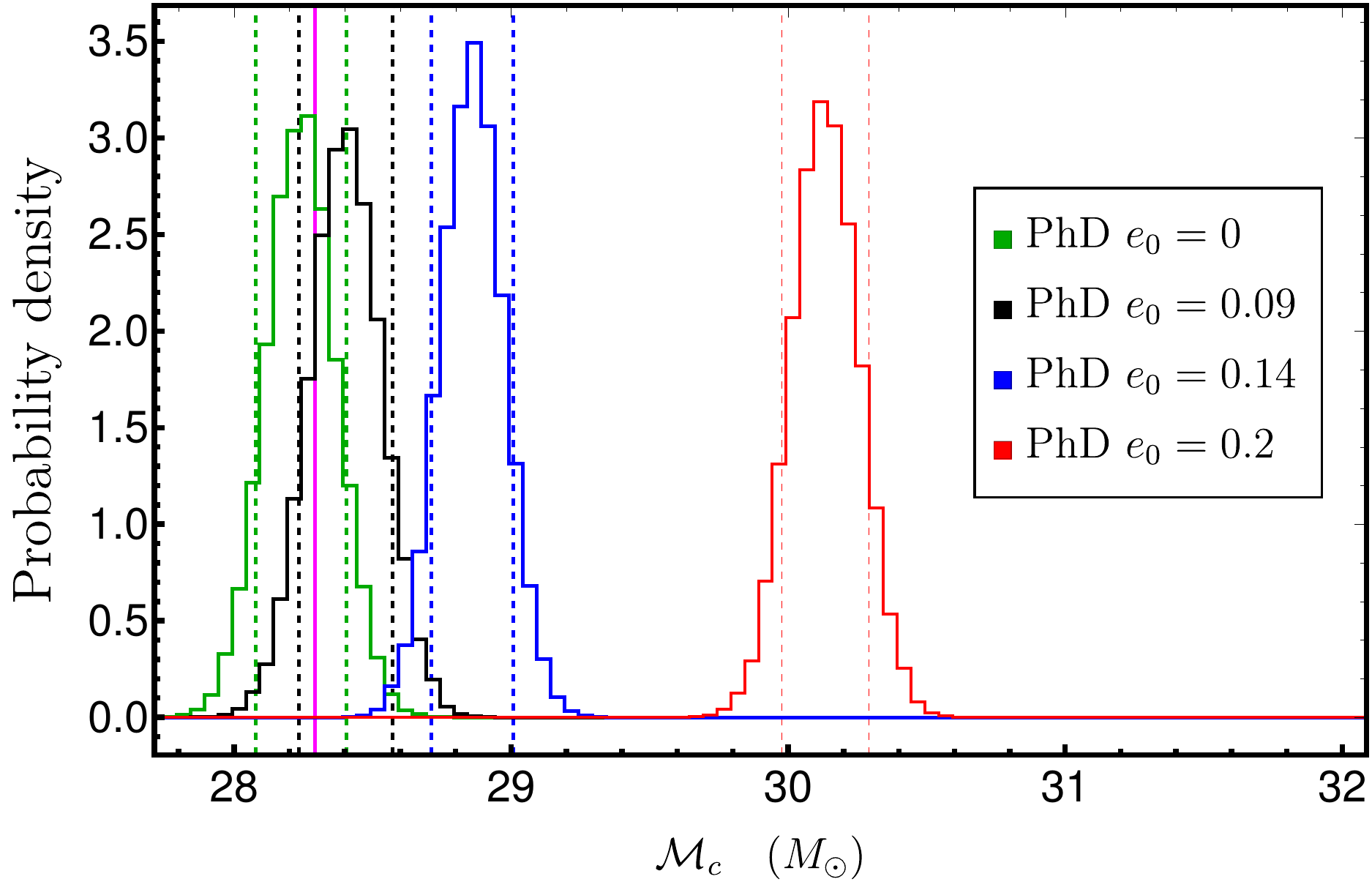}
\end{minipage} 
%\hfill
\hspace{1cm}
\begin{minipage}[h]{.45\textwidth}
\includegraphics[scale=0.42]{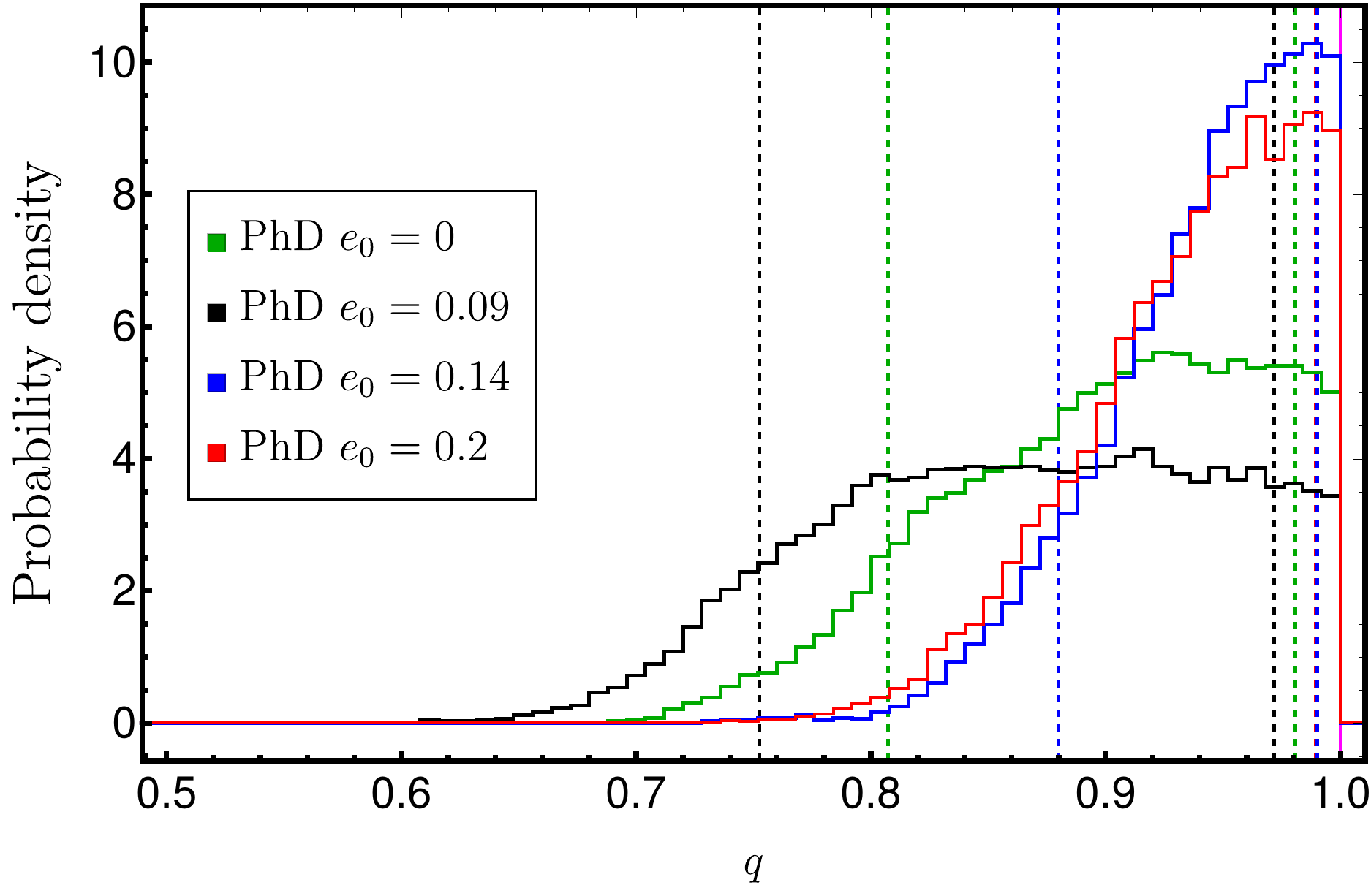}
\end{minipage}

\end{minipage}

%\vspace{5ex}

%B
\noindent\begin{minipage}{\textwidth}

\noindent\begin{minipage}[h]{.45\textwidth}
\includegraphics[scale=0.42]{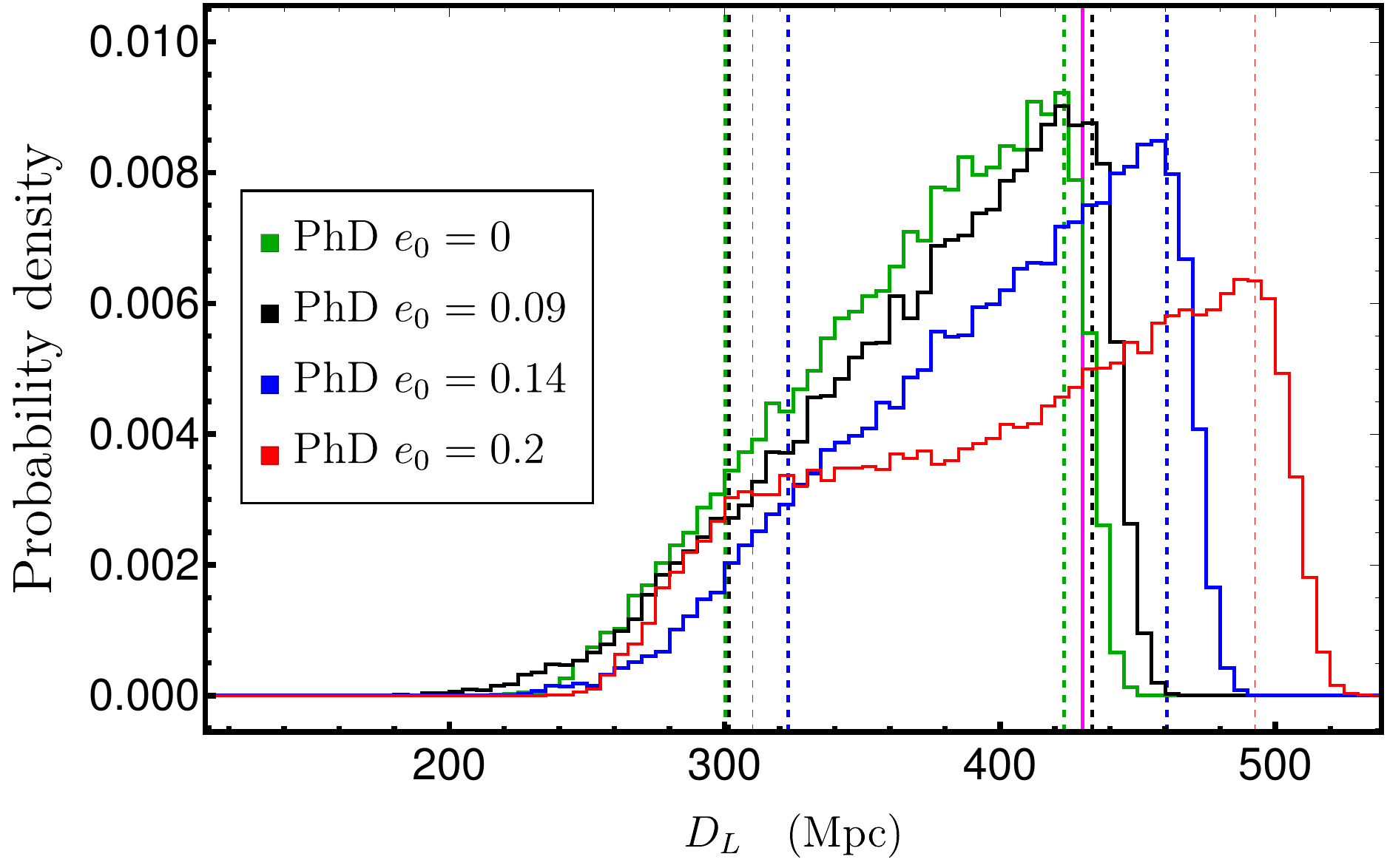}
\end{minipage} 
%\hfill
\hspace{1cm}
\begin{minipage}[h]{.45\textwidth}
\includegraphics[scale=0.42]{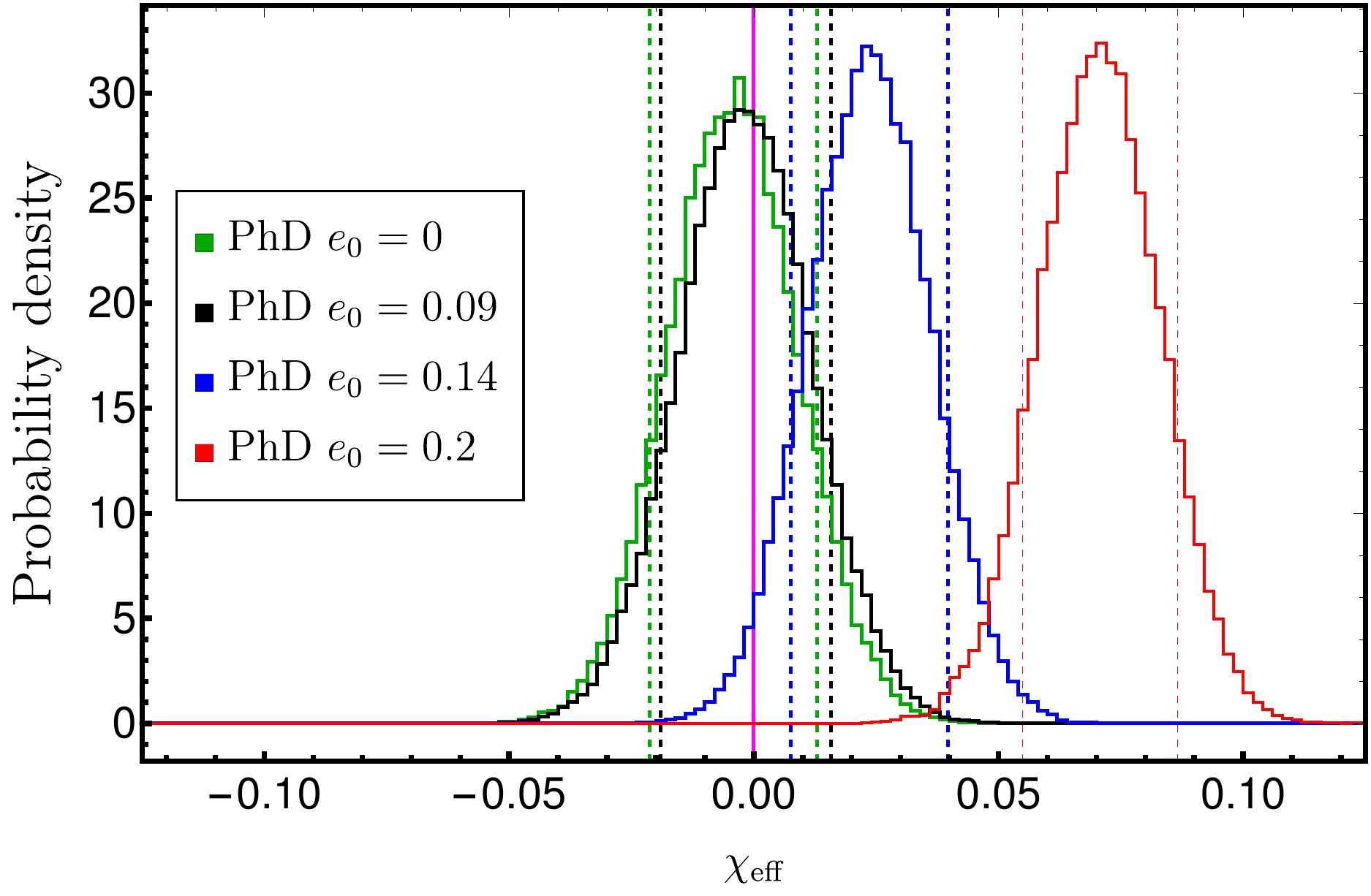}
\end{minipage}

\end{minipage}

\caption{Posterior probability distributions for the injected NR simulations of Table \ref{tab:tabNRPE} and a zero-eccentricity injection using the NRHybSur3dq8 model. The vertical dashed lines correspond to $90 \%$ credible regions. The magenta thick vertical line represents the injected value. The green, black, blue and red curves represent distributions sampled using the IMRPhenomD approximant with injected initial eccentricities $e_0= 0.0 ,0.09,0.14,0.2$, respectively.  }

\label{fig:posteriorsNRPhD}
\end{figure*}

We plot the posterior probability distribution for the chirp mass, mass ratio, effective spin parameter and luminosity distance for the PhenomD approximant in Fig. \ref{fig:posteriorsNRPhD} with $90 \%$ credible intervals specified by the dashed vertical lines and the injected values by the  thick vertical magenta lines. As a control case, we also show in Fig. \ref{fig:posteriorsNRPhD} the posterior distribution of an equal mass nonspinning zero-eccentricity injection performed using the hybridized surrogate model NRHybSur3dq8 \cite{PhysRevD.99.064045} with the same injected parameters  as in Table \ref{tab:tabNRbias} and recovered with the PhenomD model.  The NRHybSur3dq8 injected waveform contains all higher order modes up to $l=4$, which in this case seems to cause the small bias one observes in the luminosity distance when recovering with the PhenomD model, which contains only the $(2,|2|)$ modes. For the rest of the parameters, like the mass ratio, the chirp mass and the effective spin parameter, we obtain results consistent with the accuracy of the PhenomD model for parameter estimation of injected signals as shown in \cite{phenX}.

\begin{figure*}[!]

\noindent\begin{minipage}{\textwidth}

%A
\noindent\begin{minipage}[h]{.45\textwidth}
\includegraphics[scale=0.42]{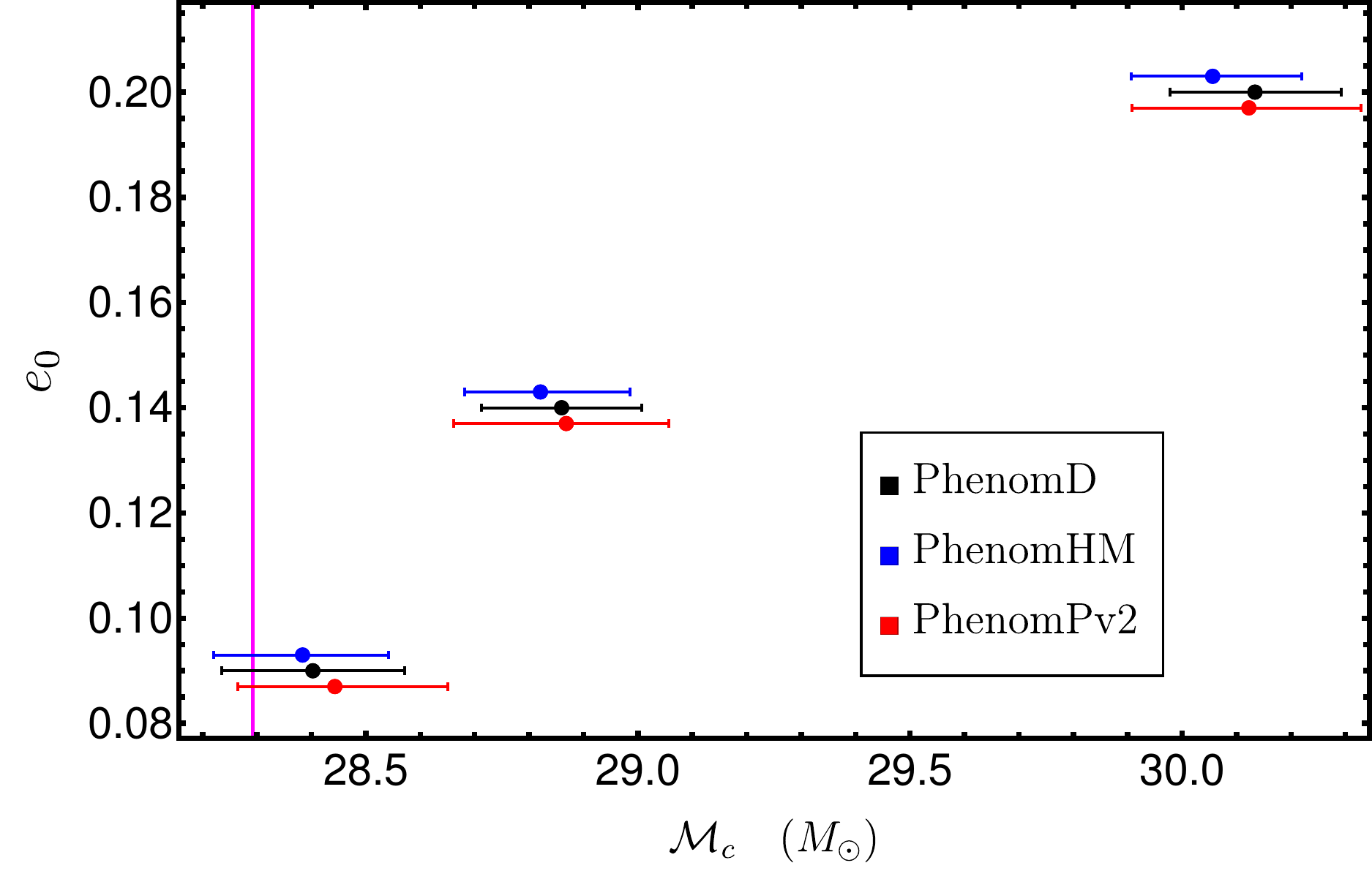}
\end{minipage} 
%\hfill
\hspace{1cm}
\begin{minipage}[h]{.45\textwidth}
\includegraphics[scale=0.42]{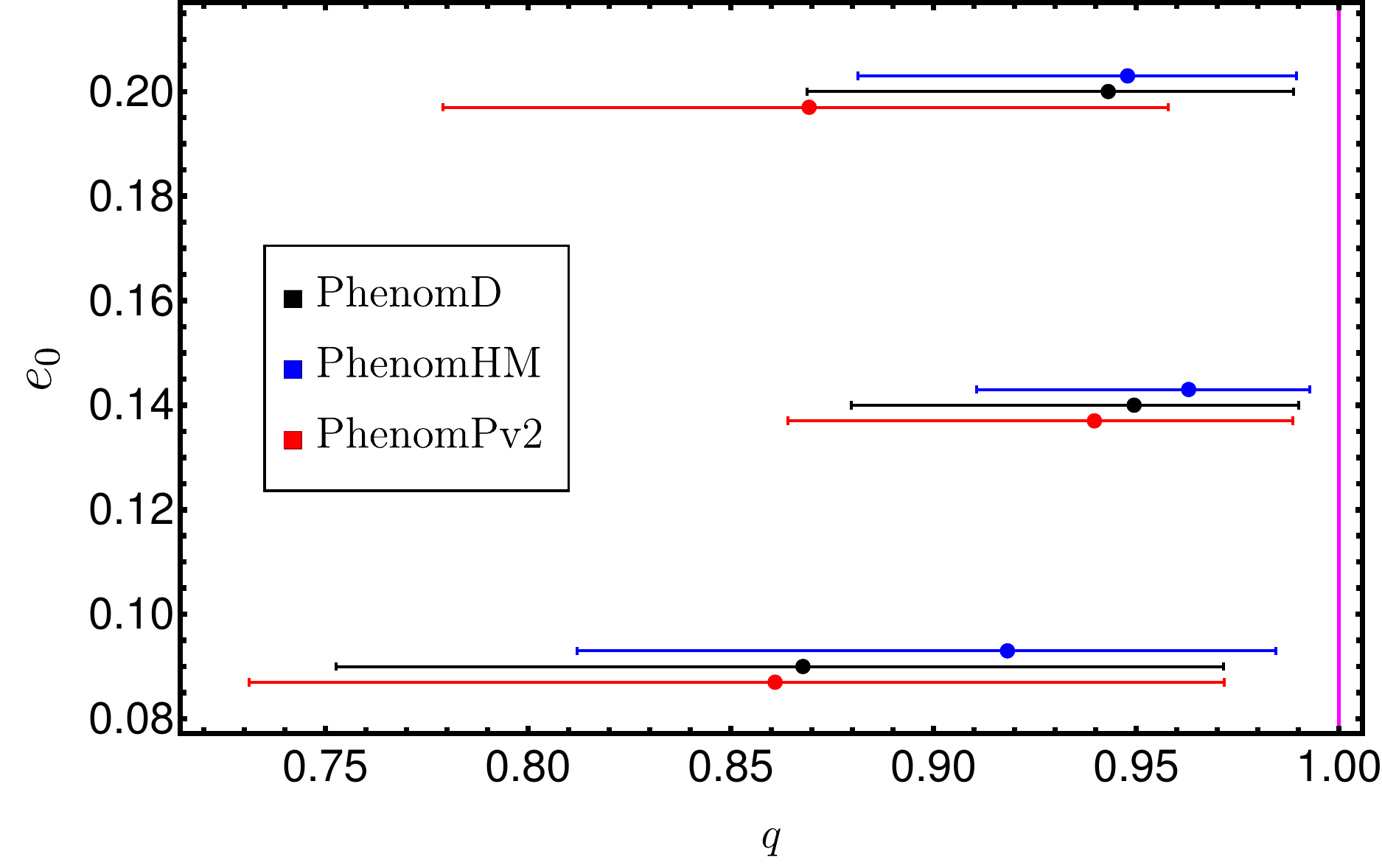}
\end{minipage}

\end{minipage}

%\vspace{5ex}

%B
\noindent\begin{minipage}{\textwidth}

\noindent\begin{minipage}[h]{.45\textwidth}
\includegraphics[scale=0.42]{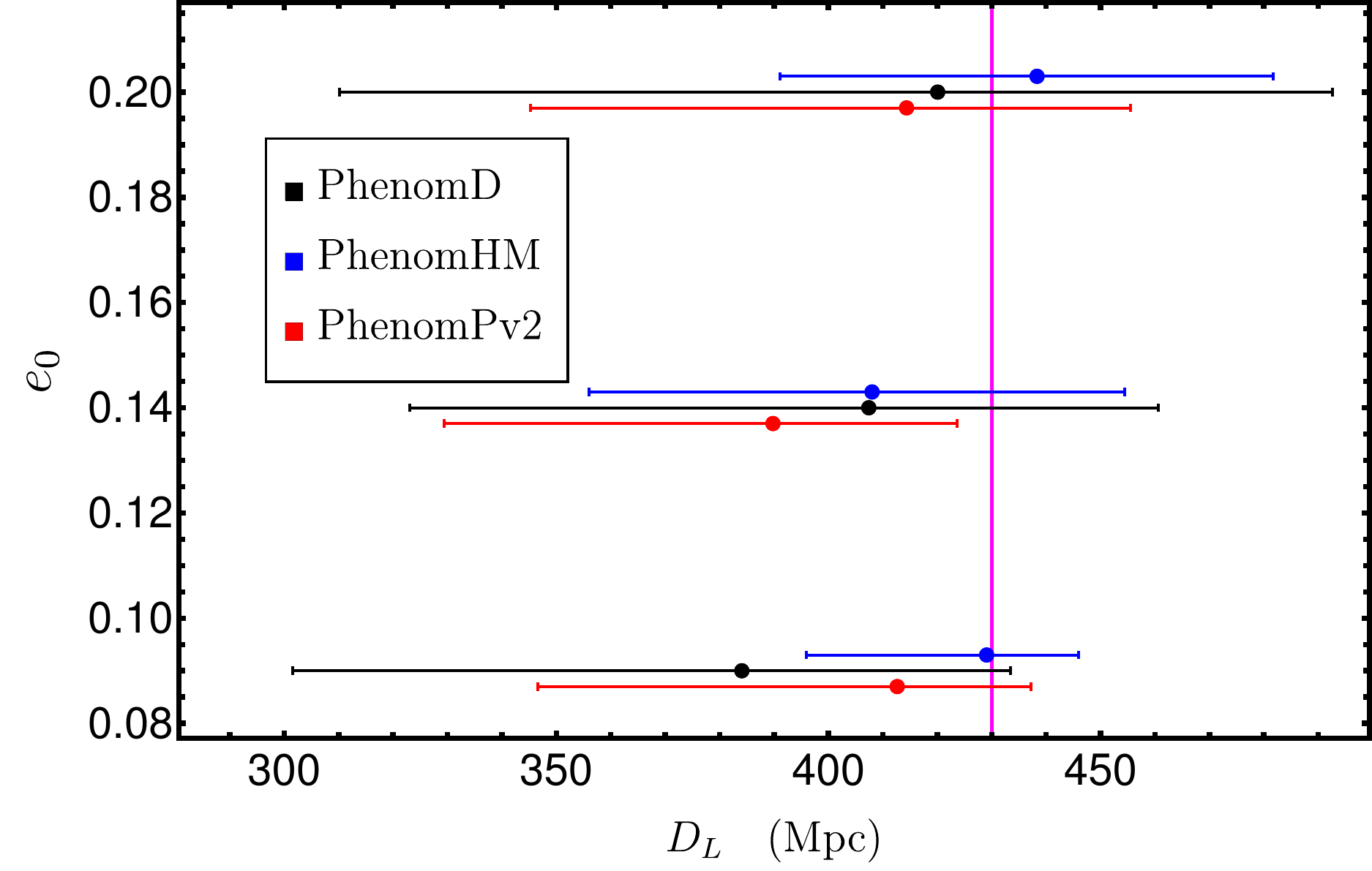}
\end{minipage} 
%\hfill
\hspace{1cm}
\begin{minipage}[h]{.45\textwidth}
\includegraphics[scale=0.42]{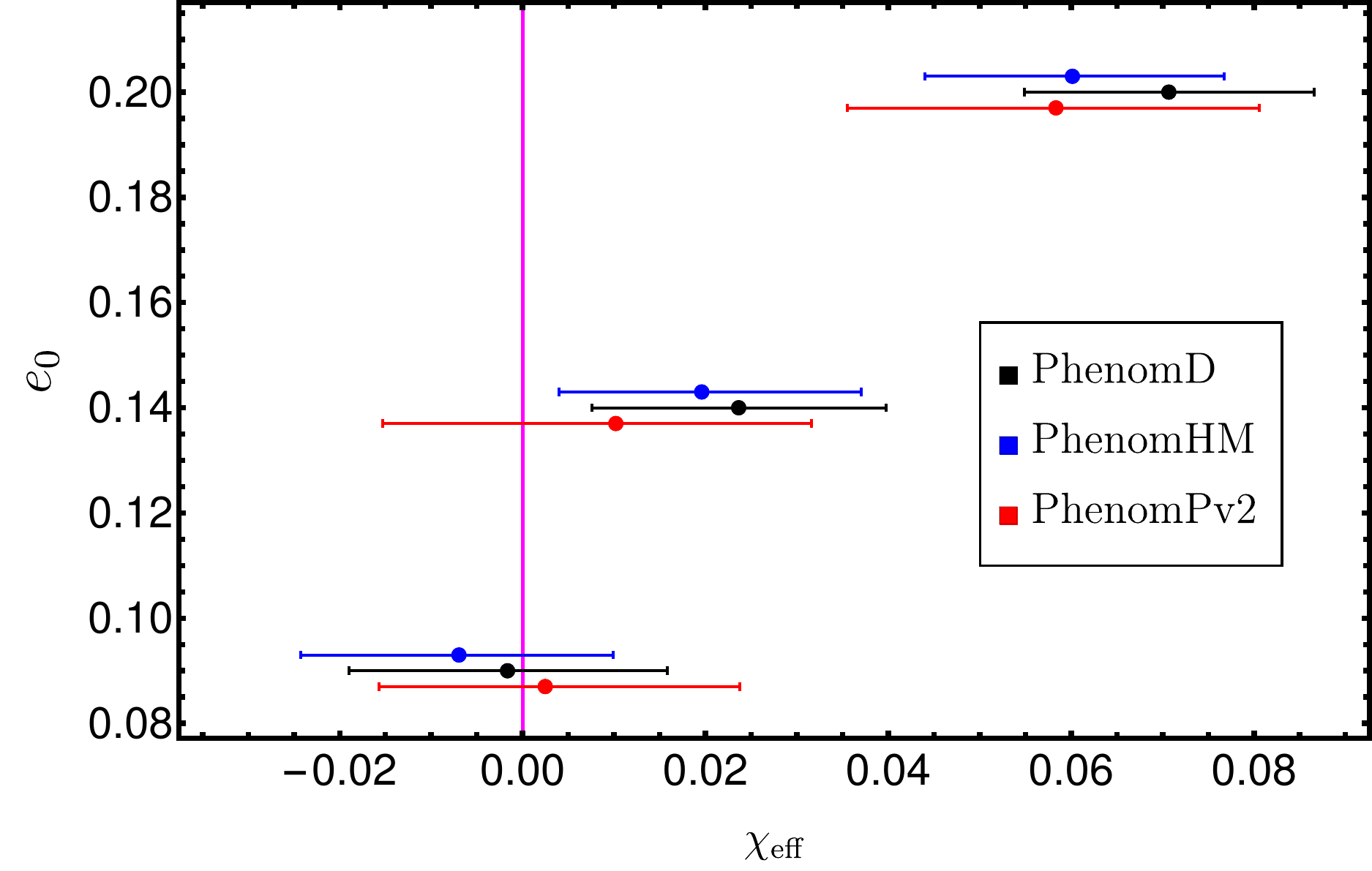}
\end{minipage}

\end{minipage}

\caption{Median values and error bars corresponding to $90 \%$ credible regions  of the posterior probabibility distributions for the injected NR simulations of Table \ref{tab:tabNRPE}. The vertical magenta line represents the injected value. The black, blue and red segments represent the median values and errors bars of the distributions sampled using the IMRPhenomD, IMRPhenomHM and IMRPhenomPv2 approximants, respectively. The cases are represented for three initial eccentricities of the injected signal, $e_0=0.09,0.14,0.2$. To ease the visualization of the horizontal bars, cases with the same initial eccentricity and run with different approximants have been separated a $\Delta e =0.003$.}

\label{fig:errorbarsPE}
\end{figure*}

The posterior distributions for the rest of the approximants are shown in Fig. \ref{fig:posteriorsNR}.  The same information is summarized in Fig. \ref{fig:errorbarsPE}, where the median and the error bars corresponding to the $90 \%$ credible intervals of the posterior distribution are shown as a function of the initial eccentricity. Note that the bars corresponding to the same initial eccentricity but different approximants have been separated by a small amount to ease the visualization of the results. For the lowest initial eccentricity, $e_0=0.09$, the results for the four quantities are pretty different. The chirp mass and the effective spin parameter produce similar distributions for the three approximants, while for the mass ratio and the luminosity distance, PhenomHM distributions are closer to the injected values than PhenomD and PhenomPv2. 

Furthermore, for $e_0=0.14$ and $e_0=0.2$ we observe increasingly poor agreement with the injected values, except for the mass ratio where the lowest initial eccentricity signal produces wider distributions than those with higher initial eccentricities. This can also be checked in Table \ref{tab:tabNRbias}, where the recovered parameters, median values and $90 \%$ credible intervals, are compared with the injected values. Regarding the effective spin parameter and the chirp mass, the increase of initial eccentricity in the injected signal shifts the posteriors for the three quasicircular models, while for the mass ratio the increase of initial eccentricity reduces the bias on the measurement of the mass ratio, probably as a consequence of the shift in the chirp mass distribution, as displayed in Fig. \ref{fig:q1PhDcontourPlot}  for the all injections recovered with the PhenomD model.

One observes also that PhenomHM recovers the injected parameters more effectively than PhenomD and PhenomPv2. For the luminosity distance the probability densities tend to flatten and be closer to the prior distributions for high initial eccentricities. One notes again that PhenomHM has less parameter bias than PhenomD and PhenomPv2. Injected values of the sky position like the right ascension $\alpha=1.375$ rad and  $\delta=-1.21$ rad are well recovered for all nine runs, probably due to the expensive parameter estimation (PE) settings described in Appendix \ref{sec:AppendixC} : $\alpha=1.37^{+0.01}_{-0.01}$ rad and $\delta=-1.21^{+0.01}_{-0.01}$ rad.

Furthermore, we have computed the recovered matched-filter SNR for the detector network, $\rho_{\text{Match}}$, for each simulation. This quantity,  $\rho_{\text{Match}}$, is computed calculating the matched filter between the detector data with the eccentric signal injected and the waveform of the approximant waveform model with the parameters corresponding to the highest log-likelihood value of the posterior distribution. The results of such a calculation are shown in Table \ref{tab:tabNRPE}. As expected, we observe that the zero-eccentricity injection recovers much more SNR than the eccentric injections, with decreasing values of the recovered SNR with increasing eccentricity.

Additionally, we display the values of the log Bayes factor for each simulation. The Bayes factor is computed here as the ratio between the signal and null evidence [see Eq. (13) of \cite{Thrane_2019}]. One can observe that both the recovered matched-filter SNR and the log Bayes factor decrease more the higher the initial eccentricity of the injected signal is. The matched-filter SNR produces similar values between models for simulations with the same initial eccentricity. However, the log Bayes factor tends to be slightly higher for the aligned-spin waveform models, PhenomD and PhenomHM, for the lowest initial eccentric injected signal, while for higher initial eccentricities the precessing model IMRPhenomPv2 shows slightly greater log Bayes factors than the aligned-spin ones. The highest log Bayes factor is obtained for the zero-eccentricity injection.

\begin{table*}
\begin{center}
%\resizebox{13.cm}{!}{
 \def\arraystretch{1.4}
\begin{tabular}{  c  c  c   c   c  c  c  c   c  c c c c }
\hline
\hline
$e_0$& Model &  $m_1/M_\odot$&  $m_2/M_\odot$& $\mathcal{M}_c/M_\odot$&  $q$ &     $D_L/$Mpc&  $\chi_{\text{eff}}$ &$\psi$ (rad) &$\iota$ (rad)& $\rho_{\text{Match}}$  & $\log \mathcal{B}$ \\ %& $1-\mathcal{M}$ &$\rho_{\text{Net}}$ \\
\hline
\multirow{3}{*}{$0.09$} & PhenomD   &  $35.06^{+2.55}_{-1.92}$&  $31.29^{+1.10}_{-1.33}$ &   $28.40^{+0.17}_{-0.17}$&  $0.87_{-0.12}^{+0.10}$&   $384^{+49}_{-82}$ &  $0.00^{+0.02}_{-0.02}$  & $1.60^{+1.28}_{-1.34}$  &$0.54^{+0.32}_{-0.32}$    & 89.40  & 3463.79 \\ % & 0.013 & \multirow{3}{*}{$85.11$}\\
%\cline{3-10}
   & PhenomHM   &  $34.05^{+2.14}_{-1.16}$&  $31.79^{+0.69}_{-1.17}$ &   $28.38^{+0.16}_{-0.16}$&  $0.92_{-0.11}^{+0.07}$&   $429^{+16}_{-33}$ &  $-0.01^{+0.02}_{-0.02}$ & $2.01^{+0.97}_{-1.82}$ & $0.28^{+0.21}_{-0.17}$    & 89.28 & 3463.78 \\ % & 0.017  & \\
   & PhenomPv2   &  $35.26^{+2.97}_{-2.06}$&  $31.28^{+1.18}_{-1.53}$ &   $28.44^{+0.21}_{-0.18}$&  $0.86^{+0.11}_{-0.13}$&   $412^{+24}_{-66}$ &  $0.00^{+0.02}_{-0.02}$   & $1.65^{+1.22}_{-1.32}$& $0.39^{+0.32}_{-0.22}$   & 89.19  & 3459.54\\ %&  0.013  & \\
 \hline

\multirow{3}{*}{$0.14$} & PhenomD   &  $34.03^{+1.34}_{-0.72}$&  $32.63^{+0.44}_{-0.73}$ &   $28.86^{+0.15}_{-0.15}$&  $0.95^{+0.04}_{-0.07}$&   $407^{+53}_{-84}$ &  $0.02^{+0.02}_{-0.02}$ & $1.58^{+1.21}_{-1.23}$ & $0.54^{+0.32}_{-0.32}$   & 84.87   & 3288.25 \\ %& 0.053 & \multirow{3}{*}{$85.58$} \\
%\cline{3-10}
   & PhenomHM   &  $33.76^{+0.96}_{-0.54}$&  $32.73^{+0.35}_{-0.56}$ &   $28.82^{+0.16}_{-0.14}$&  $0.96^{+0.03}_{-0.05}$&   $408^{+46}_{-52}$ &  $0.02^{+0.02}_{-0.02}$  & $1.91^{+0.46}_{-0.58}$ & $0.54^{+0.19}_{-0.25}$     & 84.74  & 3283.61 \\ %& 0.053  & \\
   & PhenomPv2   &  $34.22^{+1.48}_{-0.89}$&  $32.54^{+0.54}_{-0.82}$ &   $28.87^{+0.19}_{-0.21}$&  $0.94^{+0.05}_{-0.08}$&   $389^{+33}_{-60}$ &  $0.01^{+0.02}_{-0.03}$  & $1.70^{+1.09}_{-1.07}$  &$0.64^{+0.25}_{-0.18}$ & 85.08  & 3302.37 \\ %& 0.048  & \\
 %\clineB{1-12}{3}
 \hline
 
\multirow{3}{*}{$0.2$} & PhenomD   &  $35.65^{+1.52}_{-0.85}$&  $34.01^{+0.51}_{-0.82}$ &   $30.13^{+0.16}_{-0.16}$&  $0.94^{+0.05}_{-0.07}$&   $420^{+72}_{-109}$ &  $0.07^{+0.02}_{-0.02}$ & $1.57^{+1.36}_{-1.18}$  & $0.61^{+0.41}_{-0.37}$   &   81.88  & 3102.70 \\ %& 0.090 & \multirow{3}{*}{$86.27$}\\
%\cline{3-10}
   & PhenomHM    &  $35.47^{+1.36}_{-0.78}$&  $33.97^{+0.46}_{-0.72}$ &   $30.06^{+0.16}_{-0.15}$&  $0.95^{+0.04}_{-0.07}$&   $438^{+43}_{-47}$ &  $0.06^{+0.02}_{-0.02}$ & $0.42^{+0.90}_{-0.29}$ & $0.54^{+0.16}_{-0.20}$    & 81.97  &  3101.79 \\ %& 0.091  & \\
   & PhenomPv2   &  $37.13^{+2.11}_{-1.76}$&  $33.20^{+1.00}_{-1.12}$ &   $30.12^{+0.21}_{-0.22}$&  $0.87^{0.09}_{-0.09}$&   $414^{+41}_{-69}$ &  $0.06^{+0.02}_{-0.02}$  & $1.62^{+0.98}_{-1.29}$ & $0.66^{+0.25}_{-0.18}$ &  82.05  & 3112.97  \\ %& 0.088  &  \\
 %\clineB{1-12}{3}
 \hline
0 & PhenomD &  $34.07^{+2.05}_{-1.29}$&  $30.91^{+1.24}_{-1.77}$ &   $28.24^{+0.16}_{-0.16}$&  $0.91^{+0.07}_{-0.10}$&   $375^{+48}_{-75}$ &  $0.0^{+0.02}_{-0.02}$ & $1.58^{+1.28}_{-1.24}$  & $0.53^{+0.33}_{-0.32}$   &  173.16   &  3632.19 \\
 \hline
& Injected &  $32.5$&  $32.5$& $28.29$&  $1$ &      430 &  $0$ &$0.33$  & $0.3$ & \\
%\hline
\hline
\hline
 \end{tabular}
 %}
\end{center}
\caption{Black hole binary recovered parameters for the three NR simulations from Table \ref{tab:tabNRPE}. The last row corresponds to the injected parameters. In the penultimate row we show the recovered parameters of the zero-eccentricity injection performed with the NRHybSur3dq8 model. The first column describes the initial eccentricity of the injected signal. Then we specify the approximant, the component masses, the chirp mass, the mass ratio, the luminosity distance, the effective spin parameter, the polarization angle, the inclination, the recovered matched-filter SNR for the detector network, and the log of the Bayes factor.}
\label{tab:tabNRbias}
\end{table*}

%\begin{widetext}

%\end{widetext}

%\onecolumngrid
%\begin{widetext}

\begin{figure*}[!]

\noindent\begin{minipage}{\textwidth}

%A
\noindent\begin{minipage}[h]{.45\textwidth}
\includegraphics[scale=0.42]{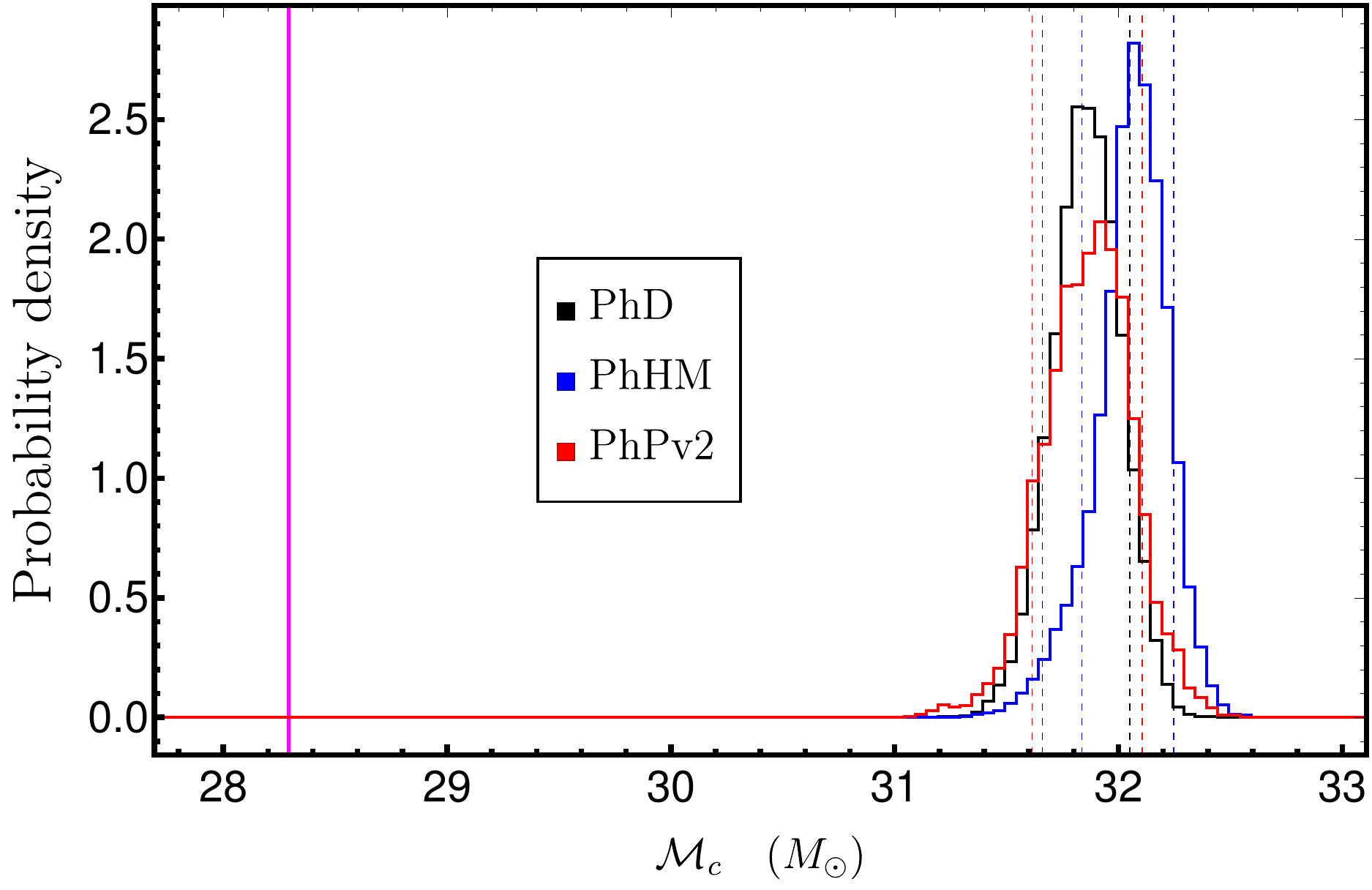}
\end{minipage} 
%\hfill
\hspace{1cm}
\begin{minipage}[h]{.45\textwidth}
\includegraphics[scale=0.42]{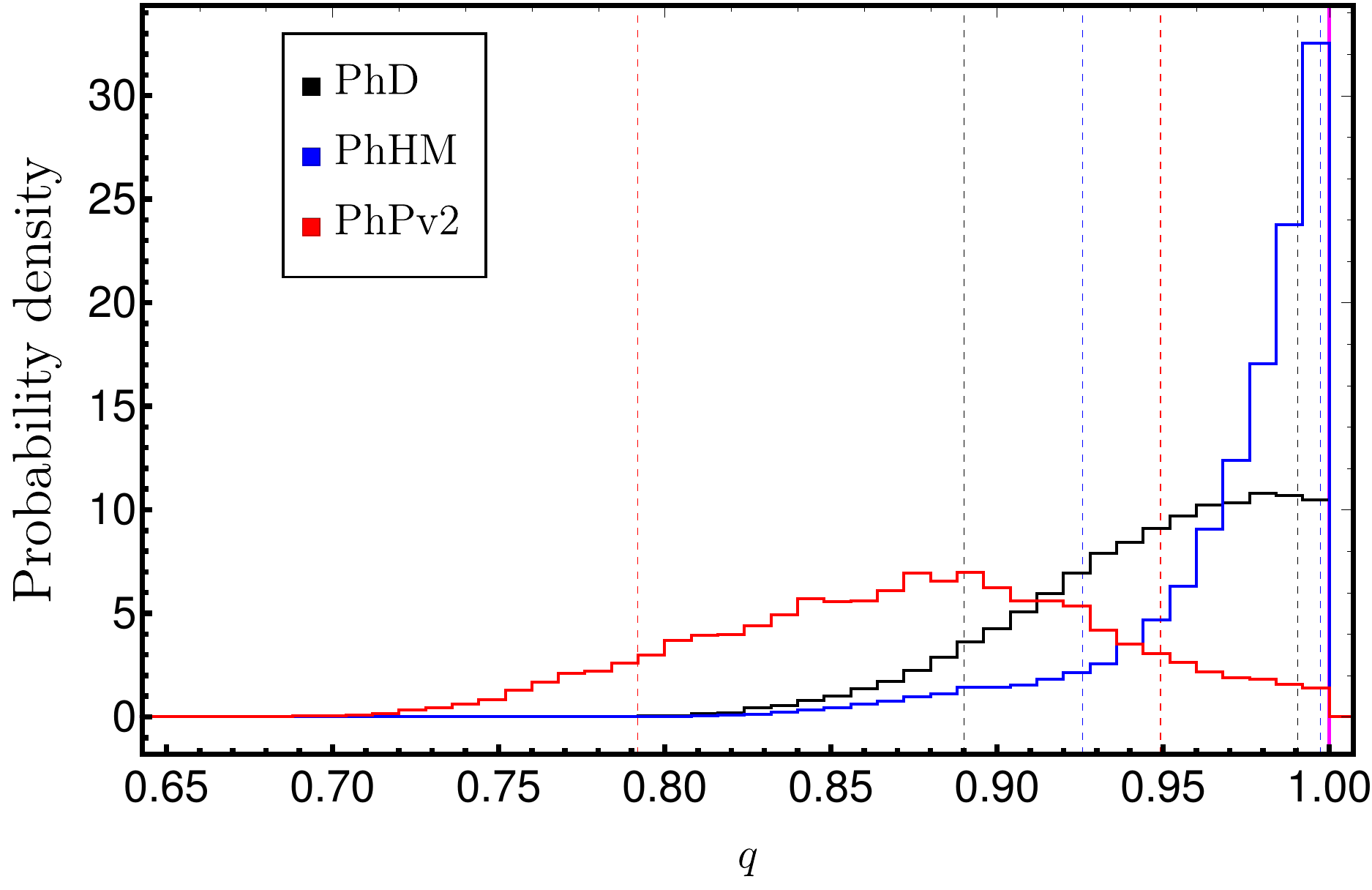}
\end{minipage}

\end{minipage}

%\vspace{5ex}

%B
\noindent\begin{minipage}{\textwidth}

\noindent\begin{minipage}[h]{.45\textwidth}
\includegraphics[scale=0.42]{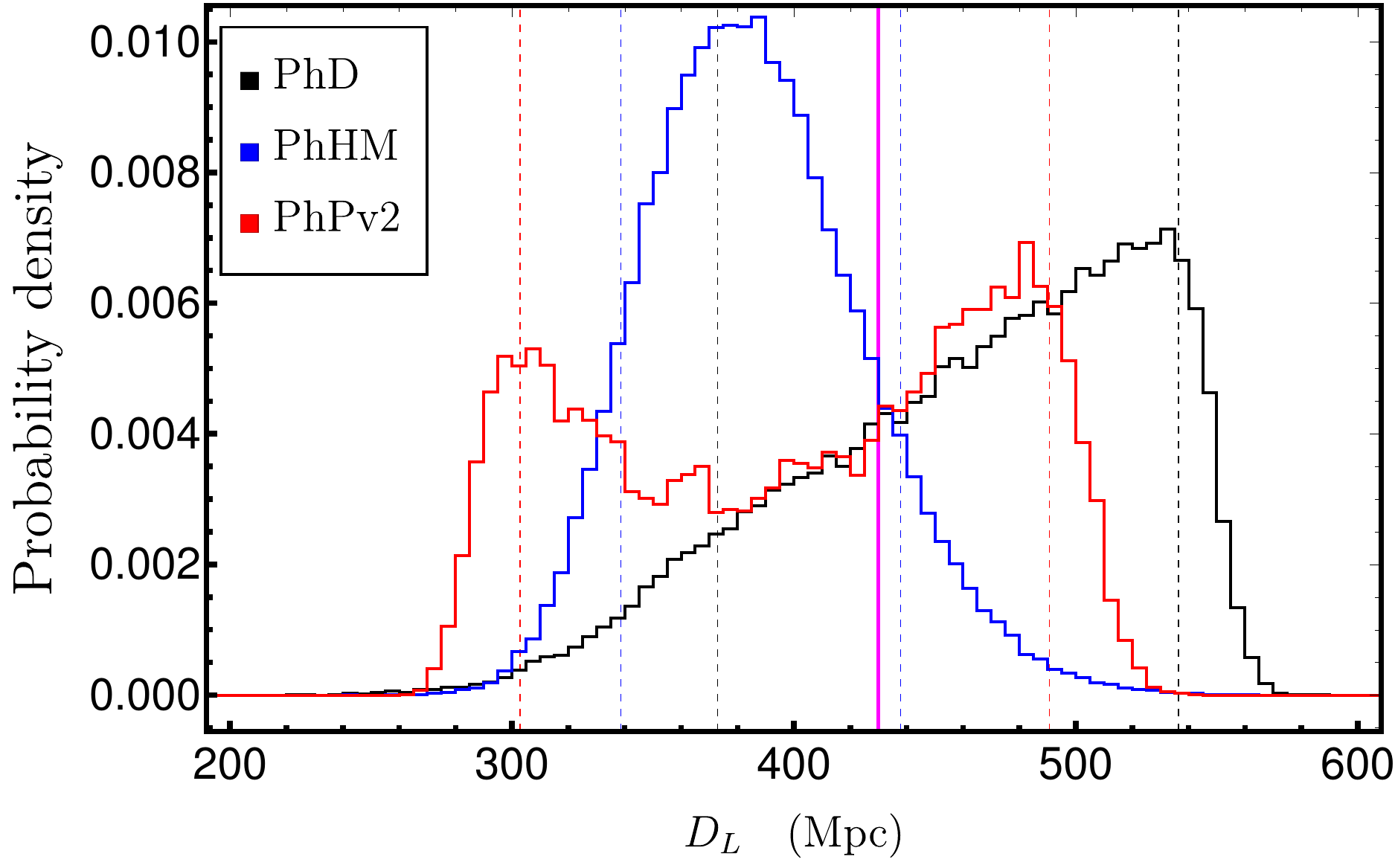}
\end{minipage} 
%\hfill
\hspace{1cm}
\begin{minipage}[h]{.45\textwidth}
\includegraphics[scale=0.42]{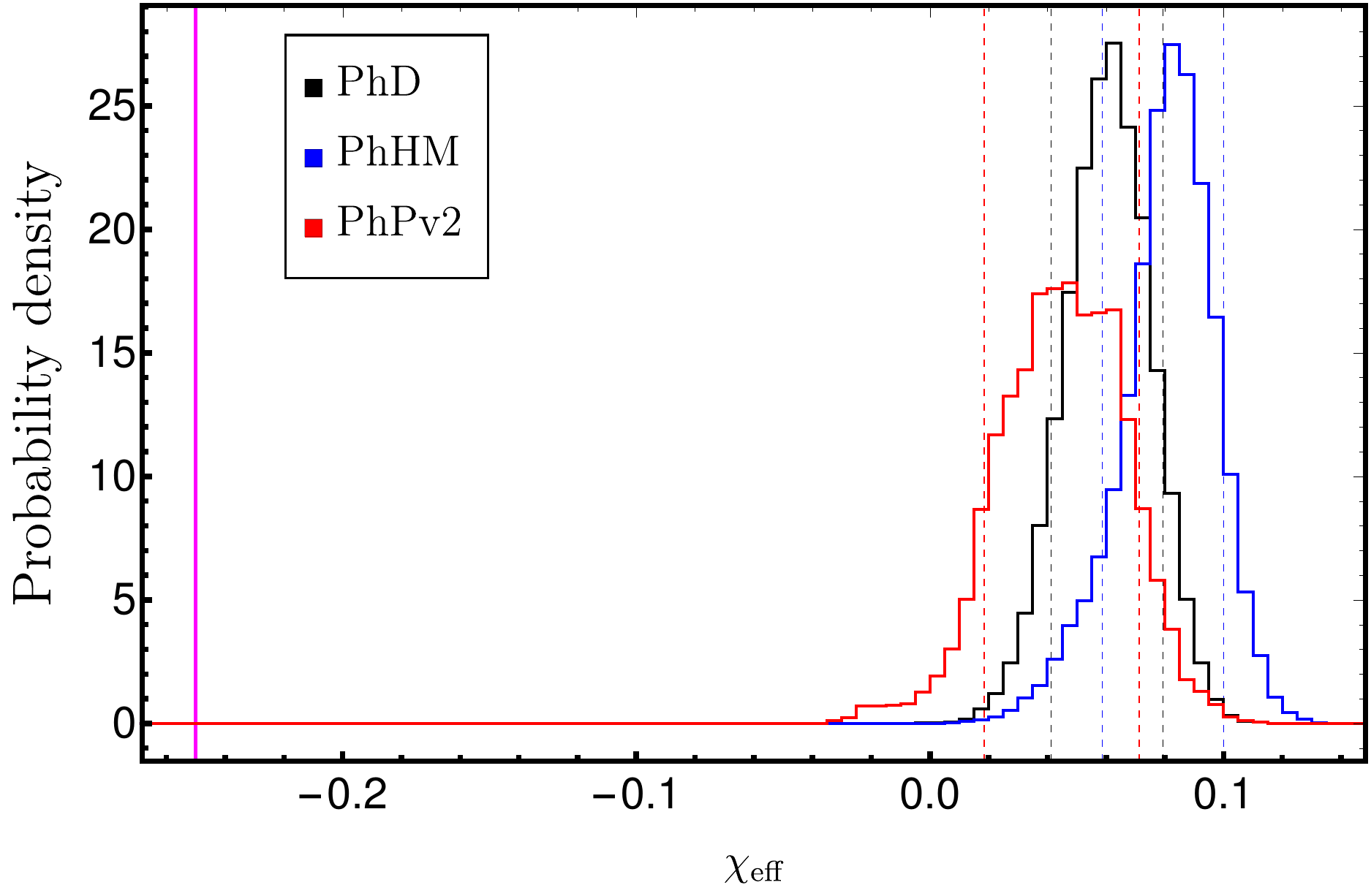}
\end{minipage}

\end{minipage}

\caption{ Posterior probability distributions for the injected spinning eccentric hybrid waveform, with initial eccentricity   $e_0=0.420\pm 0.006$. The vertical dashed lines correspond to $90 \%$ credible regions. The thick vertical magenta line represents the injected value. The black, blue and red curves represent distributions sampled using the IMRPhenomD, IMRPhenomHM and IMRPhenomPv2 approximants, respectively.}
\label{fig:posteriorsHybrid}
\end{figure*}

%\end{widetext}

We repeat the same procedure while injecting a hybrid waveform, including only the $(l,m)=(2,\pm 2)$ modes, of an eccentric spinning waveform with ID 8 of Table \ref{tab:tabNR3}. This is an equal mass with the z component of the dimensionless spin vectors $\chi_{1z}=\chi_{2z}=-0.25$ and initial eccentricity $e_0=0.420\pm 0.006$. The posterior distribution for the chirp mass, mass ratio, luminosity distance and $\chi_{\text{eff}}$ are shown in Fig. \ref{fig:posteriorsHybrid} for IMRPhenomD, IMRPhenomHM,  and IMRPhenomPv2 as waveform models. In this case, the parameter biases are much higher than in the previous injection study mainly due to the fact that the injected signal has a much higher initial eccentricity. 

The values of the recovered parameters as well as the injected values are shown in Table \ref{tab:tabHybridbias}. The injected values of the sky position like the right ascension $\alpha=1.375$ rad and  $\delta=-1.21$ rad are again well recovered parameters for the three runs $\alpha=1.37^{+0.01}_{-0.01}$ rad and $\delta=-1.21^{+0.01}_{-0.01}$ rad. The bias in the chirp mass is $\sim 4 M_\odot$ for the three models. Here one again observes  the correlation between chirp mass and mass ratio. The shift in chirp mass posteriors with respect to the injected value translates into a better determined mass ratio distribution, which is clearly the case for PhenomHM, which performs unexpectedly well in recovering the mass ratio parameter, while PhenomD and PhenomPv2 show much wider distributions and much larger credible intervals. 

The posteriors of the luminosity distance show also large error bars for the three models, where PhenomHM again reduces the bias with respect to PhenomD and PhenomPv2. The recovered effective spin parameter is completely off with respect to the injected value for the three approximants. The recovered $\chi_{\text{eff}}$ is positive, while the injected one is negative. The bias in the effective spin parameter is  approximately $-0.3$ for the three models, indicating the inability of the quasicircular models to estimate the spin parameter of highly eccentric spinning binaries with quasicircular models. Regarding the  recovered matched-filter SNR and the log Bayes factor displayed in Table \ref{tab:tabHybridbias}, one can observe that while the SNR provides comparable values among models, the values of the log Bayes factor indicate that PhenomPv2 fits  the data scarcely better than PhenomHM and PhenomD.

\begin{table*}
\begin{center}
%\resizebox{13.cm}{!}{
 \def\arraystretch{1.4 }
\begin{tabular}{  c   c   c   c   c  c  c  c   c  c c c c }
\hline
\hline
$e_0$& Model &  $m_1/M_\odot$&  $m_2/M_\odot$& $\mathcal{M}_c/M_\odot$&  $q$ &     $D_L/$Mpc&  $\chi_{\text{eff}}$ &$\psi$ (rad) &$\iota$ (rad) & $\rho_{\text{Match}}$ & $\log \mathcal{B}$\\ %& $1-\mathcal{M}$ &$\rho_{\text{Net}}$\\
\hline
\multirow{3}{*}{$0.42$} & PhenomD   &  $37.52^{+1.30}_{-0.76}$&  $36.04^{+0.49}_{-0.73}$ &   $31.86^{+0.19}_{-0.2}$&  $0.95^{+0.04}_{-0.06}$&   $474^{+62}_{-101}$ &  $0.06^{+0.02}_{-0.02}$ &$2.60^{+0.31}_{-0.33}$  & $1.54^{+1.22}_{-1.19}$   &  82.68  & 2895.91 \\ %& 0.071 & \multirow{3}{*}{$81.72$}\\
%\cline{3-10}
   & PhenomHM   &  $37.23^{+0.95}_{-0.37}$&  $36.62^{+0.30}_{-0.81}$ &   $32.07^{+0.18}_{-0.23}$&  $0.98^{+0.02}_{-0.06}$&   $384^{+54}_{-45}$ &  $0.08^{+0.02}_{-0.02}$ & $2.28^{+0.18}_{-0.16}$  & $1.04^{+1.12}_{-0.26}$  &  82.54 & 2894.17 \\ %&  0.072  &\\
   & PhenomPv2   &  $39.15^{+2.08}_{-1.62}$&  $35.20^{+0.84}_{-1.06}$ &   $31.87^{+0.23}_{-0.26}$&  $0.88^{+0.07}_{-0.08}$&   $413^{+77}_{-110}$ &  $0.05^{+0.03}_{-0.03}$ &$2.33^{+0.33}_{-0.44}$  & $1.46^{+1.36}_{-0.45}$ &  82.62  & 2910.28 \\ %&  0.070  & \\
 \hline
& Injected &  $32.5$&  $32.5$& $28.29$&  $1$ &  $430$ &  $-0.25$ &$0.33$  & $0.3$ & & \\ %&\\
%\hline
\hline
\hline
 \end{tabular}
 %}
\end{center}
\caption{Black hole binary recovered parameters for the spinning hybrid waveform from Fig. 	\ref{fig:posteriorsHybrid}. The last row corresponds to the injected parameters. The first column describes the initial eccentricity of the injected signal. Then we specify the approximant, the component masses, the chirp mass, the mass ratio,  the luminosity distance, the effective spin parameter, the polarization angle, the inclination, the recovered matched-filter SNR for the detector network, and the log of the Bayes factor.}
\label{tab:tabHybridbias}
\end{table*}

This section shows examples of the kind of study that one is able to perform with the current eccentric waveform dataset.  We have shown the limitations of the current IMR quasicircular to estimate the parameters of moderately eccentric waveforms including a moderately spinning case. For the cases studied in this section, we have found that although the use of quasicircular models to estimate parameters of eccentric signals leads to inevitable biases, aligned-spin quasicircular models with higher order modes leverage the impact of these biases for the mass ratio and the luminosity distance when compared to aligned-spin models with only the $(2,\pm 2)$ modes or precessing models. Because of the computational cost of the PE runs and the amount of eccentric waveforms available, we leave for future work a detailed study of the whole dataset using not only quasicircular models but also eccentric waveform approximants.

\section{Summary and conclusions} \label{sec:summary}
%%%%%%%%%%%%%%%%%%% NR set up
In this paper we have presented the first parameter study of numerical relativity simulations of eccentric spinning black hole binaries.
We have presented a simple procedure to set up the initial parameters of eccentric simulations. The higher the initial eccentricity of the simulation, the longer the initial separation has to be  to avoid the immediate plunge of the binary due to the strong interactions at the periastron. This increases the computational cost of the simulations of Table \ref{tab:tabNR3} with $e_0 \sim 0.4$, which is roughly double the one with $e_0 \sim 0.2$, as can be observed in their merger times. Additionally, longer initial separations produce long enough waveforms, which allows one to avoid the breakdown of the post-Newtonian approximation and ease the posterior construction of PN-NR hybrid waveforms. As part of the postprocessing step, we have computed the final mass and final spin of the 60 new simulations presented in Table \ref{tab:tabNR3}. We have compared the final mass and final spin of those simulations with quasicircular NR fits   \cite{PhysRevD.95.064024} and found that relative differences are as high as $1 \%$, which is completely consistent with the inaccuracies of the fitting formulas and gauge transient in the apparent horizon quantities. Therefore, we have extended previous work \cite{PhysRevD.77.081502} on the circularization of eccentric nonspinning numerical relativity simulations to the eccentric spinning case. Note that the eccentricities of the simulations presented in this communication have more moderate values than the ones presented in \cite{PhysRevD.77.081502}, although ours are much longer and include spins.

%%%%%%%%%%%%%%%%%%%%% Eccentricity measurement
A crucial part of this  work has been to extend the low eccentric procedure to measure the eccentricity in NR \cite{PhysRevD.99.023003} to the arbitrary high eccentric limit. We have shown that the eccentricity estimator used in \cite{PhysRevD.99.023003} cannot be used for high eccentricities because it does not reduce to the Newtonian definition of the eccentricity. Additionally, its reliance on a noneccentric fit makes it numerically inaccurate, and it can produce eccentricity values higher than 1. As a consequence, we have decided to use another eccentricity estimator \cite{PhysRevD.66.101501} that is also constructed upon the orbital frequency and which does not rely on any noneccentric fit. This eccentricity estimator reduces to the Newtonian definition of eccentricity for arbitrarily high eccentricities. We have shown that with this eccentricity estimator we are able to robustly measure the eccentricity for the whole evolution, which will be a key result for generating a future eccentric waveform model. 

%%%%%%%%%%%%%%%%%%%%%%%%%% Hybridization with PN waveforms
We have then taken the NR waveforms and hybridized the $(2,2)$ mode with PN waveforms. The production of the eccentric PN waveforms has required to solve the point particle 3.5PN equations of motion in ADMTT coordinates  \cite{PhysRevD.99.023003} enhanced with the eccentric contribution to the energy flux from \cite{PhysRevD.80.124018}. The absence of complete generic PN expressions for the waveform modes has caused the inaccuracy of the PN waveforms to dominate the error in the hybridization procedure. The use of the instantaneous terms at 3PN order \cite{PhysRevD.91.084040} produces inaccurate waveforms due to the lack of low order tail terms, while the full 3PN expressions in  \cite{Boetzel:2019nfw} are restricted to the QK parametrization and rely on a certain decomposition of the dynamical variables which complicates their combination with the generic numerical solution of the equations of motion \eqref{eq:eq10}. Therefore, we have restricted to the use of the quadrupole formula with a correction procedure for the initial orbital separation. We have developed a procedure which corrects the initial orbital separation of the PN evolution code for a certain $\delta r$, such that it minimizes the difference in amplitude between the PN and NR $(2,2)$  waveform modes. We have shown that with that procedure we are left with relative errors in the amplitude and phase below $1 \%$ in the hybridization region. These errors in amplitude and phase are high compared to the quasicircular ones \cite{QCHybrids}, where the PN knowledge is wider. Therefore, we expect that in the future an improvement in  knowledge about post-Newtonian waveforms will allow us to construct more accurate hybrid waveforms, not only for the $(2,2)$ mode  but also for the higher order modes. 

%%%%%%%%%%%%%%%%%%%%%%%%%% Mismatches and Parameter Estimation

We have also compared the hybrid waveforms with quasicircular IMR waveform models. This has been done by first computing the mismatch of the eccentric hybrid dataset against the quasicircular nonprecessing  PhenomX model \cite{phenX,phenXHM}. We find that the mismatches become much higher than $3 \%$ for binaries with a total mass lower than $100 M_\odot$, while for total masses higher than $150 M_\odot$ the mismatch lowers below $3 \%$ due to the fact that most of the eccentric waveforms in the frequency band of the detector are in the merger-ringdown  parts, which as shown in Sec. \ref{sec:NRcatalogB}, due to circularization agrees really well with the quasicircular model. 

Additionally, we have made a set of injections into Gaussian detector noise colored to match the LIGO and Virgo design detector sensitivities. We have studied the parameter biases on recovered parameters when using quasicircular models as approximants. We have used three different quasicircular models to recover the parameters and have shown that, although the use of quasicircular models leads to inevitable biases in parameters like the effective spin parameter or the chirp mass, where the biases are similar among the three models, others like the mass ratio and the luminosity distance present lower biases when using quasicircular aligned-spin models including higher order modes. Another important feature is the correlation between chirp mass and mass ratio, the better the measurement of the chirp mass the worse the determination of the mass ratio, and vice versa. This can be clearly observed in Figs. \ref{fig:posteriorsHybrid}  and \ref{fig:posteriorsNR}  where for initial eccentricities $0.09$ the chirp mass is well measured for the three models  but the mass ratio distributions are not. As the initial eccentricity increases, so does the shift in the chirp mass distribution, and  the mass ratio is generally better determined. In the case of the spinning eccentric hybrid, the high initial eccentricity produces clear biases in all quantities and, unexpectedly, PhenomHM recovers the injected value of the mass ratio well and performs the best for the luminosity distance. The study of this phenomenology for the different cases that we have available is ongoing and we leave for a future communication the extension of these results to the whole parameter space.

The work presented in this communication is a natural extension of \cite{PhysRevD.99.023003}. We have set up the current infrastructure  of our group for quasicircular waveform modeling to the eccentric case. As shown in this paper, we have developed new methods to produce a set of spinning eccentric hybrid waveforms which can actually be used for data analysis purposes. The next natural step is to use this hybrid dataset to produce a calibrated eccentric IMR waveform, which can be used for the detection and parameter estimation of eccentric blackhole binaries.

\section{Acknowledgements}  \label{sec:Acknowledgements}

We thank Frank Ohme for the useful comments about manuscript and Maria Haney for the valuable discussions. This work was supported by European Union FEDER funds, the Ministry of Science, Innovation and Universities, and  Spanish Agencia Estatal de Investigación Grants No. FPA2016-76821-P, FPA2017-90687-REDC, FPA2017-90566-REDC. FIS2016-81770-REDT, FPA2015-68783-REDT,  Spanish Ministry of Education, Culture and Sport grants  FPU15/03344, FPU15/01319, Vicepresid`encia i Conselleria d’Innovació, Recerca i Turisme, Conselleria d’Educació, i Universitats del Govern de les Illes Balears i Fons Social Europeu, Generalitat Valenciana (PROMETEO/2019/071),  EU COST Actions CA18108 , CA17137, CA16214, and CA16104, H2020-MSCA-IF-2016. Marie Skłodowska-Curie individual fellowships Proposal No. 751492. The authors thankfully acknowledge the computer resources at MareNostrum and the technical support provided by Barcelona Supercomputing Center (BSC) through Grants No. AECT-2019-2-0010, AECT-2019-1-0022, AECT-2018-3-0017, AECT-2018-2-0022, AECT-2018-1-0009, AECT-2017-3-0013, AECT-2017-2-0017, AECT-2017-1-0017, AECT-2016-3-0014, AECT2016-2-0009,  from the Red Española de Supercomputación (RES) and PRACE (Grant No. 2015133131). \texttt{BAM} and \texttt{ET} simulations were carried out on the BSC MareNostrum computer under PRACE and RES  allocations and on the FONER computer at the University of the Balearic Islands.

\appendix

\section{Numerical Relativity Simulations} \label{sec:AppendixA}
The numerical setup for the \texttt{BAM} and the EinsteinToolkit codes is the same as that described in Appendix C of \cite{PhysRevD.99.023003}. We present in Table \ref{tab:tabNR3} the NR simulations we have produced for this publication. 
 In Table \ref{tab:tabNR3} we show the main properties of the NR simulations: from left to right we start providing an identifier to the simulations, the simulation name, the mass ratio $q=m_1/m_2\geq 1 $, the code used to  produce it, the z component of the dimensionless spin vectors  $\chi_{1,z}$, $\chi_{2,z}$  of each black hole, the initial orbital separation $D/M$, where $M$ is the total mass of the system, the initial eccentricity $e_0$ corresponding to the eccentricity value used in Eq. \eqref{eq:eq3} to compute the perturbation factors of the initial linear momenta of the simulations, the initial orbital eccentricity $e_\omega$  measured with Eq. \eqref{eq:eq9} from the orbital frequency computed from the motion of the black holes, an eccentricity error estimate, $\delta e_\omega$, computed using Eq. \eqref{eq:eq91}, the time to merger $T_{\text{merger}}/M$, calculated as the time elapsed from the start of the simulation until the peak of the amplitude of the $(l,m)=(2,2)$ mode, the number of orbits, $N_{\text{orbits}}= \phi^M_{22}/(4\pi)$, where $\phi^M_{22}$ is the value of the phase of the $(2,2)$ mode at merger, the final mass, $M_f$, as defined in Eq. \eqref{eq:eq4} and the magnitude of the dimensionless final spin, $\chi_f=S/M_f$, where $S$ is specified in Eq. \eqref{eq:eq5}.

\begin{figure}[h]
\centering
\captionsetup{justification=centering}
\includegraphics[scale=0.42]{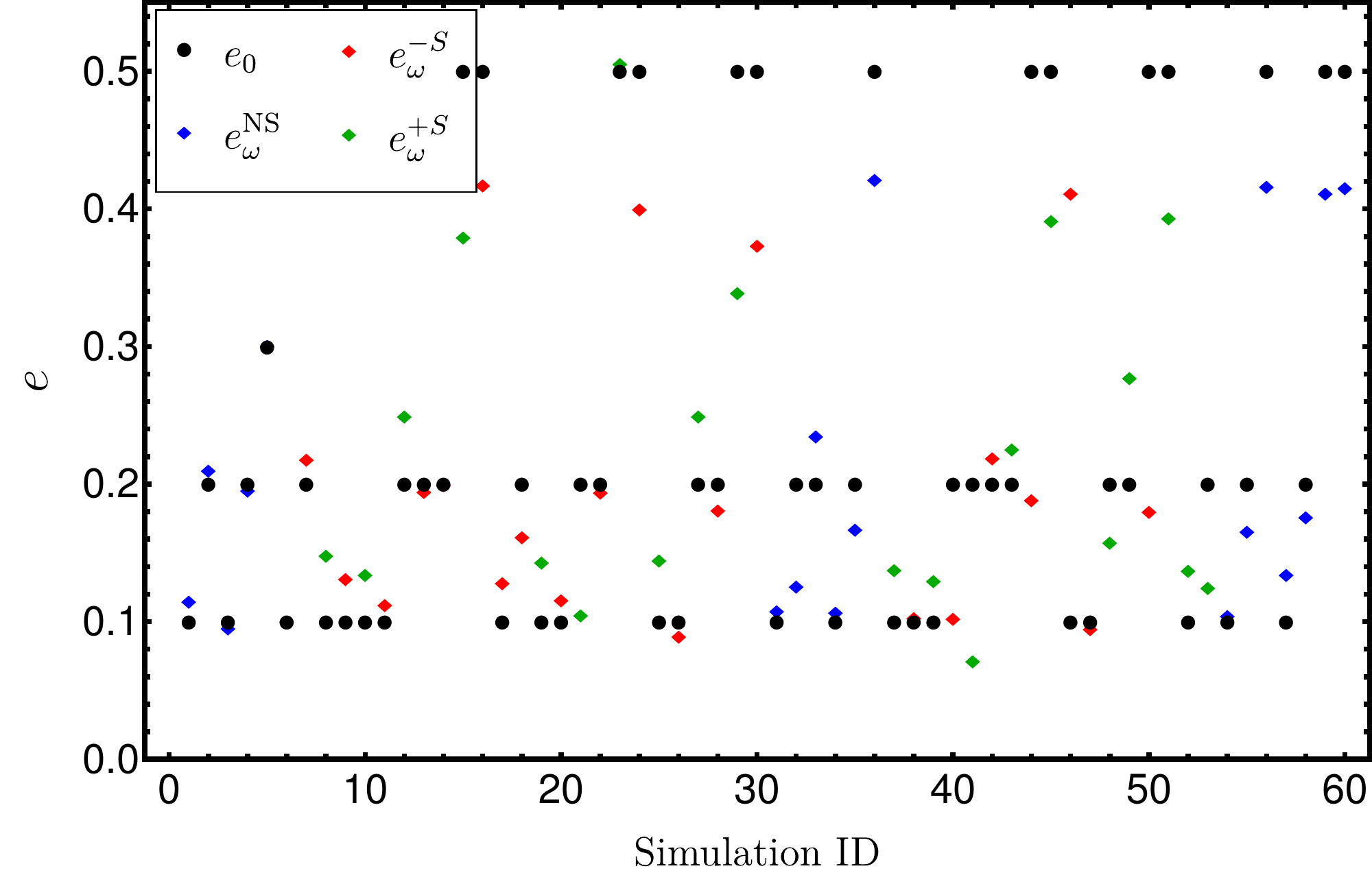}
\caption{ Eccentricity measured from the orbital frequency $e^X_\omega$ with $X=\text{NS},-S,+S$ corresponding to nonspinning and positive and negative spin simulations, for all of the simulations in Table \ref{tab:tabNR3} compared to the PN eccentricity $e_0$ specified in Eq. \eqref{eq:eq3} to compute the perturbation factors for the initial linear momentum of the simulations. The black dots represent $e_0$, the eccentricity value prescribed in Eq.  \eqref{eq:eq3}, while the diamonds represent the actual measured eccentricity $e^X_\omega$. For $e^X_\omega$ we distinguish nonspinning (X=NS), positive spin (X=+S) and negative spin (X=-S) simulations with blue, red and green symbols, respectively.}
\label{fig:eccMeasEccDes}
\end{figure}

\begin{table*}[!]
\begin{center}
\resizebox{16.cm}{!}{
\def\arraystretch{1.1 }
\begin{tabular}{c c c l c c c c c c c c c c c c c}
\hline 
\hline  
ID & & & Simulation &  Code &  q  &${\chi}_{1,z}$ & $\chi_{2,z}$ & $ \chi_{\text{eff}}$ &$D/M$ & $e_0$ & $e_\omega \pm \delta e_\omega$   & $T_{\text{merger}}/M$  & $N_{\text{orbits}}$ & $M_f$ & $\chi_f$\\
\hline
%1 & & & \texttt{q1.\_\_0.\_\_0.\_\_D12.23\_lam96}& \texttt{BAM}  &  1.& $0.$ & $0.$& $0.$ & 12.23 &  $0.114\pm 0.002$ & 1256.42 & 4.5 & 0.9527 & 0.6871\\
1 & & & \texttt{q1.\_\_0.\_\_0.\_\_et0.1\_\_D12.23}& \texttt{BAM}  &  1 & $0$ & $0$& $0$ & 12.23 & 0.1 &  $0.114\pm 0.002$ & 1256.42 & 4.5 & 0.9527 & 0.6871\\
%\hline
%2 & & &  \texttt{q1.\_\_0.\_\_0.\_\_D15\_lam915}& \texttt{BAM}  &  1.& $0.$ & $ 0.$& $0.$ & 15.0 &  $0.210 \pm 0.002$ & 1682.01 & 5.2 & 0.9533 & 0.6895 \\
2 & & &  \texttt{q1.\_\_0.\_\_0.\_\_et0.2\_\_D15}& \texttt{BAM}  &  1& $0$ & $ 0$& $0$ & 15.0 &  0.2 &  $0.210 \pm 0.002$ & 1682.01 & 5.2 & 0.9533 & 0.6895 \\
%\hline
%3 & & & \texttt{q1.\_\_0.\_\_0.\_\_D15\_lam96}& \texttt{BAM}  &  1.& $0.$ & $ 0.$& $0.$ & 15.0 &  $0.095 \pm 0.002$ & 2961.01 & 8.3  & 0.9525 & 0.6869 \\
3 & & & \texttt{q1.\_\_0.\_\_0.\_\_et0.1\_\_D15}& \texttt{BAM}  &  1& $0$ & $ 0$& $0$ & 15.0 &  0.1 &  $0.095 \pm 0.002$ & 2961.01 & 8.3  & 0.9525 & 0.6869 \\
%\hline
%4 & &  & \texttt{q1.\_\_0.\_\_0.\_\_D17\_lam915}& \texttt{BAM}  &  1.& $0.$ & $ 0.$& $0.$ & 17.0 & $0.195 \pm 0.003$ & 2917.42 & 8.2  & 0.9535 & 0.6889 \\
4 & &  & \texttt{q1.\_\_0.\_\_0.\_\_et0.2\_\_D17}& \texttt{BAM}  &  1& $0$ & $ 0$& $0$ & 17.0 & 0.2 &  $0.195 \pm 0.003$ & 2917.42 & 8.2  & 0.9535 & 0.6889 \\
%\hline
5 & &  & \texttt{q1.\_\_0.\_\_0.\_\_et0.3\_\_D20}& \texttt{BAM}  &  1& $0$ & $ 0$& $0$ & 20.0 & 0.3 &  $ 0.301 \pm 0.001$ & 497.48 &  1.5 &  0.9548 & 0.6950 \\
%\hline
6 &  & & \texttt{Eccq1.\_\_0.\_\_0.25\_\_et0.1\_D14}& \texttt{ET}  &  1& $0$ & $ 0.25$& 0.125 & 14.0   &  0.1 &   $0.100\pm 0.002$  & 2319.85 & 6.4 & 0.9480  & 0.7249\\
%\hline
7 &  & & \texttt{Eccq1.\_\_0.\_\_0.25\_\_et0.2\_D16}& \texttt{ET}  &  1& $0$ & $ 0.25$&  0.125 &  16.0 &  0.2 &  $0.217 \pm 0.003$ &  2449.84 & 5.8  &  0.9474 & 0.7243 \\
%\hline
8 &  & & \texttt{Eccq1.\_\_-0.25\_\_-0.25\_\_et0.1\_D12}& \texttt{ET}  &  1& $-0.25$ & $-0.25$& $-0.25$  & 12.0 &   0.1 & $0.148 \pm 0.002$ & 939.87 & 2.8   &  0.9579 & 0.6080 \\
%\hline
9 &  & & \texttt{Eccq1.\_\_0.25\_\_0.25\_\_et0.1\_D12}& \texttt{ET}  &  1& $0.25$ & $ 0.25$& 0.25 & 12.0 & 0.1 &  $0.131 \pm 0.002$  &  1347.59 & 4.8  & 0.9440  & 0.7605 \\
%\hline
10 &  & & \texttt{Eccq1.\_\_-0.25\_\_-0.25\_\_et0.1\_D14}& \texttt{ET}  &  1& $-0.25$ & $-0.25$&  $-0.25$ & 14.0 & 0.1 &  $0.134 \pm 0.002$ & 1897.26 &  5.3 & 0.9573  & 0.6091 \\
%\hline
11 &  & & \texttt{Eccq1.\_\_0.25\_\_0.25\_\_et0.1\_D14}& \texttt{ET}  &  1& $0.25$ & $0.25$ & $0.25$ & 14.0 &  0.1 &  $0.112 \pm 0.003$ & 2464.75 & 7.6  & 0.9440  & 0.7607  \\
%\hline
12 &  & & \texttt{Eccq1.\_\_-0.25\_\_-0.25\_\_et0.2\_D14}& \texttt{ET}  &  1& $-0.25$ & $-0.25$&  $-0.25$ &14.0  &  0.2 & $0.249 \pm 0.002 $  & 1067.25 &  3.8  & 0.9578  & 0.6109 \\
%\hline
13 &  & & \texttt{Eccq1.\_\_0.25\_\_0.25\_\_et0.2\_D14}& \texttt{ET}  &  1& $0.25$ & $0.25$ & $0.25$ & 14.0  &   0.2 & $0.194 \pm 0.002$&  1499.92 &  5.0  & 0.9432  & 0.7620 \\
%\hline
14 &  & & \texttt{Eccq1.\_\_0.25\_\_0.25\_\_et0.2\_D16}& \texttt{ET}  &  1& $0.25$ & $0.25$ & $0.25$ & 16.0 &  0.2 & $0.199 \pm 0.003 $  & 2599.90 &  8.9  & 0.9437  & 0.7624 \\
%\hline
15 &  & & \texttt{Eccq1.\_\_-0.25\_\_-0.25\_\_et0.5\_D26}& \texttt{ET}  &  1&  $-0.25$ & $-0.25$&  $-0.25$  &  26.0 &  0.5 & $0.38 \pm 0.004 $  & 3287.31 & 7.7  & 0.9566  & 0.6080 \\
%\hline
16 &  & & \texttt{Eccq1.\_\_0.25\_\_0.25\_\_et0.5\_D26}& \texttt{ET}  &  1& $0.25$ & $0.25$ & $0.25$ & 26.0 &  0.5 & $0.418 \pm 0.004 $  & 4613.02 & 11.3  & 0.9428  & 0.7604\\
%\hline
17 &  & & \texttt{Eccq1.\_\_0.25\_\_0.\_\_et0.1\_D14}& \texttt{ET}  &  1& $0.25$ & $ 0$& 0.125 &  14.0 & 0.1 & $0.128 \pm 0.003$  & 2302.69 & 7.2  & 0.9480  & 0.7249\\
%\hline
18 &  & & \texttt{Eccq1.\_\_0.25\_\_0.\_\_et0.2\_D16}& \texttt{ET}  &  1& $0.25$ & $ 0$& 0.125 & 16.0 & 0.2 & $0.161 \pm 0.002 $  &  2411.27 & 7.4  & 0.9474  & 0.7242 \\
%\hline
19  & & & \texttt{Eccq1.\_\_-0.5\_\_-0.5\_\_et0.1\_D13}& \texttt{ET}  &  1& $-0.5$ & $-0.5$&  $-0.5$&  13.0  &  0.1 &  $0.143 \pm 0.002$ & 1131.58 & 3.2  & 0.9623  & 0.5286\\
%\hline
20  & & & \texttt{Eccq1.\_\_0.5\_\_0.5\_\_et0.1\_D13}& \texttt{ET}  &  1& $0.5$ & $0.5$& $0.5$ & 13.0 &  0.1 &  $0.116 \pm 0.002$  & 2071.02 & 7.3 & 0.9323  & 0.8309 \\
%\hline
21 & &  & \texttt{Eccq1.\_\_-0.5\_\_-0.5\_\_et0.2\_D15}& \texttt{ET}  &  1& $-0.5$ & $-0.5$& $-0.5$ &  15.0  &  0.2 &  $0.104 \pm 0.001$ & 1170.51 & 3.3  & 0.9624  & 0.5298 \\
%\hline
22 & &  & \texttt{Eccq1.\_\_0.5\_\_0.5\_\_et0.2\_D15}& \texttt{ET}  &  1& $0.5$ & $0.5$& $0.5$ & 15.0 &  0.2 &  $0.194 \pm 0.002$  & 2290.43  & 7.7  & 0.9329  & 0.8323 \\
%\hline
23 & &  & \texttt{Eccq1.\_\_-0.5\_\_-0.5\_\_et0.5\_D26}& \texttt{ET}  &  1& $0.5$ & $0.5$& $0.5$ &  26.0  & 0.5 &  $0.505 \pm 0.005 $ & 2675.44 & 6.1  & 0.9622  & 0.5230 \\
 %\hline
24 &  & & \texttt{Eccq1.\_\_0.5\_\_0.5\_\_et0.5\_D26}& \texttt{ET}  &  1& $0.5$ & $0.5$&  $0.5$ &  26.0 &  0.5 &  $0.400 \pm 0.004$  & 5307.53 &  13.4  & 0.9322  & 0.8294 \\
 %\hline
25 &  & & \texttt{Eccq1.\_\_-0.75\_\_-0.75\_\_et0.1\_D13}& \texttt{ET}  &  1& $-0.75$ & $-0.75$&  $-0.75$  & 13.0 &  0.1 & $0.144 \pm 0.002$ & 907.44 & 2.5   & 0.9654  & 0.4458 \\
 %\hline
26 &  & & \texttt{Eccq1.\_\_0.75\_\_0.75\_\_et0.1\_D13}& \texttt{ET}  &  1 & $0.75$ & $0.75$& $0.75$ &  13.0  &  0.1 & $ 0.089 \pm 0.002$  & 2307.95  & 8.3  & 0.9156  & 0.8934 \\
% \hline
27 & &  & \texttt{Eccq1.\_\_-0.75\_\_-0.75\_\_et0.2\_D15}& \texttt{ET}  &  1 & $-0.75$ & $-0.75$& $-0.75$   & 15.0 &  0.2 &  $0.249 \pm 0.002 $ & 902.561 & 2.6  & 0.9657  & 0.4475 \\
 %\hline
28 & &  & \texttt{Eccq1.\_\_0.75\_\_0.75\_\_et0.2\_D15}& \texttt{ET}  &  1 & $0.75$ & $0.75$&  $0.75$ & 15.0  & 0.2 &  $0.181 \pm 0.002$ & 2629.47 &  9.5  & 0.9149  & 0.8904\\
 %\hline
29 & &  & \texttt{Eccq1.\_\_-0.75\_\_-0.75\_\_et0.5\_D26}& \texttt{ET}  &  1 & $-0.75$ & $-0.75$&  $-0.75$  & 26.0  &  0.5 &  $0.339 \pm 0.003 $ & 2079.87 & 4.1  & 0.9655  & 0.4506 \\
 %\hline
30 & & &  \texttt{Eccq1.\_\_0.75\_\_0.75\_\_et0.5\_D26}& \texttt{ET}  &  1 & $0.75$ & $0.75$& $0.75$ & 26.0 &  0.5 &  $0.373 \pm 0.004 $ & 5907.6 &  15.1  & 0.9158  & 0.8843 \\
 %\hline
 31  & & & \texttt{Eccq1.5\_\_0.\_\_0.\_\_et0.1\_D13}& \texttt{ET}  &  1.5& $0$ & $0$& $0$ & 13.0 &  0.1 &  $ 0.108 \pm  0.002$ & 1606.33 & 5.2  & 0.9552  & 0.6651  \\
% \hline
32 & & &  \texttt{Eccq1.5\_\_0.\_\_0.\_\_et0.2\_D13.5}& \texttt{ET}  &  1.5& $0$ & $0$& $0$ & 13.5   &  0.2 & $ 0.126 \pm  0.001$ & 1142.56 &  3.8  & 0.9553  & 0.6619 \\
% \hline
33 &  & & \texttt{Eccq1.5\_\_0.\_\_0.\_\_et0.2\_D15}& \texttt{ET}  &  1.5& $0$ & $0$& $0$ & 15.0 &  0.2 & $0.245 \pm 0.002$  & 1809.34 &  5.4 & 0.9548  & 0.6636 \\
% \hline
34 &  & & \texttt{Eccq2.\_\_0.\_\_0.\_\_et0.1\_D13}& \texttt{ET}  &  2& $0$ & $0$& $0$ & 13.0 &  0.1 & $ 0.106 \pm 0.002$  & 1738.71 & 5.3  & 0.9610  & 0.6232 \\
% \hline
35 &  & & \texttt{Eccq2.\_\_0.\_\_0.\_\_et0.2\_D16}& \texttt{ET}  &  2 & $0$ & $0$& $0$ &  16.0  &  0.2 &  $0.167 \pm 0.002$ &  2499.02 & 7.5  & 0.9610  & 0.6249  \\
 %\hline
36 &  & & \texttt{Eccq2.\_\_0.\_\_0.\_\_et0.5\_D26}& \texttt{ET}  &  2& $0$ & $0$& $0$ &  26.0  &  0.5 &  $ 0.422 \pm 0.004$ &  4380.33 & 10.4   & 0.9609  & 0.6262   \\
 %\hline
37 &  & & \texttt{Eccq2.\_\_-0.25\_\_-0.25\_\_et0.1\_D12}& \texttt{ET}  &  2& $-0.25$ & $-0.25$&  $-0.25$ & 12.0  &  0.1 & $0.138 \pm 0.002 $ & 1026.39 &  3.2  & 0.9664  & 0.5283  \\
% \hline
38 &  & & \texttt{Eccq2.\_\_0.25\_\_0.25\_\_et0.1\_D12}& \texttt{ET}  &  2& $0.25$ & $0.25$& $0.25$ &  12.0  &   0.1 &  $0.103 \pm 0.002$ & 1435.07 &  5.1  & 0.9544  & 0.7170  \\
 %\hline
39 &  & & \texttt{Eccq2.\_\_-0.25\_\_-0.25\_\_et0.1\_D14}& \texttt{ET}  &  2& $-0.25$ & $-0.25$& $-0.25$  & 14.0 &  0.1 & $ 0.130 \pm 0.002$ & 2001.7 & 5.6  & 0.9663  & 0.5261  \\
 %\hline
40 &  & & \texttt{Eccq2.\_\_0.25\_\_0.25\_\_et0.1\_D14}& \texttt{ET}  &  2& $0.25$ & $0.25$& $0.25$ &  14.0  &  0.1 &  $0.103 \pm 0.002$ & 2707.25 & 8.3  & 0.9544  & 0.7155   \\
 %\hline
41 &  & & \texttt{Eccq2.\_\_-0.25\_\_-0.25\_\_et0.2\_D14}& \texttt{ET}  &  2& $-0.25$ & $-0.25$& $-0.25$ & 14.0 &  0.2 &  $0.072 \pm 0.001$ & 1123.58 &  3.5  & 0.9660  & 0.5300   \\
 %\hline
42 &  & & \texttt{Eccq2.\_\_0.25\_\_0.25\_\_et0.2\_D14}& \texttt{ET}  &  2& $0.25$ & $0.25$& $0.25$ & 14.0 &  0.2 & $0.219 \pm 0.002$ &  1708.92 & 5.6  & 0.9548  & 0.7151  \\
% \hline
43 &  & & \texttt{Eccq2.\_\_-0.25\_\_-0.25\_\_et0.2\_D16}& \texttt{ET}  &  2& $-0.25$ & $-0.25$& $-0.25$  &16.0  &  0.2 & $0.225 \pm 0.003 $ & 2085.67 &  5.8  & 0.9663  & 0.5253   \\
% \hline
44 &  & & \texttt{Eccq2.\_\_0.25\_\_0.25\_\_et0.2\_D16}& \texttt{ET}  &  2& $0.25$ & $0.25$& $0.25$ &   16.0 &  0.2 & $0.188 \pm 0.003$  & 2847.34 & 8.3  & 0.9549  & 0.7165  \\
% \hline
45  &  & & \texttt{Eccq2.\_\_-0.25\_\_-0.25\_\_et0.5\_D26}& \texttt{ET}  &  2& $-0.25$ & $-0.25$& $-0.25$ & 26.0 &  0.5 &$0.392 \pm 0.003$ & 3628.05 & 8.4  & 0.9665  & 0.5308  \\
% \hline
46 &  & & \texttt{Eccq2.\_\_0.25\_\_0.25\_\_et0.5\_D26}& \texttt{ET}  &  2& $0.25$ & $0.25$& $0.25$ & 26.0  & 0.5 &$0.411 \pm 0.004 $ &5203.86  & 12.5 & 0.9542  & 0.7140   \\
% \hline
47 &  & & \texttt{Eccq2.\_\_0.5\_\_0.5\_\_et0.1\_D14}& \texttt{ET}  &  2& $0.5$ & $0.5$& $0.5$ & 14.0  & 0.1 & $ 0.095 \pm 0.002$ & 2985.28 &  9.1  & 0.9448  & 0.8052  \\
% \hline
48 &  & & \texttt{Eccq2.\_\_-0.5\_\_-0.5\_\_et0.1\_D14}& \texttt{ET}  &  2& $-0.5$ & $-0.5$& $-0.5$ & 14.0  & 0.1 & $0.158 \pm 0.003 $ & 1714.88 & 4.2  & 0.9698  & 0.4279  \\
% \hline
49 &  & & \texttt{Eccq2.\_\_-0.5\_\_-0.5\_\_et0.2\_D16}& \texttt{ET}  &  2& $-0.5$ & $-0.5$& $-0.5$ & 16.0  & 0.2 & $0.277 \pm 0.003$ &  1712.98&  4.2  & 0.9696  & 0.4300  \\
% \hline
50 &  & & \texttt{Eccq2.\_\_0.5\_\_0.5\_\_et0.2\_D16}& \texttt{ET}  &  2& $0.5$ & $0.5$& $0.5$ & 16.0  & 0.2 &  $0.180 \pm 0.003$ & 3294.21 &  10.5  & 0.9451  & 0.8035  \\
 %\hline
51 &  & & \texttt{Eccq2.\_\_-0.5\_\_-0.5\_\_et0.5\_D27}& \texttt{ET}  &  2& $-0.5$ & $-0.5$& $-0.5$ & 27.0  & 0.5 & $0.393 \pm 0.004$  & 3522.66 &  7.2 & 0.9696  & 0.4328 \\
 %\hline
52 &  & & \texttt{Eccq2.\_\_-0.75\_\_-0.75\_\_et0.1\_D14}& \texttt{ET}  &  2& $-0.75$ & $-0.75$& $-0.75$ & 14.0  & 0.1 & $0.137 \pm 0.002$  & 1386.95 &  3.2 & 0.9725  & 0.3273 \\
%\hline
53 &  & & \texttt{Eccq2.\_\_-0.75\_\_-0.75\_\_et0.2\_D16}& \texttt{ET}  &  2& $-0.75$ & $-0.75$& $-0.75$ & 16.0  &  0.2 & $0.125 \pm 0.002$   & 1353.72 & 3.4 & 0.9728 & 0.3297  \\
%\hline
54 &  & & \texttt{Eccq3.\_\_0.\_\_0.\_\_et0.1\_D13}& \texttt{ET}  &  3& $0$ & $0$& $0$ & 13.0 & 0.1 & $ 0.104 \pm 0.002$ & 1978.55 & 6.1 & 0.9713  & 0.5414 \\
 %\hline
55 &  & & \texttt{Eccq3.\_\_0.\_\_0.\_\_et0.2\_D15}& \texttt{ET}  &  3& $0$ & $0$& $0$ & 15.0 &  0.2 &  $0.166 \pm 0.002$ & 2156.21 & 6.2 & 0.9710  & 0.5401  \\
 %\hline
56 &  & & \texttt{Eccq3.\_\_0.\_\_0.\_\_et0.5\_D26}& \texttt{ET}  &  3& $0$ & $0$& $0$ & 26.0 & 0.5 & $ 0.416 \pm 0.004$ & 5029.06 & 11.5  & 0.9710  & 0.5385 \\
 %\hline
57 &  & & \texttt{Eccq4.\_\_0.\_\_0.\_\_et0.1\_D12}& \texttt{ET}  &  4& $0$ & $0$& $0$ & 12.0 & 0.1 & $ 0.134 \pm 0.002$ & 1609.06 &  5.3  & 0.9780  & 0.4725 \\
 %\hline
58 &  & & \texttt{Eccq4.\_\_0.\_\_0.\_\_et0.2\_D15}& \texttt{ET}  &  4& $0$ & $0$& $0$ & 15.0 &  0.2 & $ 0.176 \pm 0.002 $ & 2412.73 & 7.4  & 0.9779  & 0.4731 \\
% \hline
59 &  & & \texttt{Eccq4.\_\_0.\_\_0.\_\_et0.5\_D27}& \texttt{ET}  &  4& $0$ & $0$& $0$  &27.0  & 0.5 & $0.412 \pm 0.004 $  & 6698.64 &  15.2   & 0.9779  & 0.4739 \\
%\hline
60 &  & & \texttt{Eccq4.\_\_0.\_\_0.\_\_et0.5\_D27.5}& \texttt{ET}  &  4& $0$ & $0$& $0$  &27.5  & 0.5 &$0.415 \pm 0.005$  &  7422.59 &  16.4    & 0.9784  & 0.4717\\
\hline
\hline
\end{tabular}
 }
\end{center}
\caption{Summary of the eccentric NR simulations used in this work. In the first column we indicate the identifier of the simulation. Additionally, each simulation is specified by its mass ratio $q=m_1/m_2\geq 1 $, the code with it was produced, the z component of the dimensionless spin vectors  $\chi_{1,z}$, $\chi_{2,z}$  the orbital separation $D/M$, the desired initial eccentricity $e_0$ used in Eq. \eqref{eq:eq3} and the actual measured initial orbital eccentricity $e_\omega$ and its error $\delta e_\omega$, the time to merger $T_{\text{merger}}/M$, the number of orbits $N_{\text{orbits}}$, the final mass $M_f$, and the magnitude of the dimensionless final spin $\chi_f$.}
\label{tab:tabNR3}
\end{table*}
%%%%%%%%%%%%%%%%%%%%%%%%%%%%%%%555
 In Fig. \ref{fig:eccMeasEccDes} we display the values of the eccentricity measured from the orbital frequency $e_\omega$ defined in  Eq. \eqref{eq:eq9}, and the value of the eccentricity  $e_0$ used in Eq.  \eqref{eq:eq3} to compute the perturbation factors of the initial linear momenta of the simulations in Table \ref{tab:tabNR3}. Moreover, we distinguish for $e_\omega$  with blue, red, and green symbols indicating nonspinning and positive and negative spin simulations, respectively.  As expected, we observe an increase in the differences between $e_\omega$ and $e_0$ with higher initial eccentricities and with high spins as the formula for $\lambda_t(r,e_0,\eta, \text{sign})$ where $e_0$ is used. Equation \eqref{eq:eq1}  is a  1PN expression derived for nonspinning binaries in the low eccentricity limit. To obtain better control on the initial eccentricity of the NR simulations for high eccentricities, higher order corrections, including spin effects, should be taken into account in the derivation of the correction factors for the initial linear momenta of the black holes. We leave for future work an extension of the current PN formulas to the high eccentricity limit.

\section{Eccentricity estimators in highly eccentric systems} \label{sec:AppendixB}
In this section we briefly show the form of the eccentricity estimators of Eqs. \eqref{eq:eq8} and \eqref{eq:eq9} in the Newtonian limit. We start analyzing the eccentricity estimator
\begin{equation}
e_\omega(t)= \frac{\omega (t)-\omega(e=0)}{2 \omega(e=0)}.
\label{eq:eqb1}
\end{equation}
In the Keplerian parametrization the orbital frequency can be written as
\begin{equation}
\omega (t)= \frac{n_t \sqrt{1-e^2}}{(1- e \cos u)^2},
\label{eq:eqb2}
\end{equation}
where $n_t=2 \pi/T_{\text{orb}}$ is the mean motion, $T_{orb}$ is the orbital period, $e$ is the eccentricity and $u$ is the eccentric anomaly. In the low eccentric limit, Eq. \eqref{eq:eqb1} reduces to 
\begin{equation}
\omega (t) \approx n_t \left[1+2e \cos u\right] + \mathcal{O}\left(e^2\right).
\label{eq:eqb3}
\end{equation}
Replacing Eq. \eqref{eq:eqb3} in Eq. \eqref{eq:eqb1} one obtains $e_\omega = e$. However, if one substitutes Eq. \eqref{eq:eqb2} into Eq. \eqref{eq:eqb1} one gets
\begin{equation}
e_\omega (t) = \frac{1}{2} \left(\frac{\sqrt{1-e^2}}{[e \cos (u)-1]^2}-1\right),
\label{eq:eqb4}
\end{equation}
which  does not reduce  to the Newtonian definition of eccentricity. Moreover, one can show that the estimator of Eq. \eqref{eq:eqb4} is not normalized for a certain combination of values of $u$ and $e$. For example, if $u$ vanishes, then
\begin{equation}
e_\omega \geq 1 \quad \text{ for }\quad  e \geq 0.455212.
\label{eq:eqb5}
\end{equation}
This shows that the eccentricity estimator given by Eq. \eqref{eq:eqb1} has to be taken with caution in the high eccentric limit because it can go above 1. On the other hand, the eccentricity estimator
\begin{equation}
e_{\Omega_{a,p}}(t)= \frac{\omega^{1/2}_p - \omega^{1/2}_a}{\omega^{1/2}_p + \omega^{1/2}_a},
\label{eq:eqb6}
\end{equation}
where $\omega_a,\omega_p$ are the orbital frequencies at the apastron and periastron, respectively. This eccentricity estimator has the property such that, even for high eccentricities, it reduces to the Newtonian definition of eccentricity, i.e., $e_{\Omega_{a,p}}= e$.

\vspace{1cm}

\section{Posterior distributions} \label{sec:AppendixC}
 In this appendix we provide further information about the parameter estimation methods used and posterior distributions of several relevant quantities. The settings of the \texttt{CPNEST} sampler \cite{JohnVeitchCpnest}  are a number of live points $N_{\text{live}}=16824$ and a maximum number of Markov-chain Monte Carlo  (MCMC) steps to take $\texttt{max-mcmc}=5000$. We refer the reader to \cite{SkillingNestedsamplinggeneral2006} for details on the meaning of those parameters in the context of nested sampling. This is a computationally expensive setup aiming to ensure an accurate sampling of the likelihood given the complexity of the signal. 

\begin{figure}[h!]  
\centering
\captionsetup{justification=centering}
\includegraphics[scale=1.7]{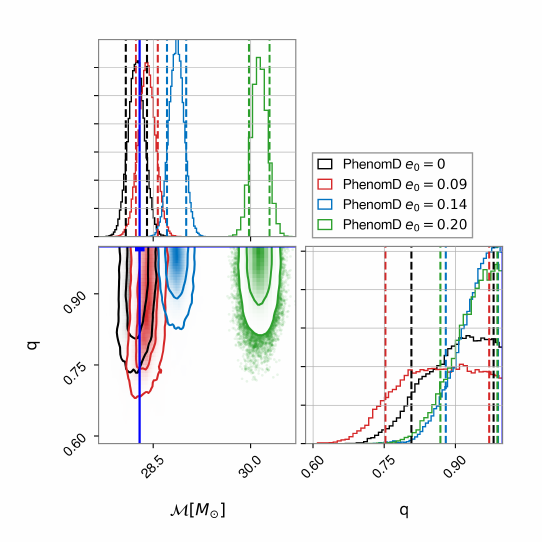}
\caption{ Posterior probability distributions of the mass ratio and the chirp mass for the injected  eccentric NR simulations in Table \ref{tab:tabNRPE} and the  zero-eccentricity injection with the NRHybSur3dq8 model, using IMRPhenomD as the approximant. The vertical dashed lines correspond to $90 \%$ credible regions. The dark blue thick vertical line represents the injected value. The black, red, blue and green curves represent injections with initial eccentricities, $e_0=0,0.09,0.14,0.2$, respectively. }
\label{fig:q1PhDcontourPlot}
\end{figure}

We show in Fig. \ref{fig:q1PhDcontourPlot}  a contour plot of the mass ratio and chirp mass posterior distributions for the injected  eccentric NR simulations in Table \ref{tab:tabNRPE} and the zero-eccentricity injection with the NRHybSur3dq8 model recovered with PhenomD. This plot explicitly exhibits the correlation between the bias in the measurement of the chirp mass and the narrowing of the posterior for the mass ratio with increasing initial eccentricity.

\begin{figure}[h!]
\centering
\captionsetup{justification=centering}
\includegraphics[scale=0.42]{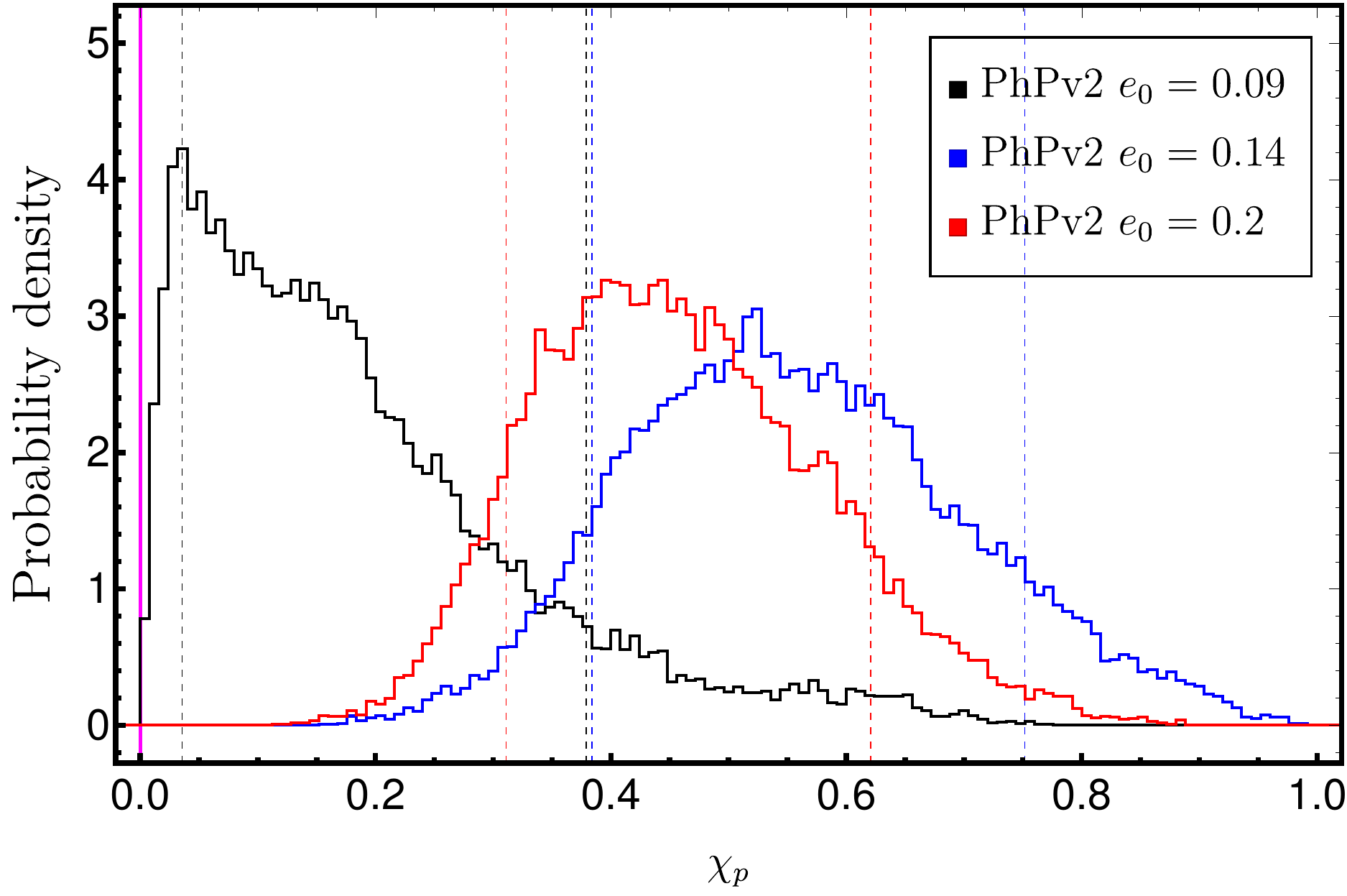}
\caption{ Posterior probability distributions of $\chi_p$ for the injected NR simulations in Table \ref{tab:tabNRPE}. The vertical dashed lines correspond to $90 \%$ credible regions. The magenta thick vertical line represents the injected value. The black, blue, and red curves represent injections with initial eccentricities $e_0=0.09,0.14,0.2$. All cases are sampled using as the approximant IMRPhenomPv2.}
\label{fig:q1NSchip}
\end{figure}

For completion we also show in Fig. \ref{fig:q1NSchip} the posterior distribution of the $\chi_p$ parameter for the NR simulations in Fig. \ref{fig:NRsims} run with IMRPhenomPv2. This parameter, defined in \cite{PhysRevD.91.024043}, accounts for the spin components orthogonal to the direction of the orbital angular momentum vector of the system. Therefore, for nonprecessing configurations $\chi_p=0$,  and for precessing configurations  $\chi_p$  ranges between 0 and 1. In Fig. \ref{fig:q1NSchip} one can observe an increase in $\chi_p$ with increasing initial eccentricity of the injected signal. This result means that the precessing waveform IMRPhenomPv2 is trying to compensate for the inability to reproduce the eccentric signal incrementing the value of the $\chi_p$ parameter, i.e., increasing the precession.

\begin{figure*}[!]

\noindent\begin{minipage}{\textwidth}

%A
\noindent\begin{minipage}[h]{.45\textwidth}
\includegraphics[scale=0.42]{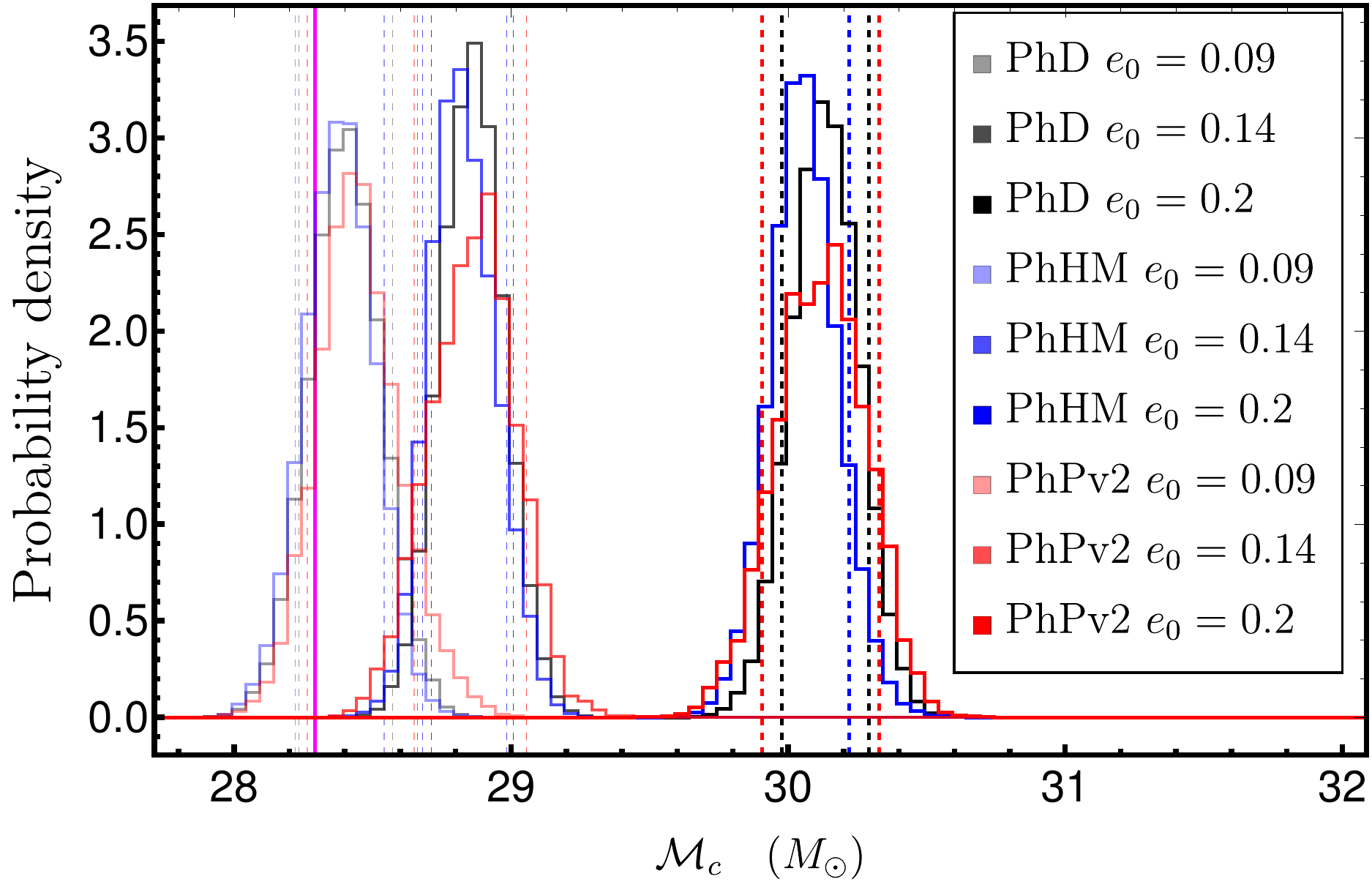}
\end{minipage} 
%\hfill
\hspace{1cm}
\begin{minipage}[h]{.45\textwidth}
\includegraphics[scale=0.42]{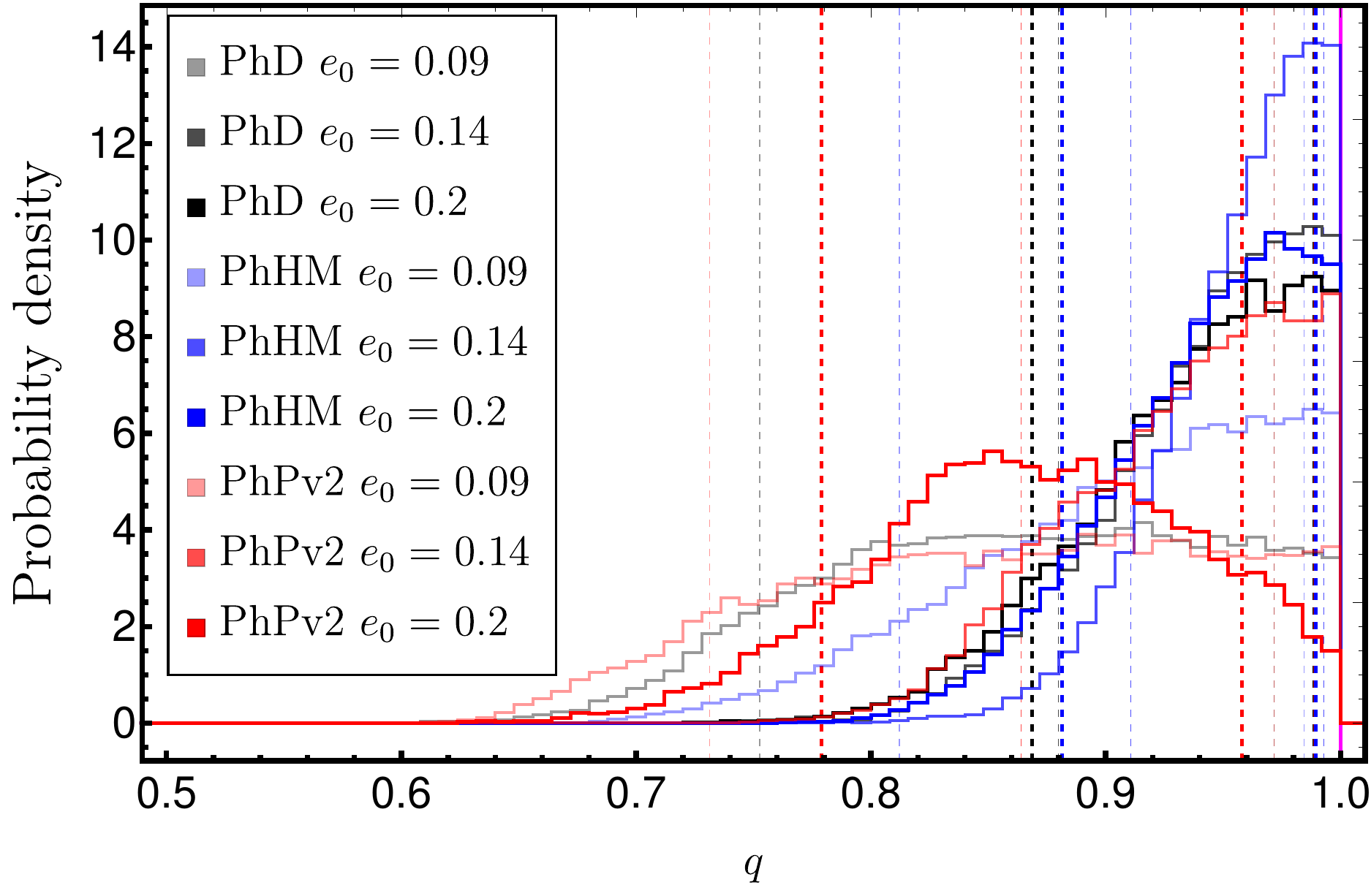}
\end{minipage}

\end{minipage}

%\vspace{5ex}

%B
\noindent\begin{minipage}{\textwidth}

\noindent\begin{minipage}[h]{.45\textwidth}
\includegraphics[scale=0.42]{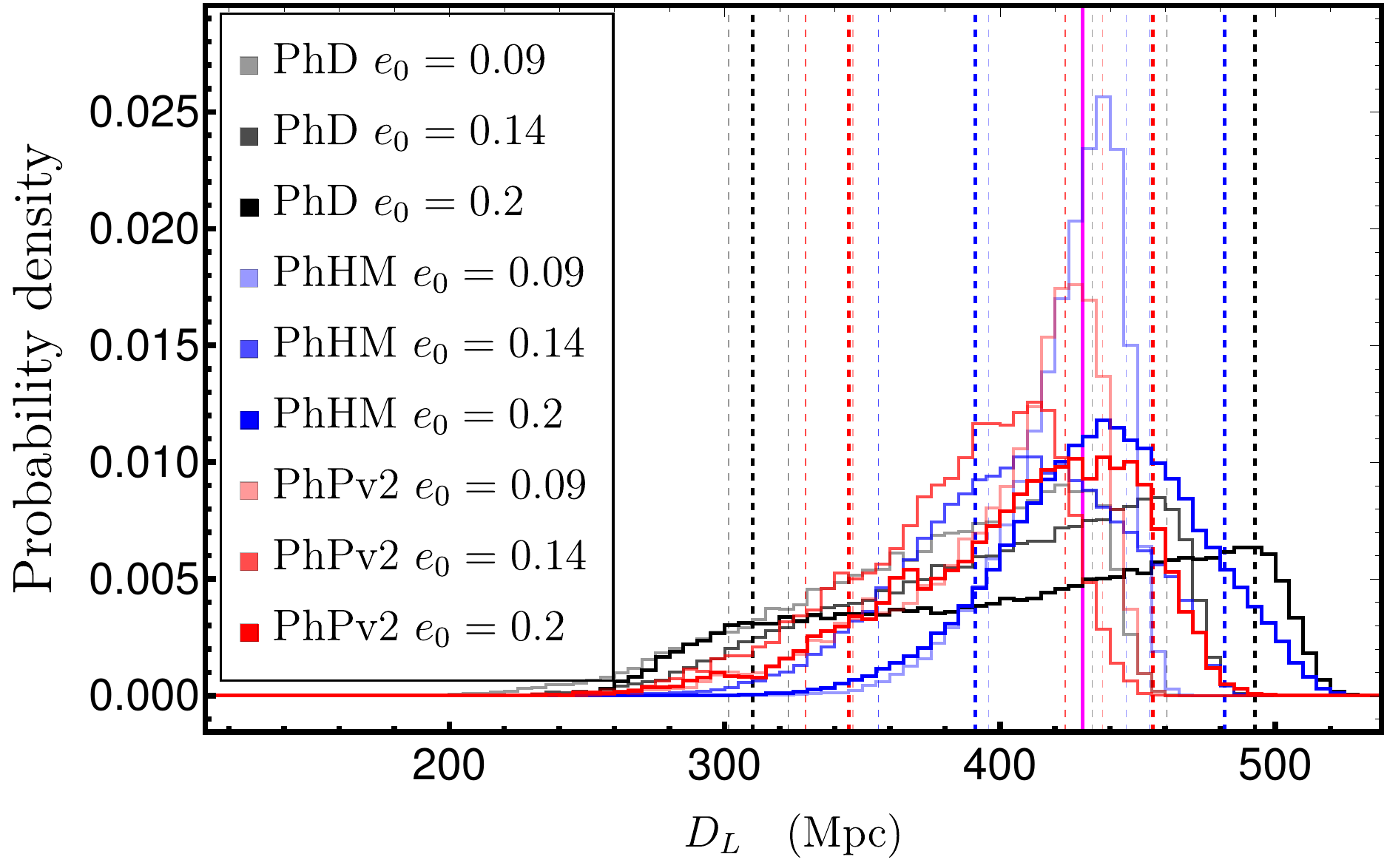}
\end{minipage} 
%\hfill
\hspace{1cm}
\begin{minipage}[h]{.45\textwidth}
\includegraphics[scale=0.42]{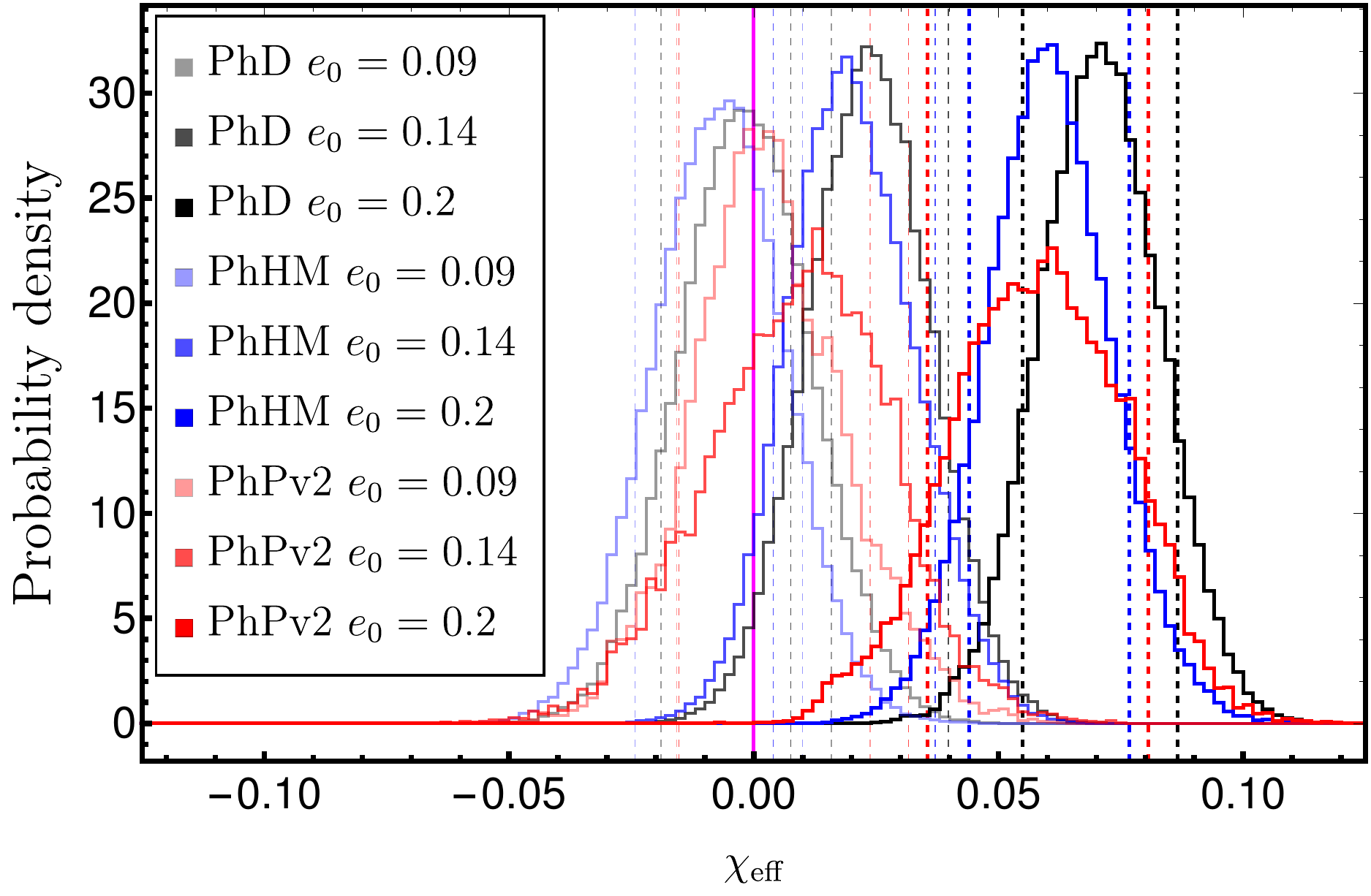}
\end{minipage}

\end{minipage}

\caption{ Posterior probabibility distributions for the eccentric injected NR simulations of Table \ref{tab:tabNRPE}. The vertical dashed lines correspond to $90 \%$ credible regions. The magenta thick vertical line represents the injected value. The black, blue and red curves represent distributions sampled using the IMRPhenomD, IMRPhenomHM and IMRPhenomPv2 approximants, respectively. With increasingly higher opacity are represented injections with initial eccentricities, $e_0=0.09,0.14,0.2$. }

\label{fig:posteriorsNR}
\end{figure*}

In Fig. \ref{fig:posteriorsNR} we display the posterior probability distributions of the chirp mass, mass ratio, effective spin parameter and luminosity distance for the eccentric injected NR simulations in Table \ref{tab:tabNRPE} recovered with  the IMRPhenomD, IMRPhenomHM, and IMRPhenomPv2 approximants with $90 \%$ credible intervals specified by the dashed lines and the injected values by the magenta thick vertical lines. The fainter the color of the posterior distributions, the lower the initial eccentricity ($e_0=0.09,0.14,0.20$).

%\begin{widetext}

%\end{widetext}

\clearpage

\bibliography{uib}

\begin{thebibliography}{105}%
\makeatletter
\providecommand \@ifxundefined [1]{%
 \@ifx{#1\undefined}
}%
\providecommand \@ifnum [1]{%
 \ifnum #1\expandafter \@firstoftwo
 \else \expandafter \@secondoftwo
 \fi
}%
\providecommand \@ifx [1]{%
 \ifx #1\expandafter \@firstoftwo
 \else \expandafter \@secondoftwo
 \fi
}%
\providecommand \natexlab [1]{#1}%
\providecommand \enquote  [1]{``#1''}%
\providecommand \bibnamefont  [1]{#1}%
\providecommand \bibfnamefont [1]{#1}%
\providecommand \citenamefont [1]{#1}%
\providecommand \href@noop [0]{\@secondoftwo}%
\providecommand \href [0]{\begingroup \@sanitize@url \@href}%
\providecommand \@href[1]{\@@startlink{#1}\@@href}%
\providecommand \@@href[1]{\endgroup#1\@@endlink}%
\providecommand \@sanitize@url [0]{\catcode `\\12\catcode `\$12\catcode
  `\&12\catcode `\#12\catcode `\^12\catcode `\_12\catcode `\%12\relax}%
\providecommand \@@startlink[1]{}%
\providecommand \@@endlink[0]{}%
\providecommand \url  [0]{\begingroup\@sanitize@url \@url }%
\providecommand \@url [1]{\endgroup\@href {#1}{\urlprefix }}%
\providecommand \urlprefix  [0]{URL }%
\providecommand \Eprint [0]{\href }%
\providecommand \doibase [0]{http://dx.doi.org/}%
\providecommand \selectlanguage [0]{\@gobble}%
\providecommand \bibinfo  [0]{\@secondoftwo}%
\providecommand \bibfield  [0]{\@secondoftwo}%
\providecommand \translation [1]{[#1]}%
\providecommand \BibitemOpen [0]{}%
\providecommand \bibitemStop [0]{}%
\providecommand \bibitemNoStop [0]{.\EOS\space}%
\providecommand \EOS [0]{\spacefactor3000\relax}%
\providecommand \BibitemShut  [1]{\csname bibitem#1\endcsname}%
\let\auto@bib@innerbib\@empty
%</preamble>
\bibitem [{\citenamefont {Abbott}\ \emph
  {et~al.}(2016{\natexlab{a}})\citenamefont {Abbott} \emph
  {et~al.}}]{PhysRevLett.116.061102}%
  \BibitemOpen
  \bibfield  {author} {\bibinfo {author} {\bibfnamefont {B.~P.}\ \bibnamefont
  {Abbott}} \emph {et~al.} (\bibinfo {collaboration} {LIGO Scientific
  Collaboration and Virgo Collaboration}),\ }\href {\doibase
  10.1103/PhysRevLett.116.061102} {\bibfield  {journal} {\bibinfo  {journal}
  {Phys. Rev. Lett.}\ }\textbf {\bibinfo {volume} {116}},\ \bibinfo {pages}
  {061102} (\bibinfo {year} {2016}{\natexlab{a}})}\BibitemShut {NoStop}%
\bibitem [{\citenamefont {Abbott}\ \emph
  {et~al.}(2016{\natexlab{b}})\citenamefont {Abbott} \emph
  {et~al.}}]{PhysRevLett.116.241103}%
  \BibitemOpen
  \bibfield  {author} {\bibinfo {author} {\bibfnamefont {B.~P.}\ \bibnamefont
  {Abbott}} \emph {et~al.} (\bibinfo {collaboration} {LIGO Scientific
  Collaboration and Virgo Collaboration}),\ }\href {\doibase
  10.1103/PhysRevLett.116.241103} {\bibfield  {journal} {\bibinfo  {journal}
  {Phys. Rev. Lett.}\ }\textbf {\bibinfo {volume} {116}},\ \bibinfo {pages}
  {241103} (\bibinfo {year} {2016}{\natexlab{b}})}\BibitemShut {NoStop}%
\bibitem [{\citenamefont {Abbott}\ \emph
  {et~al.}(2017{\natexlab{a}})\citenamefont {Abbott} \emph
  {et~al.}}]{Abbott:2017vtc}%
  \BibitemOpen
  \bibfield  {author} {\bibinfo {author} {\bibfnamefont {B.~P.}\ \bibnamefont
  {Abbott}} \emph {et~al.} (\bibinfo {collaboration} {VIRGO, LIGO
  Scientific}),\ }\href {\doibase 10.1103/PhysRevLett.118.221101} {\bibfield
  {journal} {\bibinfo  {journal} {Phys. Rev. Lett.}\ }\textbf {\bibinfo
  {volume} {118}},\ \bibinfo {pages} {221101} (\bibinfo {year}
  {2017}{\natexlab{a}})},\ \Eprint {http://arxiv.org/abs/1706.01812}
  {arXiv:1706.01812 [gr-qc]} \BibitemShut {NoStop}%
%%CITATION = ARXIV:1706.01812;%%
\bibitem [{\citenamefont {Abbott}\ \emph
  {et~al.}(2017{\natexlab{b}})\citenamefont {Abbott} \emph
  {et~al.}}]{Abbott:2017gyy}%
  \BibitemOpen
  \bibfield  {author} {\bibinfo {author} {\bibfnamefont {B.~P.}\ \bibnamefont
  {Abbott}} \emph {et~al.} (\bibinfo {collaboration} {Virgo, LIGO
  Scientific}),\ }\href {\doibase 10.3847/2041-8213/aa9f0c} {\bibfield
  {journal} {\bibinfo  {journal} {Astrophys. J.}\ }\textbf {\bibinfo {volume}
  {851}},\ \bibinfo {pages} {L35} (\bibinfo {year} {2017}{\natexlab{b}})},\
  \Eprint {http://arxiv.org/abs/1711.05578} {arXiv:1711.05578 [astro-ph.HE]}
  \BibitemShut {NoStop}%
%%CITATION = ARXIV:1711.05578;%%
\bibitem [{\citenamefont {Abbott}\ \emph
  {et~al.}(2017{\natexlab{c}})\citenamefont {Abbott} \emph
  {et~al.}}]{Abbott:2017oio}%
  \BibitemOpen
  \bibfield  {author} {\bibinfo {author} {\bibfnamefont {B.~P.}\ \bibnamefont
  {Abbott}} \emph {et~al.} (\bibinfo {collaboration} {Virgo, LIGO
  Scientific}),\ }\href {\doibase 10.1103/PhysRevLett.119.141101} {\bibfield
  {journal} {\bibinfo  {journal} {Phys. Rev. Lett.}\ }\textbf {\bibinfo
  {volume} {119}},\ \bibinfo {pages} {141101} (\bibinfo {year}
  {2017}{\natexlab{c}})}\BibitemShut {NoStop}%
%%CITATION = ARXIV:1709.09660;%%
\bibitem [{\citenamefont {Abbott}\ \emph
  {et~al.}(2017{\natexlab{d}})\citenamefont {Abbott} \emph
  {et~al.}}]{TheLIGOScientific:2017qsa}%
  \BibitemOpen
  \bibfield  {author} {\bibinfo {author} {\bibfnamefont {B.~P.}\ \bibnamefont
  {Abbott}} \emph {et~al.} (\bibinfo {collaboration} {Virgo, LIGO
  Scientific}),\ }\href {\doibase 10.1103/PhysRevLett.119.161101} {\bibfield
  {journal} {\bibinfo  {journal} {Phys. Rev. Lett.}\ }\textbf {\bibinfo
  {volume} {119}},\ \bibinfo {pages} {161101} (\bibinfo {year}
  {2017}{\natexlab{d}})}\BibitemShut {NoStop}%
%%CITATION = ARXIV:1710.05832;%%
\bibitem [{\citenamefont {Abbott}\ \emph {et~al.}(2018)\citenamefont {Abbott}
  \emph {et~al.}}]{LIGOScientific:2018mvr}%
  \BibitemOpen
  \bibfield  {author} {\bibinfo {author} {\bibfnamefont {B.~P.}\ \bibnamefont
  {Abbott}} \emph {et~al.} (\bibinfo {collaboration} {LIGO Scientific,
  Virgo}),\ }\href@noop {} {\  (\bibinfo {year} {2018})},\ \Eprint
  {http://arxiv.org/abs/1811.12907} {arXiv:1811.12907 [astro-ph.HE]}
  \BibitemShut {NoStop}%
%%CITATION = ARXIV:1811.12907;%%
\bibitem [{\citenamefont {Romero-Shaw}\ \emph {et~al.}(2019)\citenamefont
  {Romero-Shaw}, \citenamefont {Lasky},\ and\ \citenamefont
  {Thrane}}]{Romero-Shaw:2019itr}%
  \BibitemOpen
  \bibfield  {author} {\bibinfo {author} {\bibfnamefont {I.~M.}\ \bibnamefont
  {Romero-Shaw}}, \bibinfo {author} {\bibfnamefont {P.~D.}\ \bibnamefont
  {Lasky}}, \ and\ \bibinfo {author} {\bibfnamefont {E.}~\bibnamefont
  {Thrane}},\ }\href@noop {} {\  (\bibinfo {year} {2019})},\ \Eprint
  {http://arxiv.org/abs/1909.05466} {arXiv:1909.05466 [astro-ph.HE]}
  \BibitemShut {NoStop}%
%%CITATION = ARXIV:1909.05466;%%
\bibitem [{\citenamefont {Peters}\ and\ \citenamefont
  {Mathews}(1963)}]{PhysRev.131.435}%
  \BibitemOpen
  \bibfield  {author} {\bibinfo {author} {\bibfnamefont {P.~C.}\ \bibnamefont
  {Peters}}\ and\ \bibinfo {author} {\bibfnamefont {J.}~\bibnamefont
  {Mathews}},\ }\href {\doibase 10.1103/PhysRev.131.435} {\bibfield  {journal}
  {\bibinfo  {journal} {Phys. Rev.}\ }\textbf {\bibinfo {volume} {131}},\
  \bibinfo {pages} {435} (\bibinfo {year} {1963})}\BibitemShut {NoStop}%
\bibitem [{\citenamefont {Peters}(1964)}]{PhysRev.136.B1224}%
  \BibitemOpen
  \bibfield  {author} {\bibinfo {author} {\bibfnamefont {P.~C.}\ \bibnamefont
  {Peters}},\ }\href {\doibase 10.1103/PhysRev.136.B1224} {\bibfield  {journal}
  {\bibinfo  {journal} {Phys. Rev.}\ }\textbf {\bibinfo {volume} {136}},\
  \bibinfo {pages} {B1224} (\bibinfo {year} {1964})}\BibitemShut {NoStop}%
\bibitem [{\citenamefont {Husa}\ \emph {et~al.}(2015)\citenamefont {Husa},
  \citenamefont {Khan}, \citenamefont {Hannam}, \citenamefont {P\"{u}rrer},
  \citenamefont {Ohme}, \citenamefont {Jim\'{e}nez~Forteza},\ and\
  \citenamefont {Boh\'{e}}}]{Husa2015}%
  \BibitemOpen
  \bibfield  {author} {\bibinfo {author} {\bibfnamefont {S.}~\bibnamefont
  {Husa}}, \bibinfo {author} {\bibfnamefont {S.}~\bibnamefont {Khan}}, \bibinfo
  {author} {\bibfnamefont {M.}~\bibnamefont {Hannam}}, \bibinfo {author}
  {\bibfnamefont {M.}~\bibnamefont {P\"{u}rrer}}, \bibinfo {author}
  {\bibfnamefont {F.}~\bibnamefont {Ohme}}, \bibinfo {author} {\bibfnamefont
  {F.}~\bibnamefont {Jim\'{e}nez~Forteza}}, \ and\ \bibinfo {author}
  {\bibfnamefont {A.}~\bibnamefont {Boh\'{e}}},\ }\href@noop {} {\  (\bibinfo
  {year} {2015})}\BibitemShut {NoStop}%
\bibitem [{\citenamefont {Khan}\ \emph {et~al.}(2016)\citenamefont {Khan},
  \citenamefont {Husa}, \citenamefont {Hannam}, \citenamefont {Ohme},
  \citenamefont {P\"urrer}, \citenamefont {Forteza},\ and\ \citenamefont
  {Boh\'e}}]{PhysRevD.93.044007}%
  \BibitemOpen
  \bibfield  {author} {\bibinfo {author} {\bibfnamefont {S.}~\bibnamefont
  {Khan}}, \bibinfo {author} {\bibfnamefont {S.}~\bibnamefont {Husa}}, \bibinfo
  {author} {\bibfnamefont {M.}~\bibnamefont {Hannam}}, \bibinfo {author}
  {\bibfnamefont {F.}~\bibnamefont {Ohme}}, \bibinfo {author} {\bibfnamefont
  {M.}~\bibnamefont {P\"urrer}}, \bibinfo {author} {\bibfnamefont {X.~J.}\
  \bibnamefont {Forteza}}, \ and\ \bibinfo {author} {\bibfnamefont
  {A.}~\bibnamefont {Boh\'e}},\ }\href {\doibase 10.1103/PhysRevD.93.044007}
  {\bibfield  {journal} {\bibinfo  {journal} {Phys. Rev. D}\ }\textbf {\bibinfo
  {volume} {93}},\ \bibinfo {pages} {044007} (\bibinfo {year}
  {2016})}\BibitemShut {NoStop}%
\bibitem [{\citenamefont {Hannam}\ \emph {et~al.}(2014)\citenamefont {Hannam},
  \citenamefont {Schmidt}, \citenamefont {Boh\'e}, \citenamefont {Haegel},
  \citenamefont {Husa}, \citenamefont {Ohme}, \citenamefont {Pratten},\ and\
  \citenamefont {P\"urrer}}]{phenomp}%
  \BibitemOpen
  \bibfield  {author} {\bibinfo {author} {\bibfnamefont {M.}~\bibnamefont
  {Hannam}}, \bibinfo {author} {\bibfnamefont {P.}~\bibnamefont {Schmidt}},
  \bibinfo {author} {\bibfnamefont {A.}~\bibnamefont {Boh\'e}}, \bibinfo
  {author} {\bibfnamefont {L.}~\bibnamefont {Haegel}}, \bibinfo {author}
  {\bibfnamefont {S.}~\bibnamefont {Husa}}, \bibinfo {author} {\bibfnamefont
  {F.}~\bibnamefont {Ohme}}, \bibinfo {author} {\bibfnamefont {G.}~\bibnamefont
  {Pratten}}, \ and\ \bibinfo {author} {\bibfnamefont {M.}~\bibnamefont
  {P\"urrer}},\ }\href {\doibase 10.1103/PhysRevLett.113.151101} {\bibfield
  {journal} {\bibinfo  {journal} {Phys. Rev. Lett.}\ }\textbf {\bibinfo
  {volume} {113}},\ \bibinfo {pages} {151101} (\bibinfo {year}
  {2014})}\BibitemShut {NoStop}%
\bibitem [{\citenamefont {Boh\'e}\ \emph {et~al.}(2017)\citenamefont {Boh\'e},
  \citenamefont {Shao}, \citenamefont {Taracchini}, \citenamefont {Buonanno},
  \citenamefont {Babak}, \citenamefont {Harry}, \citenamefont {Hinder},
  \citenamefont {Ossokine}, \citenamefont {P\"urrer}, \citenamefont {Raymond},
  \citenamefont {Chu}, \citenamefont {Fong}, \citenamefont {Kumar},
  \citenamefont {Pfeiffer}, \citenamefont {Boyle}, \citenamefont {Hemberger},
  \citenamefont {Kidder}, \citenamefont {Lovelace}, \citenamefont {Scheel},\
  and\ \citenamefont {Szil\'agyi}}]{seobnrv4}%
  \BibitemOpen
  \bibfield  {author} {\bibinfo {author} {\bibfnamefont {A.}~\bibnamefont
  {Boh\'e}}, \bibinfo {author} {\bibfnamefont {L.}~\bibnamefont {Shao}},
  \bibinfo {author} {\bibfnamefont {A.}~\bibnamefont {Taracchini}}, \bibinfo
  {author} {\bibfnamefont {A.}~\bibnamefont {Buonanno}}, \bibinfo {author}
  {\bibfnamefont {S.}~\bibnamefont {Babak}}, \bibinfo {author} {\bibfnamefont
  {I.~W.}\ \bibnamefont {Harry}}, \bibinfo {author} {\bibfnamefont
  {I.}~\bibnamefont {Hinder}}, \bibinfo {author} {\bibfnamefont
  {S.}~\bibnamefont {Ossokine}}, \bibinfo {author} {\bibfnamefont
  {M.}~\bibnamefont {P\"urrer}}, \bibinfo {author} {\bibfnamefont
  {V.}~\bibnamefont {Raymond}}, \bibinfo {author} {\bibfnamefont
  {T.}~\bibnamefont {Chu}}, \bibinfo {author} {\bibfnamefont {H.}~\bibnamefont
  {Fong}}, \bibinfo {author} {\bibfnamefont {P.}~\bibnamefont {Kumar}},
  \bibinfo {author} {\bibfnamefont {H.~P.}\ \bibnamefont {Pfeiffer}}, \bibinfo
  {author} {\bibfnamefont {M.}~\bibnamefont {Boyle}}, \bibinfo {author}
  {\bibfnamefont {D.~A.}\ \bibnamefont {Hemberger}}, \bibinfo {author}
  {\bibfnamefont {L.~E.}\ \bibnamefont {Kidder}}, \bibinfo {author}
  {\bibfnamefont {G.}~\bibnamefont {Lovelace}}, \bibinfo {author}
  {\bibfnamefont {M.~A.}\ \bibnamefont {Scheel}}, \ and\ \bibinfo {author}
  {\bibfnamefont {B.}~\bibnamefont {Szil\'agyi}},\ }\href {\doibase
  10.1103/PhysRevD.95.044028} {\bibfield  {journal} {\bibinfo  {journal} {Phys.
  Rev. D}\ }\textbf {\bibinfo {volume} {95}},\ \bibinfo {pages} {044028}
  (\bibinfo {year} {2017})}\BibitemShut {NoStop}%
\bibitem [{\citenamefont {London}\ \emph {et~al.}(2018)\citenamefont {London},
  \citenamefont {Khan}, \citenamefont {Fauchon-Jones}, \citenamefont
  {Garc\'{\i}a}, \citenamefont {Hannam}, \citenamefont {Husa}, \citenamefont
  {Jim\'enez-Forteza}, \citenamefont {Kalaghatgi}, \citenamefont {Ohme},\ and\
  \citenamefont {Pannarale}}]{london2018}%
  \BibitemOpen
  \bibfield  {author} {\bibinfo {author} {\bibfnamefont {L.}~\bibnamefont
  {London}}, \bibinfo {author} {\bibfnamefont {S.}~\bibnamefont {Khan}},
  \bibinfo {author} {\bibfnamefont {E.}~\bibnamefont {Fauchon-Jones}}, \bibinfo
  {author} {\bibfnamefont {C.}~\bibnamefont {Garc\'{\i}a}}, \bibinfo {author}
  {\bibfnamefont {M.}~\bibnamefont {Hannam}}, \bibinfo {author} {\bibfnamefont
  {S.}~\bibnamefont {Husa}}, \bibinfo {author} {\bibfnamefont {X.}~\bibnamefont
  {Jim\'enez-Forteza}}, \bibinfo {author} {\bibfnamefont {C.}~\bibnamefont
  {Kalaghatgi}}, \bibinfo {author} {\bibfnamefont {F.}~\bibnamefont {Ohme}}, \
  and\ \bibinfo {author} {\bibfnamefont {F.}~\bibnamefont {Pannarale}},\ }\href
  {\doibase 10.1103/PhysRevLett.120.161102} {\bibfield  {journal} {\bibinfo
  {journal} {Phys. Rev. Lett.}\ }\textbf {\bibinfo {volume} {120}},\ \bibinfo
  {pages} {161102} (\bibinfo {year} {2018})}\BibitemShut {NoStop}%
\bibitem [{\citenamefont {Cotesta}\ \emph {et~al.}(2018)\citenamefont
  {Cotesta}, \citenamefont {Buonanno}, \citenamefont {Boh\'e}, \citenamefont
  {Taracchini}, \citenamefont {Hinder},\ and\ \citenamefont
  {Ossokine}}]{PhysRevD.98.084028}%
  \BibitemOpen
  \bibfield  {author} {\bibinfo {author} {\bibfnamefont {R.}~\bibnamefont
  {Cotesta}}, \bibinfo {author} {\bibfnamefont {A.}~\bibnamefont {Buonanno}},
  \bibinfo {author} {\bibfnamefont {A.}~\bibnamefont {Boh\'e}}, \bibinfo
  {author} {\bibfnamefont {A.}~\bibnamefont {Taracchini}}, \bibinfo {author}
  {\bibfnamefont {I.}~\bibnamefont {Hinder}}, \ and\ \bibinfo {author}
  {\bibfnamefont {S.}~\bibnamefont {Ossokine}},\ }\href {\doibase
  10.1103/PhysRevD.98.084028} {\bibfield  {journal} {\bibinfo  {journal} {Phys.
  Rev. D}\ }\textbf {\bibinfo {volume} {98}},\ \bibinfo {pages} {084028}
  (\bibinfo {year} {2018})}\BibitemShut {NoStop}%
\bibitem [{\citenamefont {Blackman}\ \emph {et~al.}(2017)\citenamefont
  {Blackman}, \citenamefont {Field}, \citenamefont {Scheel}, \citenamefont
  {Galley}, \citenamefont {Hemberger}, \citenamefont {Schmidt},\ and\
  \citenamefont {Smith}}]{Blackman:2017dfb}%
  \BibitemOpen
  \bibfield  {author} {\bibinfo {author} {\bibfnamefont {J.}~\bibnamefont
  {Blackman}}, \bibinfo {author} {\bibfnamefont {S.~E.}\ \bibnamefont {Field}},
  \bibinfo {author} {\bibfnamefont {M.~A.}\ \bibnamefont {Scheel}}, \bibinfo
  {author} {\bibfnamefont {C.~R.}\ \bibnamefont {Galley}}, \bibinfo {author}
  {\bibfnamefont {D.~A.}\ \bibnamefont {Hemberger}}, \bibinfo {author}
  {\bibfnamefont {P.}~\bibnamefont {Schmidt}}, \ and\ \bibinfo {author}
  {\bibfnamefont {R.}~\bibnamefont {Smith}},\ }\href {\doibase
  10.1103/PhysRevD.95.104023} {\bibfield  {journal} {\bibinfo  {journal} {Phys.
  Rev.}\ }\textbf {\bibinfo {volume} {D95}},\ \bibinfo {pages} {104023}
  (\bibinfo {year} {2017})},\ \Eprint {http://arxiv.org/abs/1701.00550}
  {arXiv:1701.00550 [gr-qc]} \BibitemShut {NoStop}%
%%CITATION = ARXIV:1701.00550;%%
\bibitem [{\citenamefont {Varma}\ \emph
  {et~al.}(2019{\natexlab{a}})\citenamefont {Varma}, \citenamefont {Field},
  \citenamefont {Scheel}, \citenamefont {Blackman}, \citenamefont {Gerosa},
  \citenamefont {Stein}, \citenamefont {Kidder},\ and\ \citenamefont
  {Pfeiffer}}]{PhysRevResearch.1.033015}%
  \BibitemOpen
  \bibfield  {author} {\bibinfo {author} {\bibfnamefont {V.}~\bibnamefont
  {Varma}}, \bibinfo {author} {\bibfnamefont {S.~E.}\ \bibnamefont {Field}},
  \bibinfo {author} {\bibfnamefont {M.~A.}\ \bibnamefont {Scheel}}, \bibinfo
  {author} {\bibfnamefont {J.}~\bibnamefont {Blackman}}, \bibinfo {author}
  {\bibfnamefont {D.}~\bibnamefont {Gerosa}}, \bibinfo {author} {\bibfnamefont
  {L.~C.}\ \bibnamefont {Stein}}, \bibinfo {author} {\bibfnamefont {L.~E.}\
  \bibnamefont {Kidder}}, \ and\ \bibinfo {author} {\bibfnamefont {H.~P.}\
  \bibnamefont {Pfeiffer}},\ }\href {\doibase 10.1103/PhysRevResearch.1.033015}
  {\bibfield  {journal} {\bibinfo  {journal} {Phys. Rev. Research}\ }\textbf
  {\bibinfo {volume} {1}},\ \bibinfo {pages} {033015} (\bibinfo {year}
  {2019}{\natexlab{a}})}\BibitemShut {NoStop}%
\bibitem [{\citenamefont {Varma}\ \emph
  {et~al.}(2019{\natexlab{b}})\citenamefont {Varma}, \citenamefont {Field},
  \citenamefont {Scheel}, \citenamefont {Blackman}, \citenamefont {Kidder},\
  and\ \citenamefont {Pfeiffer}}]{PhysRevD.99.064045}%
  \BibitemOpen
  \bibfield  {author} {\bibinfo {author} {\bibfnamefont {V.}~\bibnamefont
  {Varma}}, \bibinfo {author} {\bibfnamefont {S.~E.}\ \bibnamefont {Field}},
  \bibinfo {author} {\bibfnamefont {M.~A.}\ \bibnamefont {Scheel}}, \bibinfo
  {author} {\bibfnamefont {J.}~\bibnamefont {Blackman}}, \bibinfo {author}
  {\bibfnamefont {L.~E.}\ \bibnamefont {Kidder}}, \ and\ \bibinfo {author}
  {\bibfnamefont {H.~P.}\ \bibnamefont {Pfeiffer}},\ }\href {\doibase
  10.1103/PhysRevD.99.064045} {\bibfield  {journal} {\bibinfo  {journal} {Phys.
  Rev. D}\ }\textbf {\bibinfo {volume} {99}},\ \bibinfo {pages} {064045}
  (\bibinfo {year} {2019}{\natexlab{b}})}\BibitemShut {NoStop}%
\bibitem [{\citenamefont {Khan}\ \emph {et~al.}(2018)\citenamefont {Khan},
  \citenamefont {Chatziioannou}, \citenamefont {Hannam},\ and\ \citenamefont
  {Ohme}}]{Khan:2018fmp}%
  \BibitemOpen
  \bibfield  {author} {\bibinfo {author} {\bibfnamefont {S.}~\bibnamefont
  {Khan}}, \bibinfo {author} {\bibfnamefont {K.}~\bibnamefont {Chatziioannou}},
  \bibinfo {author} {\bibfnamefont {M.}~\bibnamefont {Hannam}}, \ and\ \bibinfo
  {author} {\bibfnamefont {F.}~\bibnamefont {Ohme}},\ }\href@noop {} {\
  (\bibinfo {year} {2018})},\ \Eprint {http://arxiv.org/abs/1809.10113}
  {arXiv:1809.10113 [gr-qc]} \BibitemShut {NoStop}%
%%CITATION = ARXIV:1809.10113;%%
\bibitem [{\citenamefont {Khan}\ \emph {et~al.}(2019)\citenamefont {Khan},
  \citenamefont {Ohme}, \citenamefont {Chatziioannou},\ and\ \citenamefont
  {Hannam}}]{Khan:2019kot}%
  \BibitemOpen
  \bibfield  {author} {\bibinfo {author} {\bibfnamefont {S.}~\bibnamefont
  {Khan}}, \bibinfo {author} {\bibfnamefont {F.}~\bibnamefont {Ohme}}, \bibinfo
  {author} {\bibfnamefont {K.}~\bibnamefont {Chatziioannou}}, \ and\ \bibinfo
  {author} {\bibfnamefont {M.}~\bibnamefont {Hannam}},\ }\href@noop {} {\
  (\bibinfo {year} {2019})},\ \Eprint {http://arxiv.org/abs/1911.06050}
  {arXiv:1911.06050 [gr-qc]} \BibitemShut {NoStop}%
%%CITATION = ARXIV:1911.06050;%%
\bibitem [{\citenamefont {Pratten}\ \emph {et~al.}(2020)\citenamefont
  {Pratten}, \citenamefont {Husa}, \citenamefont {Garc{\'i}a-Quir{\'o}s},
  \citenamefont {Colleoni}, \citenamefont {Ramos-Buades}, \citenamefont
  {Estell{\'e}s},\ and\ \citenamefont {Jaume}}]{phenX}%
  \BibitemOpen
  \bibfield  {author} {\bibinfo {author} {\bibfnamefont {G.}~\bibnamefont
  {Pratten}}, \bibinfo {author} {\bibfnamefont {S.}~\bibnamefont {Husa}},
  \bibinfo {author} {\bibfnamefont {C.}~\bibnamefont {Garc{\'i}a-Quir{\'o}s}},
  \bibinfo {author} {\bibfnamefont {M.}~\bibnamefont {Colleoni}}, \bibinfo
  {author} {\bibfnamefont {A.}~\bibnamefont {Ramos-Buades}}, \bibinfo {author}
  {\bibfnamefont {H.}~\bibnamefont {Estell{\'e}s}}, \ and\ \bibinfo {author}
  {\bibfnamefont {R.}~\bibnamefont {Jaume}},\ }\href@noop {} {\  (\bibinfo
  {year} {2020})},\ \Eprint {http://arxiv.org/abs/2001.11412} {arXiv:2001.11412
  [gr-qc]} \BibitemShut {NoStop}%
%%CITATION = ARXIV:2001.11412;%%
\bibitem [{\citenamefont {Garc{\'i}a-Quir{\'o}s}\ \emph
  {et~al.}(2020)\citenamefont {Garc{\'i}a-Quir{\'o}s}, \citenamefont
  {Colleoni}, \citenamefont {Husa}, \citenamefont {Estell{\'e}s}, \citenamefont
  {Pratten}, \citenamefont {Ramos-Buades}, \citenamefont {Mateu-Lucena},\ and\
  \citenamefont {Jaume}}]{phenXHM}%
  \BibitemOpen
  \bibfield  {author} {\bibinfo {author} {\bibfnamefont {C.}~\bibnamefont
  {Garc{\'i}a-Quir{\'o}s}}, \bibinfo {author} {\bibfnamefont {M.}~\bibnamefont
  {Colleoni}}, \bibinfo {author} {\bibfnamefont {S.}~\bibnamefont {Husa}},
  \bibinfo {author} {\bibfnamefont {H.}~\bibnamefont {Estell{\'e}s}}, \bibinfo
  {author} {\bibfnamefont {G.}~\bibnamefont {Pratten}}, \bibinfo {author}
  {\bibfnamefont {A.}~\bibnamefont {Ramos-Buades}}, \bibinfo {author}
  {\bibfnamefont {M.}~\bibnamefont {Mateu-Lucena}}, \ and\ \bibinfo {author}
  {\bibfnamefont {R.}~\bibnamefont {Jaume}},\ }\href@noop {} {\  (\bibinfo
  {year} {2020})},\ \Eprint {http://arxiv.org/abs/2001.10914} {arXiv:2001.10914
  [gr-qc]} \BibitemShut {NoStop}%
%%CITATION = ARXIV:2001.10914;%%
\bibitem [{\citenamefont {Estell{\'e}s}\ \emph {et~al.}(2020)\citenamefont
  {Estell{\'e}s}, \citenamefont {Ramos-Buades}, \citenamefont {Husa},
  \citenamefont {Garc{\'i}a-Quir{\'o}s},\ and\ \citenamefont {Haegel}}]{phenT}%
  \BibitemOpen
  \bibfield  {author} {\bibinfo {author} {\bibfnamefont {H.}~\bibnamefont
  {Estell{\'e}s}}, \bibinfo {author} {\bibfnamefont {A.}~\bibnamefont
  {Ramos-Buades}}, \bibinfo {author} {\bibfnamefont {S.}~\bibnamefont {Husa}},
  \bibinfo {author} {\bibfnamefont {C.}~\bibnamefont {Garc{\'i}a-Quir{\'o}s}},
  \ and\ \bibinfo {author} {\bibfnamefont {L.}~\bibnamefont {Haegel}},\
  }\href@noop {} {\  (\bibinfo {year} {2020})},\ \bibinfo {note} {in
  preparation}\BibitemShut {NoStop}%
\bibitem [{\citenamefont {Belczynski}\ \emph {et~al.}(2016)\citenamefont
  {Belczynski}, \citenamefont {Holz}, \citenamefont {Bulik},\ and\
  \citenamefont {O'Shaughnessy}}]{Belczynski:2016obo}%
  \BibitemOpen
  \bibfield  {author} {\bibinfo {author} {\bibfnamefont {K.}~\bibnamefont
  {Belczynski}}, \bibinfo {author} {\bibfnamefont {D.~E.}\ \bibnamefont
  {Holz}}, \bibinfo {author} {\bibfnamefont {T.}~\bibnamefont {Bulik}}, \ and\
  \bibinfo {author} {\bibfnamefont {R.}~\bibnamefont {O'Shaughnessy}},\ }\href
  {\doibase 10.1038/nature18322} {\bibfield  {journal} {\bibinfo  {journal}
  {Nature}\ }\textbf {\bibinfo {volume} {534}},\ \bibinfo {pages} {512}
  (\bibinfo {year} {2016})},\ \Eprint {http://arxiv.org/abs/1602.04531}
  {arXiv:1602.04531 [astro-ph.HE]} \BibitemShut {NoStop}%
%%CITATION = ARXIV:1602.04531;%%
\bibitem [{\citenamefont {Park}\ \emph {et~al.}(2017)\citenamefont {Park},
  \citenamefont {Kim}, \citenamefont {Lee}, \citenamefont {Bae},\ and\
  \citenamefont {Belczynski}}]{Park:2017zgj}%
  \BibitemOpen
  \bibfield  {author} {\bibinfo {author} {\bibfnamefont {D.}~\bibnamefont
  {Park}}, \bibinfo {author} {\bibfnamefont {C.}~\bibnamefont {Kim}}, \bibinfo
  {author} {\bibfnamefont {H.~M.}\ \bibnamefont {Lee}}, \bibinfo {author}
  {\bibfnamefont {Y.-B.}\ \bibnamefont {Bae}}, \ and\ \bibinfo {author}
  {\bibfnamefont {K.}~\bibnamefont {Belczynski}},\ }\href {\doibase
  10.1093/mnras/stx1015} {\bibfield  {journal} {\bibinfo  {journal} {Mon. Not.
  Roy. Astron. Soc.}\ }\textbf {\bibinfo {volume} {469}},\ \bibinfo {pages}
  {4665} (\bibinfo {year} {2017})},\ \Eprint {http://arxiv.org/abs/1703.01568}
  {arXiv:1703.01568 [astro-ph.HE]} \BibitemShut {NoStop}%
%%CITATION = ARXIV:1703.01568;%%
\bibitem [{\citenamefont {Samsing}(2018)}]{PhysRevD.97.103014}%
  \BibitemOpen
  \bibfield  {author} {\bibinfo {author} {\bibfnamefont {J.}~\bibnamefont
  {Samsing}},\ }\href {\doibase 10.1103/PhysRevD.97.103014} {\bibfield
  {journal} {\bibinfo  {journal} {Phys. Rev. D}\ }\textbf {\bibinfo {volume}
  {97}},\ \bibinfo {pages} {103014} (\bibinfo {year} {2018})}\BibitemShut
  {NoStop}%
\bibitem [{\citenamefont {Samsing}\ \emph {et~al.}(2014)\citenamefont
  {Samsing}, \citenamefont {MacLeod},\ and\ \citenamefont
  {Ramirez-Ruiz}}]{Samsing:2013kua}%
  \BibitemOpen
  \bibfield  {author} {\bibinfo {author} {\bibfnamefont {J.}~\bibnamefont
  {Samsing}}, \bibinfo {author} {\bibfnamefont {M.}~\bibnamefont {MacLeod}}, \
  and\ \bibinfo {author} {\bibfnamefont {E.}~\bibnamefont {Ramirez-Ruiz}},\
  }\href {\doibase 10.1088/0004-637X/784/1/71} {\bibfield  {journal} {\bibinfo
  {journal} {Astrophys. J.}\ }\textbf {\bibinfo {volume} {784}},\ \bibinfo
  {pages} {71} (\bibinfo {year} {2014})},\ \Eprint
  {http://arxiv.org/abs/1308.2964} {arXiv:1308.2964 [astro-ph.HE]} \BibitemShut
  {NoStop}%
%%CITATION = ARXIV:1308.2964;%%
\bibitem [{\citenamefont {Damour}\ and\ \citenamefont
  {Deruelle}(1985)}]{1985AIHS...43..107D}%
  \BibitemOpen
  \bibfield  {author} {\bibinfo {author} {\bibfnamefont {T.}~\bibnamefont
  {Damour}}\ and\ \bibinfo {author} {\bibfnamefont {N.}~\bibnamefont
  {Deruelle}},\ }\href@noop {} {\bibfield  {journal} {\bibinfo  {journal}
  {Ann.~Inst.~Henri Poincar{\'e} Phys.~Th{\'e}or., Vol.~43, No.~1, p.~107 -
  132}\ } (\bibinfo {year} {1985})}\BibitemShut {NoStop}%
\bibitem [{\citenamefont {Damour}\ \emph {et~al.}(2004)\citenamefont {Damour},
  \citenamefont {Gopakumar},\ and\ \citenamefont {Iyer}}]{PhysRevD.70.064028}%
  \BibitemOpen
  \bibfield  {author} {\bibinfo {author} {\bibfnamefont {T.}~\bibnamefont
  {Damour}}, \bibinfo {author} {\bibfnamefont {A.}~\bibnamefont {Gopakumar}}, \
  and\ \bibinfo {author} {\bibfnamefont {B.~R.}\ \bibnamefont {Iyer}},\ }\href
  {\doibase 10.1103/PhysRevD.70.064028} {\bibfield  {journal} {\bibinfo
  {journal} {Phys. Rev. D}\ }\textbf {\bibinfo {volume} {70}},\ \bibinfo
  {pages} {064028} (\bibinfo {year} {2004})}\BibitemShut {NoStop}%
\bibitem [{\citenamefont {Memmesheimer}\ \emph {et~al.}(2004)\citenamefont
  {Memmesheimer}, \citenamefont {Gopakumar},\ and\ \citenamefont
  {Sch\"afer}}]{PhysRevD.70.104011}%
  \BibitemOpen
  \bibfield  {author} {\bibinfo {author} {\bibfnamefont {R.-M.}\ \bibnamefont
  {Memmesheimer}}, \bibinfo {author} {\bibfnamefont {A.}~\bibnamefont
  {Gopakumar}}, \ and\ \bibinfo {author} {\bibfnamefont {G.}~\bibnamefont
  {Sch\"afer}},\ }\href {\doibase 10.1103/PhysRevD.70.104011} {\bibfield
  {journal} {\bibinfo  {journal} {Phys. Rev. D}\ }\textbf {\bibinfo {volume}
  {70}},\ \bibinfo {pages} {104011} (\bibinfo {year} {2004})}\BibitemShut
  {NoStop}%
\bibitem [{\citenamefont {Tanay}\ \emph {et~al.}(2016)\citenamefont {Tanay},
  \citenamefont {Haney},\ and\ \citenamefont {Gopakumar}}]{PhysRevD.93.064031}%
  \BibitemOpen
  \bibfield  {author} {\bibinfo {author} {\bibfnamefont {S.}~\bibnamefont
  {Tanay}}, \bibinfo {author} {\bibfnamefont {M.}~\bibnamefont {Haney}}, \ and\
  \bibinfo {author} {\bibfnamefont {A.}~\bibnamefont {Gopakumar}},\ }\href
  {\doibase 10.1103/PhysRevD.93.064031} {\bibfield  {journal} {\bibinfo
  {journal} {Phys. Rev. D}\ }\textbf {\bibinfo {volume} {93}},\ \bibinfo
  {pages} {064031} (\bibinfo {year} {2016})}\BibitemShut {NoStop}%
\bibitem [{\citenamefont {Moore}\ \emph {et~al.}(2018)\citenamefont {Moore},
  \citenamefont {Robson}, \citenamefont {Loutrel},\ and\ \citenamefont
  {Yunes}}]{Moore:2018kvz}%
  \BibitemOpen
  \bibfield  {author} {\bibinfo {author} {\bibfnamefont {B.}~\bibnamefont
  {Moore}}, \bibinfo {author} {\bibfnamefont {T.}~\bibnamefont {Robson}},
  \bibinfo {author} {\bibfnamefont {N.}~\bibnamefont {Loutrel}}, \ and\
  \bibinfo {author} {\bibfnamefont {N.}~\bibnamefont {Yunes}},\ }\href
  {\doibase 10.1088/1361-6382/aaea00} {\bibfield  {journal} {\bibinfo
  {journal} {Class. Quant. Grav.}\ }\textbf {\bibinfo {volume} {35}},\ \bibinfo
  {pages} {235006} (\bibinfo {year} {2018})},\ \Eprint
  {http://arxiv.org/abs/1807.07163} {arXiv:1807.07163 [gr-qc]} \BibitemShut
  {NoStop}%
%%CITATION = ARXIV:1807.07163;%%
\bibitem [{\citenamefont {Moore}\ and\ \citenamefont
  {Yunes}(2019{\natexlab{a}})}]{Moore:2019xkm}%
  \BibitemOpen
  \bibfield  {author} {\bibinfo {author} {\bibfnamefont {B.}~\bibnamefont
  {Moore}}\ and\ \bibinfo {author} {\bibfnamefont {N.}~\bibnamefont {Yunes}},\
  }\href {\doibase 10.1088/1361-6382/ab3778} {\bibfield  {journal} {\bibinfo
  {journal} {Class. Quant. Grav.}\ }\textbf {\bibinfo {volume} {36}},\ \bibinfo
  {pages} {185003} (\bibinfo {year} {2019}{\natexlab{a}})},\ \Eprint
  {http://arxiv.org/abs/1903.05203} {arXiv:1903.05203 [gr-qc]} \BibitemShut
  {NoStop}%
%%CITATION = ARXIV:1903.05203;%%
\bibitem [{\citenamefont {Loutrel}\ \emph {et~al.}(2019)\citenamefont
  {Loutrel}, \citenamefont {Liebersbach}, \citenamefont {Yunes},\ and\
  \citenamefont {Cornish}}]{Loutrel:2018ydu}%
  \BibitemOpen
  \bibfield  {author} {\bibinfo {author} {\bibfnamefont {N.}~\bibnamefont
  {Loutrel}}, \bibinfo {author} {\bibfnamefont {S.}~\bibnamefont
  {Liebersbach}}, \bibinfo {author} {\bibfnamefont {N.}~\bibnamefont {Yunes}},
  \ and\ \bibinfo {author} {\bibfnamefont {N.}~\bibnamefont {Cornish}},\ }\href
  {\doibase 10.1088/1361-6382/aaf2a9} {\bibfield  {journal} {\bibinfo
  {journal} {Class. Quant. Grav.}\ }\textbf {\bibinfo {volume} {36}},\ \bibinfo
  {pages} {025004} (\bibinfo {year} {2019})},\ \Eprint
  {http://arxiv.org/abs/1810.03521} {arXiv:1810.03521 [gr-qc]} \BibitemShut
  {NoStop}%
%%CITATION = ARXIV:1810.03521;%%
\bibitem [{\citenamefont {Moore}\ and\ \citenamefont
  {Yunes}(2019{\natexlab{b}})}]{Moore_2019}%
  \BibitemOpen
  \bibfield  {author} {\bibinfo {author} {\bibfnamefont {B.}~\bibnamefont
  {Moore}}\ and\ \bibinfo {author} {\bibfnamefont {N.}~\bibnamefont {Yunes}},\
  }\href {\doibase 10.1088/1361-6382/ab3778} {\bibfield  {journal} {\bibinfo
  {journal} {Classical and Quantum Gravity}\ }\textbf {\bibinfo {volume}
  {36}},\ \bibinfo {pages} {185003} (\bibinfo {year}
  {2019}{\natexlab{b}})}\BibitemShut {NoStop}%
\bibitem [{\citenamefont {Huerta}\ \emph {et~al.}(2018)\citenamefont {Huerta},
  \citenamefont {Moore}, \citenamefont {Kumar}, \citenamefont {George},
  \citenamefont {Chua}, \citenamefont {Haas}, \citenamefont {Wessel},
  \citenamefont {Johnson}, \citenamefont {Glennon}, \citenamefont {Rebei},
  \citenamefont {Holgado}, \citenamefont {Gair},\ and\ \citenamefont
  {Pfeiffer}}]{PhysRevD.97.024031}%
  \BibitemOpen
  \bibfield  {author} {\bibinfo {author} {\bibfnamefont {E.~A.}\ \bibnamefont
  {Huerta}}, \bibinfo {author} {\bibfnamefont {C.~J.}\ \bibnamefont {Moore}},
  \bibinfo {author} {\bibfnamefont {P.}~\bibnamefont {Kumar}}, \bibinfo
  {author} {\bibfnamefont {D.}~\bibnamefont {George}}, \bibinfo {author}
  {\bibfnamefont {A.~J.~K.}\ \bibnamefont {Chua}}, \bibinfo {author}
  {\bibfnamefont {R.}~\bibnamefont {Haas}}, \bibinfo {author} {\bibfnamefont
  {E.}~\bibnamefont {Wessel}}, \bibinfo {author} {\bibfnamefont
  {D.}~\bibnamefont {Johnson}}, \bibinfo {author} {\bibfnamefont
  {D.}~\bibnamefont {Glennon}}, \bibinfo {author} {\bibfnamefont
  {A.}~\bibnamefont {Rebei}}, \bibinfo {author} {\bibfnamefont {A.~M.}\
  \bibnamefont {Holgado}}, \bibinfo {author} {\bibfnamefont {J.~R.}\
  \bibnamefont {Gair}}, \ and\ \bibinfo {author} {\bibfnamefont {H.~P.}\
  \bibnamefont {Pfeiffer}},\ }\href {\doibase 10.1103/PhysRevD.97.024031}
  {\bibfield  {journal} {\bibinfo  {journal} {Phys. Rev. D}\ }\textbf {\bibinfo
  {volume} {97}},\ \bibinfo {pages} {024031} (\bibinfo {year}
  {2018})}\BibitemShut {NoStop}%
\bibitem [{\citenamefont {Hinder}\ \emph {et~al.}(2018)\citenamefont {Hinder},
  \citenamefont {Kidder},\ and\ \citenamefont {Pfeiffer}}]{PhysRevD.98.044015}%
  \BibitemOpen
  \bibfield  {author} {\bibinfo {author} {\bibfnamefont {I.}~\bibnamefont
  {Hinder}}, \bibinfo {author} {\bibfnamefont {L.~E.}\ \bibnamefont {Kidder}},
  \ and\ \bibinfo {author} {\bibfnamefont {H.~P.}\ \bibnamefont {Pfeiffer}},\
  }\href {\doibase 10.1103/PhysRevD.98.044015} {\bibfield  {journal} {\bibinfo
  {journal} {Phys. Rev. D}\ }\textbf {\bibinfo {volume} {98}},\ \bibinfo
  {pages} {044015} (\bibinfo {year} {2018})}\BibitemShut {NoStop}%
\bibitem [{\citenamefont {Cao}\ and\ \citenamefont
  {Han}(2017)}]{PhysRevD.96.044028}%
  \BibitemOpen
  \bibfield  {author} {\bibinfo {author} {\bibfnamefont {Z.}~\bibnamefont
  {Cao}}\ and\ \bibinfo {author} {\bibfnamefont {W.-B.}\ \bibnamefont {Han}},\
  }\href {\doibase 10.1103/PhysRevD.96.044028} {\bibfield  {journal} {\bibinfo
  {journal} {Phys. Rev. D}\ }\textbf {\bibinfo {volume} {96}},\ \bibinfo
  {pages} {044028} (\bibinfo {year} {2017})}\BibitemShut {NoStop}%
\bibitem [{\citenamefont {Hinderer}\ and\ \citenamefont
  {Babak}(2017)}]{PhysRevD.96.104048}%
  \BibitemOpen
  \bibfield  {author} {\bibinfo {author} {\bibfnamefont {T.}~\bibnamefont
  {Hinderer}}\ and\ \bibinfo {author} {\bibfnamefont {S.}~\bibnamefont
  {Babak}},\ }\href {\doibase 10.1103/PhysRevD.96.104048} {\bibfield  {journal}
  {\bibinfo  {journal} {Phys. Rev. D}\ }\textbf {\bibinfo {volume} {96}},\
  \bibinfo {pages} {104048} (\bibinfo {year} {2017})}\BibitemShut {NoStop}%
\bibitem [{\citenamefont {Chiaramello}\ and\ \citenamefont
  {Nagar}(2020)}]{Chiaramello:2020ehz}%
  \BibitemOpen
  \bibfield  {author} {\bibinfo {author} {\bibfnamefont {D.}~\bibnamefont
  {Chiaramello}}\ and\ \bibinfo {author} {\bibfnamefont {A.}~\bibnamefont
  {Nagar}},\ }\href@noop {} {\  (\bibinfo {year} {2020})},\ \Eprint
  {http://arxiv.org/abs/2001.11736} {arXiv:2001.11736 [gr-qc]} \BibitemShut
  {NoStop}%
%%CITATION = ARXIV:2001.11736;%%
\bibitem [{\citenamefont {Klein}\ \emph {et~al.}(2018)\citenamefont {Klein},
  \citenamefont {Boetzel}, \citenamefont {Gopakumar}, \citenamefont {Jetzer},\
  and\ \citenamefont {de~Vittori}}]{PhysRevD.98.104043}%
  \BibitemOpen
  \bibfield  {author} {\bibinfo {author} {\bibfnamefont {A.}~\bibnamefont
  {Klein}}, \bibinfo {author} {\bibfnamefont {Y.}~\bibnamefont {Boetzel}},
  \bibinfo {author} {\bibfnamefont {A.}~\bibnamefont {Gopakumar}}, \bibinfo
  {author} {\bibfnamefont {P.}~\bibnamefont {Jetzer}}, \ and\ \bibinfo {author}
  {\bibfnamefont {L.}~\bibnamefont {de~Vittori}},\ }\href {\doibase
  10.1103/PhysRevD.98.104043} {\bibfield  {journal} {\bibinfo  {journal} {Phys.
  Rev. D}\ }\textbf {\bibinfo {volume} {98}},\ \bibinfo {pages} {104043}
  (\bibinfo {year} {2018})}\BibitemShut {NoStop}%
\bibitem [{\citenamefont {Santamar\'{\i}a}\ \emph {et~al.}(2010)\citenamefont
  {Santamar\'{\i}a}, \citenamefont {Ohme}, \citenamefont {Ajith}, \citenamefont
  {Br\"ugmann}, \citenamefont {Dorband}, \citenamefont {Hannam}, \citenamefont
  {Husa}, \citenamefont {M\"osta}, \citenamefont {Pollney}, \citenamefont
  {Reisswig}, \citenamefont {Robinson}, \citenamefont {Seiler},\ and\
  \citenamefont {Krishnan}}]{PhysRevD.82.064016}%
  \BibitemOpen
  \bibfield  {author} {\bibinfo {author} {\bibfnamefont {L.}~\bibnamefont
  {Santamar\'{\i}a}}, \bibinfo {author} {\bibfnamefont {F.}~\bibnamefont
  {Ohme}}, \bibinfo {author} {\bibfnamefont {P.}~\bibnamefont {Ajith}},
  \bibinfo {author} {\bibfnamefont {B.}~\bibnamefont {Br\"ugmann}}, \bibinfo
  {author} {\bibfnamefont {N.}~\bibnamefont {Dorband}}, \bibinfo {author}
  {\bibfnamefont {M.}~\bibnamefont {Hannam}}, \bibinfo {author} {\bibfnamefont
  {S.}~\bibnamefont {Husa}}, \bibinfo {author} {\bibfnamefont {P.}~\bibnamefont
  {M\"osta}}, \bibinfo {author} {\bibfnamefont {D.}~\bibnamefont {Pollney}},
  \bibinfo {author} {\bibfnamefont {C.}~\bibnamefont {Reisswig}}, \bibinfo
  {author} {\bibfnamefont {E.~L.}\ \bibnamefont {Robinson}}, \bibinfo {author}
  {\bibfnamefont {J.}~\bibnamefont {Seiler}}, \ and\ \bibinfo {author}
  {\bibfnamefont {B.}~\bibnamefont {Krishnan}},\ }\href {\doibase
  10.1103/PhysRevD.82.064016} {\bibfield  {journal} {\bibinfo  {journal} {Phys.
  Rev. D}\ }\textbf {\bibinfo {volume} {82}},\ \bibinfo {pages} {064016}
  (\bibinfo {year} {2010})}\BibitemShut {NoStop}%
\bibitem [{\citenamefont {Ohme}(2012)}]{Ohme_2012}%
  \BibitemOpen
  \bibfield  {author} {\bibinfo {author} {\bibfnamefont {F.}~\bibnamefont
  {Ohme}},\ }\href {\doibase 10.1088/0264-9381/29/12/124002} {\bibfield
  {journal} {\bibinfo  {journal} {Classical and Quantum Gravity}\ }\textbf
  {\bibinfo {volume} {29}},\ \bibinfo {pages} {124002} (\bibinfo {year}
  {2012})}\BibitemShut {NoStop}%
\bibitem [{\citenamefont {Ohme}\ \emph {et~al.}(2011)\citenamefont {Ohme},
  \citenamefont {Hannam},\ and\ \citenamefont {Husa}}]{PhysRevD.84.064029}%
  \BibitemOpen
  \bibfield  {author} {\bibinfo {author} {\bibfnamefont {F.}~\bibnamefont
  {Ohme}}, \bibinfo {author} {\bibfnamefont {M.}~\bibnamefont {Hannam}}, \ and\
  \bibinfo {author} {\bibfnamefont {S.}~\bibnamefont {Husa}},\ }\href {\doibase
  10.1103/PhysRevD.84.064029} {\bibfield  {journal} {\bibinfo  {journal} {Phys.
  Rev. D}\ }\textbf {\bibinfo {volume} {84}},\ \bibinfo {pages} {064029}
  (\bibinfo {year} {2011})}\BibitemShut {NoStop}%
\bibitem [{\citenamefont {MacDonald}\ \emph
  {et~al.}(2011{\natexlab{a}})\citenamefont {MacDonald}, \citenamefont
  {Nissanke},\ and\ \citenamefont {Pfeiffer}}]{MacDonald_2011}%
  \BibitemOpen
  \bibfield  {author} {\bibinfo {author} {\bibfnamefont {I.}~\bibnamefont
  {MacDonald}}, \bibinfo {author} {\bibfnamefont {S.}~\bibnamefont {Nissanke}},
  \ and\ \bibinfo {author} {\bibfnamefont {H.~P.}\ \bibnamefont {Pfeiffer}},\
  }\href {\doibase 10.1088/0264-9381/28/13/134002} {\bibfield  {journal}
  {\bibinfo  {journal} {Classical and Quantum Gravity}\ }\textbf {\bibinfo
  {volume} {28}},\ \bibinfo {pages} {134002} (\bibinfo {year}
  {2011}{\natexlab{a}})}\BibitemShut {NoStop}%
\bibitem [{\citenamefont {Ajith}\ \emph {et~al.}(2012)\citenamefont {Ajith}
  \emph {et~al.}}]{Ajith_2012}%
  \BibitemOpen
  \bibfield  {author} {\bibinfo {author} {\bibfnamefont {P.}~\bibnamefont
  {Ajith}} \emph {et~al.},\ }\href {\doibase 10.1088/0264-9381/29/12/124001}
  {\bibfield  {journal} {\bibinfo  {journal} {Classical and Quantum Gravity}\
  }\textbf {\bibinfo {volume} {29}},\ \bibinfo {pages} {124001} (\bibinfo
  {year} {2012})}\BibitemShut {NoStop}%
\bibitem [{\citenamefont {Boyle}(2011)}]{PhysRevD.84.064013}%
  \BibitemOpen
  \bibfield  {author} {\bibinfo {author} {\bibfnamefont {M.}~\bibnamefont
  {Boyle}},\ }\href {\doibase 10.1103/PhysRevD.84.064013} {\bibfield  {journal}
  {\bibinfo  {journal} {Phys. Rev. D}\ }\textbf {\bibinfo {volume} {84}},\
  \bibinfo {pages} {064013} (\bibinfo {year} {2011})}\BibitemShut {NoStop}%
\bibitem [{\citenamefont {Bustillo}\ \emph {et~al.}(2015)\citenamefont
  {Bustillo}, \citenamefont {Boh{\'e}}, \citenamefont {Husa}, \citenamefont
  {Sintes}, \citenamefont {Hannam},\ and\ \citenamefont
  {P{\"u}rrer}}]{Bustillo:2015ova}%
  \BibitemOpen
  \bibfield  {author} {\bibinfo {author} {\bibfnamefont {J.~C.}\ \bibnamefont
  {Bustillo}}, \bibinfo {author} {\bibfnamefont {A.}~\bibnamefont {Boh{\'e}}},
  \bibinfo {author} {\bibfnamefont {S.}~\bibnamefont {Husa}}, \bibinfo {author}
  {\bibfnamefont {A.~M.}\ \bibnamefont {Sintes}}, \bibinfo {author}
  {\bibfnamefont {M.}~\bibnamefont {Hannam}}, \ and\ \bibinfo {author}
  {\bibfnamefont {M.}~\bibnamefont {P{\"u}rrer}},\ }\href@noop {} {\  (\bibinfo
  {year} {2015})},\ \Eprint {http://arxiv.org/abs/1501.00918} {arXiv:1501.00918
  [gr-qc]} \BibitemShut {NoStop}%
%%CITATION = ARXIV:1501.00918;%%
\bibitem [{\citenamefont {Br\"ugmann}\ \emph {et~al.}(2008)\citenamefont
  {Br\"ugmann}, \citenamefont {Gonz\'alez}, \citenamefont {Hannam},
  \citenamefont {Husa}, \citenamefont {Sperhake},\ and\ \citenamefont
  {Tichy}}]{PhysRevD.77.024027}%
  \BibitemOpen
  \bibfield  {author} {\bibinfo {author} {\bibfnamefont {B.}~\bibnamefont
  {Br\"ugmann}}, \bibinfo {author} {\bibfnamefont {J.~A.}\ \bibnamefont
  {Gonz\'alez}}, \bibinfo {author} {\bibfnamefont {M.}~\bibnamefont {Hannam}},
  \bibinfo {author} {\bibfnamefont {S.}~\bibnamefont {Husa}}, \bibinfo {author}
  {\bibfnamefont {U.}~\bibnamefont {Sperhake}}, \ and\ \bibinfo {author}
  {\bibfnamefont {W.}~\bibnamefont {Tichy}},\ }\href {\doibase
  10.1103/PhysRevD.77.024027} {\bibfield  {journal} {\bibinfo  {journal} {Phys.
  Rev. D}\ }\textbf {\bibinfo {volume} {77}},\ \bibinfo {pages} {024027}
  (\bibinfo {year} {2008})}\BibitemShut {NoStop}%
\bibitem [{\citenamefont {Loffler}\ \emph {et~al.}(2012)\citenamefont {Loffler}
  \emph {et~al.}}]{Loffler:2011ay}%
  \BibitemOpen
  \bibfield  {author} {\bibinfo {author} {\bibfnamefont {F.}~\bibnamefont
  {Loffler}} \emph {et~al.},\ }\href {\doibase 10.1088/0264-9381/29/11/115001}
  {\bibfield  {journal} {\bibinfo  {journal} {Class. Quant. Grav.}\ }\textbf
  {\bibinfo {volume} {29}},\ \bibinfo {pages} {115001} (\bibinfo {year}
  {2012})},\ \Eprint {http://arxiv.org/abs/1111.3344} {arXiv:1111.3344 [gr-qc]}
  \BibitemShut {NoStop}%
%%CITATION = ARXIV:1111.3344;%%
\bibitem [{\citenamefont {Babiuc-Hamilton}\ \emph {et~al.}(2019)\citenamefont
  {Babiuc-Hamilton} \emph {et~al.}}]{maria_babiuc_hamilton_2019_3522086}%
  \BibitemOpen
  \bibfield  {author} {\bibinfo {author} {\bibfnamefont {M.}~\bibnamefont
  {Babiuc-Hamilton}} \emph {et~al.},\ }\href {\doibase 10.5281/zenodo.3522086}
  {\enquote {\bibinfo {title} {{The Einstein Toolkit}},}\ } (\bibinfo {year}
  {2019}),\ \bibinfo {note} {computer code Einstein Toolkit,
  http://einsteintoolkit.org}\BibitemShut {NoStop}%
\bibitem [{\citenamefont {Hinder}\ \emph {et~al.}(2008)\citenamefont {Hinder},
  \citenamefont {Vaishnav}, \citenamefont {Herrmann}, \citenamefont
  {Shoemaker},\ and\ \citenamefont {Laguna}}]{PhysRevD.77.081502}%
  \BibitemOpen
  \bibfield  {author} {\bibinfo {author} {\bibfnamefont {I.}~\bibnamefont
  {Hinder}}, \bibinfo {author} {\bibfnamefont {B.}~\bibnamefont {Vaishnav}},
  \bibinfo {author} {\bibfnamefont {F.}~\bibnamefont {Herrmann}}, \bibinfo
  {author} {\bibfnamefont {D.~M.}\ \bibnamefont {Shoemaker}}, \ and\ \bibinfo
  {author} {\bibfnamefont {P.}~\bibnamefont {Laguna}},\ }\href {\doibase
  10.1103/PhysRevD.77.081502} {\bibfield  {journal} {\bibinfo  {journal} {Phys.
  Rev. D}\ }\textbf {\bibinfo {volume} {77}},\ \bibinfo {pages} {081502}
  (\bibinfo {year} {2008})}\BibitemShut {NoStop}%
\bibitem [{\citenamefont {Pollney}\ \emph {et~al.}(2011)\citenamefont
  {Pollney}, \citenamefont {Reisswig}, \citenamefont {Schnetter}, \citenamefont
  {Dorband},\ and\ \citenamefont {Diener}}]{Pollney:2009yz}%
  \BibitemOpen
  \bibfield  {author} {\bibinfo {author} {\bibfnamefont {D.}~\bibnamefont
  {Pollney}}, \bibinfo {author} {\bibfnamefont {C.}~\bibnamefont {Reisswig}},
  \bibinfo {author} {\bibfnamefont {E.}~\bibnamefont {Schnetter}}, \bibinfo
  {author} {\bibfnamefont {N.}~\bibnamefont {Dorband}}, \ and\ \bibinfo
  {author} {\bibfnamefont {P.}~\bibnamefont {Diener}},\ }\href {\doibase
  10.1103/PhysRevD.83.044045} {\bibfield  {journal} {\bibinfo  {journal} {Phys.
  Rev.}\ }\textbf {\bibinfo {volume} {D83}},\ \bibinfo {pages} {044045}
  (\bibinfo {year} {2011})}\BibitemShut {NoStop}%
%%CITATION = ARXIV:0910.3803;%%
\bibitem [{\citenamefont {Ramos-Buades}\ \emph {et~al.}(2019)\citenamefont
  {Ramos-Buades}, \citenamefont {Husa},\ and\ \citenamefont
  {Pratten}}]{PhysRevD.99.023003}%
  \BibitemOpen
  \bibfield  {author} {\bibinfo {author} {\bibfnamefont {A.}~\bibnamefont
  {Ramos-Buades}}, \bibinfo {author} {\bibfnamefont {S.}~\bibnamefont {Husa}},
  \ and\ \bibinfo {author} {\bibfnamefont {G.}~\bibnamefont {Pratten}},\ }\href
  {\doibase 10.1103/PhysRevD.99.023003} {\bibfield  {journal} {\bibinfo
  {journal} {Phys. Rev. D}\ }\textbf {\bibinfo {volume} {99}},\ \bibinfo
  {pages} {023003} (\bibinfo {year} {2019})}\BibitemShut {NoStop}%
\bibitem [{\citenamefont {{SXS collaboration}}(2010)}]{SpEC}%
  \BibitemOpen
  \bibfield  {author} {\bibinfo {author} {\bibnamefont {{SXS collaboration}}},\
  }\href {https://www.black-holes.org/SpEC.html} {\enquote {\bibinfo {title}
  {Spec},}\ } (\bibinfo {year} {2010})\BibitemShut {NoStop}%
\bibitem [{\citenamefont {Bowen}\ and\ \citenamefont
  {York}(1980)}]{PhysRevD.21.2047}%
  \BibitemOpen
  \bibfield  {author} {\bibinfo {author} {\bibfnamefont {J.~M.}\ \bibnamefont
  {Bowen}}\ and\ \bibinfo {author} {\bibfnamefont {J.~W.}\ \bibnamefont
  {York}},\ }\href {\doibase 10.1103/PhysRevD.21.2047} {\bibfield  {journal}
  {\bibinfo  {journal} {Phys. Rev. D}\ }\textbf {\bibinfo {volume} {21}},\
  \bibinfo {pages} {2047} (\bibinfo {year} {1980})}\BibitemShut {NoStop}%
\bibitem [{\citenamefont {Healy}\ and\ \citenamefont
  {Lousto}(2018)}]{PhysRevD.97.084002}%
  \BibitemOpen
  \bibfield  {author} {\bibinfo {author} {\bibfnamefont {J.}~\bibnamefont
  {Healy}}\ and\ \bibinfo {author} {\bibfnamefont {C.~O.}\ \bibnamefont
  {Lousto}},\ }\href {\doibase 10.1103/PhysRevD.97.084002} {\bibfield
  {journal} {\bibinfo  {journal} {Phys. Rev. D}\ }\textbf {\bibinfo {volume}
  {97}},\ \bibinfo {pages} {084002} (\bibinfo {year} {2018})}\BibitemShut
  {NoStop}%
\bibitem [{\citenamefont {Jim\'enez-Forteza}\ \emph {et~al.}(2017)\citenamefont
  {Jim\'enez-Forteza}, \citenamefont {Keitel}, \citenamefont {Husa},
  \citenamefont {Hannam}, \citenamefont {Khan},\ and\ \citenamefont
  {P\"urrer}}]{PhysRevD.95.064024}%
  \BibitemOpen
  \bibfield  {author} {\bibinfo {author} {\bibfnamefont {X.}~\bibnamefont
  {Jim\'enez-Forteza}}, \bibinfo {author} {\bibfnamefont {D.}~\bibnamefont
  {Keitel}}, \bibinfo {author} {\bibfnamefont {S.}~\bibnamefont {Husa}},
  \bibinfo {author} {\bibfnamefont {M.}~\bibnamefont {Hannam}}, \bibinfo
  {author} {\bibfnamefont {S.}~\bibnamefont {Khan}}, \ and\ \bibinfo {author}
  {\bibfnamefont {M.}~\bibnamefont {P\"urrer}},\ }\href {\doibase
  10.1103/PhysRevD.95.064024} {\bibfield  {journal} {\bibinfo  {journal} {Phys.
  Rev. D}\ }\textbf {\bibinfo {volume} {95}},\ \bibinfo {pages} {064024}
  (\bibinfo {year} {2017})}\BibitemShut {NoStop}%
\bibitem [{\citenamefont {Campanelli}\ \emph {et~al.}(2007)\citenamefont
  {Campanelli}, \citenamefont {Lousto}, \citenamefont {Zlochower},
  \citenamefont {Krishnan},\ and\ \citenamefont
  {Merritt}}]{PhysRevD.75.064030}%
  \BibitemOpen
  \bibfield  {author} {\bibinfo {author} {\bibfnamefont {M.}~\bibnamefont
  {Campanelli}}, \bibinfo {author} {\bibfnamefont {C.~O.}\ \bibnamefont
  {Lousto}}, \bibinfo {author} {\bibfnamefont {Y.}~\bibnamefont {Zlochower}},
  \bibinfo {author} {\bibfnamefont {B.}~\bibnamefont {Krishnan}}, \ and\
  \bibinfo {author} {\bibfnamefont {D.}~\bibnamefont {Merritt}},\ }\href
  {\doibase 10.1103/PhysRevD.75.064030} {\bibfield  {journal} {\bibinfo
  {journal} {Phys. Rev. D}\ }\textbf {\bibinfo {volume} {75}},\ \bibinfo
  {pages} {064030} (\bibinfo {year} {2007})}\BibitemShut {NoStop}%
\bibitem [{\citenamefont {Dreyer}\ \emph {et~al.}(2003)\citenamefont {Dreyer},
  \citenamefont {Krishnan}, \citenamefont {Shoemaker},\ and\ \citenamefont
  {Schnetter}}]{PhysRevD.67.024018}%
  \BibitemOpen
  \bibfield  {author} {\bibinfo {author} {\bibfnamefont {O.}~\bibnamefont
  {Dreyer}}, \bibinfo {author} {\bibfnamefont {B.}~\bibnamefont {Krishnan}},
  \bibinfo {author} {\bibfnamefont {D.}~\bibnamefont {Shoemaker}}, \ and\
  \bibinfo {author} {\bibfnamefont {E.}~\bibnamefont {Schnetter}},\ }\href
  {\doibase 10.1103/PhysRevD.67.024018} {\bibfield  {journal} {\bibinfo
  {journal} {Phys. Rev. D}\ }\textbf {\bibinfo {volume} {67}},\ \bibinfo
  {pages} {024018} (\bibinfo {year} {2003})}\BibitemShut {NoStop}%
\bibitem [{\citenamefont {Thornburg}(1996)}]{PhysRevD.54.4899}%
  \BibitemOpen
  \bibfield  {author} {\bibinfo {author} {\bibfnamefont {J.}~\bibnamefont
  {Thornburg}},\ }\href {\doibase 10.1103/PhysRevD.54.4899} {\bibfield
  {journal} {\bibinfo  {journal} {Phys. Rev. D}\ }\textbf {\bibinfo {volume}
  {54}},\ \bibinfo {pages} {4899} (\bibinfo {year} {1996})}\BibitemShut
  {NoStop}%
\bibitem [{\citenamefont {Thornburg}(2004)}]{Thornburg:2003sf}%
  \BibitemOpen
  \bibfield  {author} {\bibinfo {author} {\bibfnamefont {J.}~\bibnamefont
  {Thornburg}},\ }\href {\doibase 10.1088/0264-9381/21/2/026} {\bibfield
  {journal} {\bibinfo  {journal} {Class. Quant. Grav.}\ }\textbf {\bibinfo
  {volume} {21}},\ \bibinfo {pages} {743} (\bibinfo {year} {2004})},\ \Eprint
  {http://arxiv.org/abs/gr-qc/0306056} {arXiv:gr-qc/0306056 [gr-qc]}
  \BibitemShut {NoStop}%
%%CITATION = GR-QC/0306056;%%
\bibitem [{\citenamefont {Purrer}\ \emph {et~al.}(2012)\citenamefont {Purrer},
  \citenamefont {Husa},\ and\ \citenamefont {Hannam}}]{Purrer:2012wy}%
  \BibitemOpen
  \bibfield  {author} {\bibinfo {author} {\bibfnamefont {M.}~\bibnamefont
  {Purrer}}, \bibinfo {author} {\bibfnamefont {S.}~\bibnamefont {Husa}}, \ and\
  \bibinfo {author} {\bibfnamefont {M.}~\bibnamefont {Hannam}},\ }\href
  {\doibase 10.1103/PhysRevD.85.124051} {\bibfield  {journal} {\bibinfo
  {journal} {Phys. Rev.}\ }\textbf {\bibinfo {volume} {D85}},\ \bibinfo {pages}
  {124051} (\bibinfo {year} {2012})},\ \Eprint {http://arxiv.org/abs/1203.4258}
  {arXiv:1203.4258 [gr-qc]} \BibitemShut {NoStop}%
%%CITATION = ARXIV:1203.4258;%%
\bibitem [{\citenamefont {Mrou\'e}\ \emph {et~al.}(2010)\citenamefont
  {Mrou\'e}, \citenamefont {Pfeiffer}, \citenamefont {Kidder},\ and\
  \citenamefont {Teukolsky}}]{PhysRevD.82.124016}%
  \BibitemOpen
  \bibfield  {author} {\bibinfo {author} {\bibfnamefont {A.~H.}\ \bibnamefont
  {Mrou\'e}}, \bibinfo {author} {\bibfnamefont {H.~P.}\ \bibnamefont
  {Pfeiffer}}, \bibinfo {author} {\bibfnamefont {L.~E.}\ \bibnamefont
  {Kidder}}, \ and\ \bibinfo {author} {\bibfnamefont {S.~A.}\ \bibnamefont
  {Teukolsky}},\ }\href {\doibase 10.1103/PhysRevD.82.124016} {\bibfield
  {journal} {\bibinfo  {journal} {Phys. Rev. D}\ }\textbf {\bibinfo {volume}
  {82}},\ \bibinfo {pages} {124016} (\bibinfo {year} {2010})}\BibitemShut
  {NoStop}%
\bibitem [{\citenamefont {Buonanno}\ \emph {et~al.}(2011)\citenamefont
  {Buonanno}, \citenamefont {Kidder}, \citenamefont {Mrou\'e}, \citenamefont
  {Pfeiffer},\ and\ \citenamefont {Taracchini}}]{PhysRevD.83.104034}%
  \BibitemOpen
  \bibfield  {author} {\bibinfo {author} {\bibfnamefont {A.}~\bibnamefont
  {Buonanno}}, \bibinfo {author} {\bibfnamefont {L.~E.}\ \bibnamefont
  {Kidder}}, \bibinfo {author} {\bibfnamefont {A.~H.}\ \bibnamefont {Mrou\'e}},
  \bibinfo {author} {\bibfnamefont {H.~P.}\ \bibnamefont {Pfeiffer}}, \ and\
  \bibinfo {author} {\bibfnamefont {A.}~\bibnamefont {Taracchini}},\ }\href
  {\doibase 10.1103/PhysRevD.83.104034} {\bibfield  {journal} {\bibinfo
  {journal} {Phys. Rev. D}\ }\textbf {\bibinfo {volume} {83}},\ \bibinfo
  {pages} {104034} (\bibinfo {year} {2011})}\BibitemShut {NoStop}%
\bibitem [{\citenamefont {Alcubierre}\ \emph {et~al.}(2003)\citenamefont
  {Alcubierre}, \citenamefont {Bruegmann}, \citenamefont {Diener},
  \citenamefont {Koppitz}, \citenamefont {Pollney}, \citenamefont {Seidel},\
  and\ \citenamefont {Takahashi}}]{Alcubierre:2002kk}%
  \BibitemOpen
  \bibfield  {author} {\bibinfo {author} {\bibfnamefont {M.}~\bibnamefont
  {Alcubierre}}, \bibinfo {author} {\bibfnamefont {B.}~\bibnamefont
  {Bruegmann}}, \bibinfo {author} {\bibfnamefont {P.}~\bibnamefont {Diener}},
  \bibinfo {author} {\bibfnamefont {M.}~\bibnamefont {Koppitz}}, \bibinfo
  {author} {\bibfnamefont {D.}~\bibnamefont {Pollney}}, \bibinfo {author}
  {\bibfnamefont {E.}~\bibnamefont {Seidel}}, \ and\ \bibinfo {author}
  {\bibfnamefont {R.}~\bibnamefont {Takahashi}},\ }\href {\doibase
  10.1103/PhysRevD.67.084023} {\bibfield  {journal} {\bibinfo  {journal} {Phys.
  Rev.}\ }\textbf {\bibinfo {volume} {D67}},\ \bibinfo {pages} {084023}
  (\bibinfo {year} {2003})},\ \Eprint {http://arxiv.org/abs/gr-qc/0206072}
  {arXiv:gr-qc/0206072 [gr-qc]} \BibitemShut {NoStop}%
%%CITATION = GR-QC/0206072;%%
\bibitem [{\citenamefont {Mora}\ and\ \citenamefont
  {Will}(2002)}]{PhysRevD.66.101501}%
  \BibitemOpen
  \bibfield  {author} {\bibinfo {author} {\bibfnamefont {T.}~\bibnamefont
  {Mora}}\ and\ \bibinfo {author} {\bibfnamefont {C.~M.}\ \bibnamefont
  {Will}},\ }\href {\doibase 10.1103/PhysRevD.66.101501} {\bibfield  {journal}
  {\bibinfo  {journal} {Phys. Rev. D}\ }\textbf {\bibinfo {volume} {66}},\
  \bibinfo {pages} {101501} (\bibinfo {year} {2002})}\BibitemShut {NoStop}%
\bibitem [{\citenamefont {Inc.}()}]{Mathematica}%
  \BibitemOpen
  \bibfield  {author} {\bibinfo {author} {\bibfnamefont {W.~R.}\ \bibnamefont
  {Inc.}},\ }\href {https://www.wolfram.com/mathematica} {\enquote {\bibinfo
  {title} {Mathematica, {V}ersion 12.0},}\ }\bibinfo {note} {Champaign, IL,
  2019}\BibitemShut {NoStop}%
\bibitem [{\citenamefont {Husa}\ \emph {et~al.}(2019)\citenamefont {Husa} \emph
  {et~al.}}]{NRUIB}%
  \BibitemOpen
  \bibfield  {author} {\bibinfo {author} {\bibfnamefont {S.}~\bibnamefont
  {Husa}} \emph {et~al.},\ }\href@noop {} {\  (\bibinfo {year}
  {2019})}\BibitemShut {NoStop}%
\bibitem [{\citenamefont {Arun}\ \emph {et~al.}(2009)\citenamefont {Arun},
  \citenamefont {Blanchet}, \citenamefont {Iyer},\ and\ \citenamefont
  {Sinha}}]{PhysRevD.80.124018}%
  \BibitemOpen
  \bibfield  {author} {\bibinfo {author} {\bibfnamefont {K.~G.}\ \bibnamefont
  {Arun}}, \bibinfo {author} {\bibfnamefont {L.}~\bibnamefont {Blanchet}},
  \bibinfo {author} {\bibfnamefont {B.~R.}\ \bibnamefont {Iyer}}, \ and\
  \bibinfo {author} {\bibfnamefont {S.}~\bibnamefont {Sinha}},\ }\href
  {\doibase 10.1103/PhysRevD.80.124018} {\bibfield  {journal} {\bibinfo
  {journal} {Phys. Rev. D}\ }\textbf {\bibinfo {volume} {80}},\ \bibinfo
  {pages} {124018} (\bibinfo {year} {2009})}\BibitemShut {NoStop}%
\bibitem [{\citenamefont {Arnowitt}\ \emph {et~al.}(1959)\citenamefont
  {Arnowitt}, \citenamefont {Deser},\ and\ \citenamefont
  {Misner}}]{PhysRev.116.1322}%
  \BibitemOpen
  \bibfield  {author} {\bibinfo {author} {\bibfnamefont {R.}~\bibnamefont
  {Arnowitt}}, \bibinfo {author} {\bibfnamefont {S.}~\bibnamefont {Deser}}, \
  and\ \bibinfo {author} {\bibfnamefont {C.~W.}\ \bibnamefont {Misner}},\
  }\href {\doibase 10.1103/PhysRev.116.1322} {\bibfield  {journal} {\bibinfo
  {journal} {Phys. Rev.}\ }\textbf {\bibinfo {volume} {116}},\ \bibinfo {pages}
  {1322} (\bibinfo {year} {1959})}\BibitemShut {NoStop}%
\bibitem [{\citenamefont {Deser}\ \emph {et~al.}(1960)\citenamefont {Deser},
  \citenamefont {Arnowitt},\ and\ \citenamefont {Misner}}]{Deser:1960zzc}%
  \BibitemOpen
  \bibfield  {author} {\bibinfo {author} {\bibfnamefont {S.}~\bibnamefont
  {Deser}}, \bibinfo {author} {\bibfnamefont {R.}~\bibnamefont {Arnowitt}}, \
  and\ \bibinfo {author} {\bibfnamefont {C.~W.}\ \bibnamefont {Misner}},\
  }\href {\doibase 10.1063/1.1703677} {\bibfield  {journal} {\bibinfo
  {journal} {J. Math. Phys.}\ }\textbf {\bibinfo {volume} {1}},\ \bibinfo
  {pages} {434} (\bibinfo {year} {1960})}\BibitemShut {NoStop}%
%%CITATION = JMAPA,1,434;%%
\bibitem [{\citenamefont {Arnowitt}\ \emph {et~al.}(1960)\citenamefont
  {Arnowitt}, \citenamefont {Deser},\ and\ \citenamefont
  {Misner}}]{PhysRev.117.1595}%
  \BibitemOpen
  \bibfield  {author} {\bibinfo {author} {\bibfnamefont {R.}~\bibnamefont
  {Arnowitt}}, \bibinfo {author} {\bibfnamefont {S.}~\bibnamefont {Deser}}, \
  and\ \bibinfo {author} {\bibfnamefont {C.~W.}\ \bibnamefont {Misner}},\
  }\href {\doibase 10.1103/PhysRev.117.1595} {\bibfield  {journal} {\bibinfo
  {journal} {Phys. Rev.}\ }\textbf {\bibinfo {volume} {117}},\ \bibinfo {pages}
  {1595} (\bibinfo {year} {1960})}\BibitemShut {NoStop}%
\bibitem [{\citenamefont {Buonanno}\ \emph {et~al.}(2006)\citenamefont
  {Buonanno}, \citenamefont {Chen},\ and\ \citenamefont
  {Damour}}]{Buonanno:2005xu}%
  \BibitemOpen
  \bibfield  {author} {\bibinfo {author} {\bibfnamefont {A.}~\bibnamefont
  {Buonanno}}, \bibinfo {author} {\bibfnamefont {Y.}~\bibnamefont {Chen}}, \
  and\ \bibinfo {author} {\bibfnamefont {T.}~\bibnamefont {Damour}},\ }\href
  {\doibase 10.1103/PhysRevD.74.104005} {\bibfield  {journal} {\bibinfo
  {journal} {Phys. Rev.}\ }\textbf {\bibinfo {volume} {D74}},\ \bibinfo {pages}
  {104005} (\bibinfo {year} {2006})}\BibitemShut {NoStop}%
%%CITATION = GR-QC/0508067;%%
\bibitem [{\citenamefont {Mishra}\ \emph {et~al.}(2015)\citenamefont {Mishra},
  \citenamefont {Arun},\ and\ \citenamefont {Iyer}}]{PhysRevD.91.084040}%
  \BibitemOpen
  \bibfield  {author} {\bibinfo {author} {\bibfnamefont {C.~K.}\ \bibnamefont
  {Mishra}}, \bibinfo {author} {\bibfnamefont {K.~G.}\ \bibnamefont {Arun}}, \
  and\ \bibinfo {author} {\bibfnamefont {B.~R.}\ \bibnamefont {Iyer}},\ }\href
  {\doibase 10.1103/PhysRevD.91.084040} {\bibfield  {journal} {\bibinfo
  {journal} {Phys. Rev. D}\ }\textbf {\bibinfo {volume} {91}},\ \bibinfo
  {pages} {084040} (\bibinfo {year} {2015})}\BibitemShut {NoStop}%
\bibitem [{\citenamefont {Boetzel}\ \emph {et~al.}(2019)\citenamefont
  {Boetzel}, \citenamefont {Mishra}, \citenamefont {Faye}, \citenamefont
  {Gopakumar},\ and\ \citenamefont {Iyer}}]{Boetzel:2019nfw}%
  \BibitemOpen
  \bibfield  {author} {\bibinfo {author} {\bibfnamefont {Y.}~\bibnamefont
  {Boetzel}}, \bibinfo {author} {\bibfnamefont {C.~K.}\ \bibnamefont {Mishra}},
  \bibinfo {author} {\bibfnamefont {G.}~\bibnamefont {Faye}}, \bibinfo {author}
  {\bibfnamefont {A.}~\bibnamefont {Gopakumar}}, \ and\ \bibinfo {author}
  {\bibfnamefont {B.~R.}\ \bibnamefont {Iyer}},\ }\href@noop {} {\  (\bibinfo
  {year} {2019})},\ \Eprint {http://arxiv.org/abs/1904.11814} {arXiv:1904.11814
  [gr-qc]} \BibitemShut {NoStop}%
%%CITATION = ARXIV:1904.11814;%%
\bibitem [{\citenamefont {Ebersold}\ \emph {et~al.}(2019)\citenamefont
  {Ebersold}, \citenamefont {Boetzel}, \citenamefont {Faye}, \citenamefont
  {Mishra}, \citenamefont {Iyer},\ and\ \citenamefont
  {Jetzer}}]{PhysRevD.100.084043}%
  \BibitemOpen
  \bibfield  {author} {\bibinfo {author} {\bibfnamefont {M.}~\bibnamefont
  {Ebersold}}, \bibinfo {author} {\bibfnamefont {Y.}~\bibnamefont {Boetzel}},
  \bibinfo {author} {\bibfnamefont {G.}~\bibnamefont {Faye}}, \bibinfo {author}
  {\bibfnamefont {C.~K.}\ \bibnamefont {Mishra}}, \bibinfo {author}
  {\bibfnamefont {B.~R.}\ \bibnamefont {Iyer}}, \ and\ \bibinfo {author}
  {\bibfnamefont {P.}~\bibnamefont {Jetzer}},\ }\href {\doibase
  10.1103/PhysRevD.100.084043} {\bibfield  {journal} {\bibinfo  {journal}
  {Phys. Rev. D}\ }\textbf {\bibinfo {volume} {100}},\ \bibinfo {pages}
  {084043} (\bibinfo {year} {2019})}\BibitemShut {NoStop}%
\bibitem [{\citenamefont {MacDonald}\ \emph
  {et~al.}(2011{\natexlab{b}})\citenamefont {MacDonald}, \citenamefont
  {Nissanke},\ and\ \citenamefont {Pfeiffer}}]{MacDonald:2011ne}%
  \BibitemOpen
  \bibfield  {author} {\bibinfo {author} {\bibfnamefont {I.}~\bibnamefont
  {MacDonald}}, \bibinfo {author} {\bibfnamefont {S.}~\bibnamefont {Nissanke}},
  \ and\ \bibinfo {author} {\bibfnamefont {H.~P.}\ \bibnamefont {Pfeiffer}},\
  }\bibfield  {booktitle} {\emph {\bibinfo {booktitle} {{Theory meets data
  analysis at comparable and extreme mass ratios. Proceedings, Conference,
  NRDA/CAPRA 2010, Waterloo, Canada, June 20-26, 2010}}},\ }\href {\doibase
  10.1088/0264-9381/28/13/134002} {\bibfield  {journal} {\bibinfo  {journal}
  {Class. Quant. Grav.}\ }\textbf {\bibinfo {volume} {28}},\ \bibinfo {pages}
  {134002} (\bibinfo {year} {2011}{\natexlab{b}})},\ \Eprint
  {http://arxiv.org/abs/1102.5128} {arXiv:1102.5128 [gr-qc]} \BibitemShut
  {NoStop}%
%%CITATION = ARXIV:1102.5128;%%
\bibitem [{\citenamefont {MacDonald}\ \emph {et~al.}(2013)\citenamefont
  {MacDonald}, \citenamefont {Mrou\'e}, \citenamefont {Pfeiffer}, \citenamefont
  {Boyle}, \citenamefont {Kidder}, \citenamefont {Scheel}, \citenamefont
  {Szil\'agyi},\ and\ \citenamefont {Taylor}}]{PhysRevD.87.024009}%
  \BibitemOpen
  \bibfield  {author} {\bibinfo {author} {\bibfnamefont {I.}~\bibnamefont
  {MacDonald}}, \bibinfo {author} {\bibfnamefont {A.~H.}\ \bibnamefont
  {Mrou\'e}}, \bibinfo {author} {\bibfnamefont {H.~P.}\ \bibnamefont
  {Pfeiffer}}, \bibinfo {author} {\bibfnamefont {M.}~\bibnamefont {Boyle}},
  \bibinfo {author} {\bibfnamefont {L.~E.}\ \bibnamefont {Kidder}}, \bibinfo
  {author} {\bibfnamefont {M.~A.}\ \bibnamefont {Scheel}}, \bibinfo {author}
  {\bibfnamefont {B.}~\bibnamefont {Szil\'agyi}}, \ and\ \bibinfo {author}
  {\bibfnamefont {N.~W.}\ \bibnamefont {Taylor}},\ }\href {\doibase
  10.1103/PhysRevD.87.024009} {\bibfield  {journal} {\bibinfo  {journal} {Phys.
  Rev. D}\ }\textbf {\bibinfo {volume} {87}},\ \bibinfo {pages} {024009}
  (\bibinfo {year} {2013})}\BibitemShut {NoStop}%
\bibitem [{\citenamefont {Tichy}\ \emph {et~al.}(2003)\citenamefont {Tichy},
  \citenamefont {Brugmann}, \citenamefont {Campanelli},\ and\ \citenamefont
  {Diener}}]{Tichy:2002ec}%
  \BibitemOpen
  \bibfield  {author} {\bibinfo {author} {\bibfnamefont {W.}~\bibnamefont
  {Tichy}}, \bibinfo {author} {\bibfnamefont {B.}~\bibnamefont {Brugmann}},
  \bibinfo {author} {\bibfnamefont {M.}~\bibnamefont {Campanelli}}, \ and\
  \bibinfo {author} {\bibfnamefont {P.}~\bibnamefont {Diener}},\ }\href
  {\doibase 10.1103/PhysRevD.67.064008} {\bibfield  {journal} {\bibinfo
  {journal} {Phys. Rev.}\ }\textbf {\bibinfo {volume} {D67}},\ \bibinfo {pages}
  {064008} (\bibinfo {year} {2003})}\BibitemShut {NoStop}%
%%CITATION = GR-QC/0207011;%%
\bibitem [{\citenamefont {Yunes}\ and\ \citenamefont
  {Tichy}(2006)}]{Yunes:2006iw}%
  \BibitemOpen
  \bibfield  {author} {\bibinfo {author} {\bibfnamefont {N.}~\bibnamefont
  {Yunes}}\ and\ \bibinfo {author} {\bibfnamefont {W.}~\bibnamefont {Tichy}},\
  }\href {\doibase 10.1103/PhysRevD.74.064013} {\bibfield  {journal} {\bibinfo
  {journal} {Phys. Rev.}\ }\textbf {\bibinfo {volume} {D74}},\ \bibinfo {pages}
  {064013} (\bibinfo {year} {2006})}\BibitemShut {NoStop}%
%%CITATION = GR-QC/0601046;%%
\bibitem [{\citenamefont {Yunes}\ \emph {et~al.}(2006)\citenamefont {Yunes},
  \citenamefont {Tichy}, \citenamefont {Owen},\ and\ \citenamefont
  {Brugmann}}]{PhysRevD.74.104011}%
  \BibitemOpen
  \bibfield  {author} {\bibinfo {author} {\bibfnamefont {N.}~\bibnamefont
  {Yunes}}, \bibinfo {author} {\bibfnamefont {W.}~\bibnamefont {Tichy}},
  \bibinfo {author} {\bibfnamefont {B.~J.}\ \bibnamefont {Owen}}, \ and\
  \bibinfo {author} {\bibfnamefont {B.}~\bibnamefont {Brugmann}},\ }\href
  {\doibase 10.1103/PhysRevD.74.104011} {\bibfield  {journal} {\bibinfo
  {journal} {Phys. Rev. D}\ }\textbf {\bibinfo {volume} {74}},\ \bibinfo
  {pages} {104011} (\bibinfo {year} {2006})}\BibitemShut {NoStop}%
\bibitem [{\citenamefont {Husa}\ \emph {et~al.}(2020)\citenamefont {Husa} \emph
  {et~al.}}]{QCHybrids}%
  \BibitemOpen
  \bibfield  {author} {\bibinfo {author} {\bibfnamefont {S.}~\bibnamefont
  {Husa}} \emph {et~al.},\ }\href@noop {} {\  (\bibinfo {year} {2020})},\
  \bibinfo {note} {in preparation}\BibitemShut {NoStop}%
\bibitem [{\citenamefont {Reisswig}\ and\ \citenamefont
  {Pollney}(2011)}]{Reisswig:2010di}%
  \BibitemOpen
  \bibfield  {author} {\bibinfo {author} {\bibfnamefont {C.}~\bibnamefont
  {Reisswig}}\ and\ \bibinfo {author} {\bibfnamefont {D.}~\bibnamefont
  {Pollney}},\ }\href {\doibase 10.1088/0264-9381/28/19/195015} {\bibfield
  {journal} {\bibinfo  {journal} {Class. Quant. Grav.}\ }\textbf {\bibinfo
  {volume} {28}},\ \bibinfo {pages} {195015} (\bibinfo {year} {2011})},\
  \Eprint {http://arxiv.org/abs/1006.1632} {arXiv:1006.1632 [gr-qc]}
  \BibitemShut {NoStop}%
%%CITATION = ARXIV:1006.1632;%%
\bibitem [{\citenamefont {{LIGO Scientific Collaboration}}(2018)}]{lalsuite}%
  \BibitemOpen
  \bibfield  {author} {\bibinfo {author} {\bibnamefont {{LIGO Scientific
  Collaboration}}},\ }\href {\doibase {10.7935/GT1W-FZ16}} {\  (\bibinfo {year}
  {{2018}}),\ {10.7935/GT1W-FZ16}}\BibitemShut {NoStop}%
\bibitem [{\citenamefont {Finn}\ and\ \citenamefont
  {Chernoff}(1993)}]{PhysRevD.47.2198}%
  \BibitemOpen
  \bibfield  {author} {\bibinfo {author} {\bibfnamefont {L.~S.}\ \bibnamefont
  {Finn}}\ and\ \bibinfo {author} {\bibfnamefont {D.~F.}\ \bibnamefont
  {Chernoff}},\ }\href {\doibase 10.1103/PhysRevD.47.2198} {\bibfield
  {journal} {\bibinfo  {journal} {Phys. Rev. D}\ }\textbf {\bibinfo {volume}
  {47}},\ \bibinfo {pages} {2198} (\bibinfo {year} {1993})}\BibitemShut
  {NoStop}%
\bibitem [{\citenamefont {Jaranowski}\ and\ \citenamefont
  {Kr{\'o}lak}(2012)}]{Jaranowski2012}%
  \BibitemOpen
  \bibfield  {author} {\bibinfo {author} {\bibfnamefont {P.}~\bibnamefont
  {Jaranowski}}\ and\ \bibinfo {author} {\bibfnamefont {A.}~\bibnamefont
  {Kr{\'o}lak}},\ }\href {\doibase 10.12942/lrr-2012-4} {\bibfield  {journal}
  {\bibinfo  {journal} {Living Reviews in Relativity}\ }\textbf {\bibinfo
  {volume} {15}},\ \bibinfo {pages} {4} (\bibinfo {year} {2012})}\BibitemShut
  {NoStop}%
\bibitem [{\citenamefont {Husa}\ \emph {et~al.}(2016)\citenamefont {Husa},
  \citenamefont {Khan}, \citenamefont {Hannam}, \citenamefont {P\"urrer},
  \citenamefont {Ohme}, \citenamefont {Forteza},\ and\ \citenamefont
  {Boh\'e}}]{PhysRevD.93.044006}%
  \BibitemOpen
  \bibfield  {author} {\bibinfo {author} {\bibfnamefont {S.}~\bibnamefont
  {Husa}}, \bibinfo {author} {\bibfnamefont {S.}~\bibnamefont {Khan}}, \bibinfo
  {author} {\bibfnamefont {M.}~\bibnamefont {Hannam}}, \bibinfo {author}
  {\bibfnamefont {M.}~\bibnamefont {P\"urrer}}, \bibinfo {author}
  {\bibfnamefont {F.}~\bibnamefont {Ohme}}, \bibinfo {author} {\bibfnamefont
  {X.~J.}\ \bibnamefont {Forteza}}, \ and\ \bibinfo {author} {\bibfnamefont
  {A.}~\bibnamefont {Boh\'e}},\ }\href {\doibase 10.1103/PhysRevD.93.044006}
  {\bibfield  {journal} {\bibinfo  {journal} {Phys. Rev. D}\ }\textbf {\bibinfo
  {volume} {93}},\ \bibinfo {pages} {044006} (\bibinfo {year}
  {2016})}\BibitemShut {NoStop}%
\bibitem [{LIG()}]{LIGOPSD}%
  \BibitemOpen
  \href {https://dcc.ligo.org/LIGO-T0900288/public.} {\enquote {\bibinfo
  {title} {{https://dcc.ligo.org/LIGO-T0900288/public.}}}\ }\BibitemShut
  {NoStop}%
\bibitem [{\citenamefont {Brown}\ and\ \citenamefont
  {Zimmerman}(2010)}]{PhysRevD.81.024007}%
  \BibitemOpen
  \bibfield  {author} {\bibinfo {author} {\bibfnamefont {D.~A.}\ \bibnamefont
  {Brown}}\ and\ \bibinfo {author} {\bibfnamefont {P.~J.}\ \bibnamefont
  {Zimmerman}},\ }\href {\doibase 10.1103/PhysRevD.81.024007} {\bibfield
  {journal} {\bibinfo  {journal} {Phys. Rev. D}\ }\textbf {\bibinfo {volume}
  {81}},\ \bibinfo {pages} {024007} (\bibinfo {year} {2010})}\BibitemShut
  {NoStop}%
\bibitem [{\citenamefont {Ashton}\ \emph {et~al.}(2019)\citenamefont {Ashton}
  \emph {et~al.}}]{Ashton:2018jfp}%
  \BibitemOpen
  \bibfield  {author} {\bibinfo {author} {\bibfnamefont {G.}~\bibnamefont
  {Ashton}} \emph {et~al.},\ }\href {\doibase 10.3847/1538-4365/ab06fc}
  {\bibfield  {journal} {\bibinfo  {journal} {Astrophys. J. Suppl.}\ }\textbf
  {\bibinfo {volume} {241}},\ \bibinfo {pages} {27} (\bibinfo {year} {2019})},\
  \Eprint {http://arxiv.org/abs/1811.02042} {arXiv:1811.02042 [astro-ph.IM]}
  \BibitemShut {NoStop}%
%%CITATION = ARXIV:1811.02042;%%
\bibitem [{\citenamefont {Veitch}\ and\ \citenamefont
  {Pozzo}(2017)}]{JohnVeitchCpnest}%
  \BibitemOpen
  \bibfield  {author} {\bibinfo {author} {\bibfnamefont {J.}~\bibnamefont
  {Veitch}}\ and\ \bibinfo {author} {\bibfnamefont {W.~D.}\ \bibnamefont
  {Pozzo}},\ }\href {\doibase 10.5281/zenodo.322819} {\enquote {\bibinfo
  {title} {{CPNEST}},}\ }\bibinfo {howpublished} {10.5281/zenodo.322819}
  (\bibinfo {year} {2017})\BibitemShut {NoStop}%
\bibitem [{\citenamefont {Galley}\ and\ \citenamefont
  {Schmidt}(2016)}]{Galley:2016mvy}%
  \BibitemOpen
  \bibfield  {author} {\bibinfo {author} {\bibfnamefont {C.~R.}\ \bibnamefont
  {Galley}}\ and\ \bibinfo {author} {\bibfnamefont {P.}~\bibnamefont
  {Schmidt}},\ }\href@noop {} {\  (\bibinfo {year} {2016})},\ \Eprint
  {http://arxiv.org/abs/1611.07529} {arXiv:1611.07529 [gr-qc]} \BibitemShut
  {NoStop}%
%%CITATION = ARXIV:1611.07529;%%
\bibitem [{\citenamefont {Schmidt}\ \emph {et~al.}(2017)\citenamefont
  {Schmidt}, \citenamefont {Harry},\ and\ \citenamefont
  {Pfeiffer}}]{Schmidt:2017btt}%
  \BibitemOpen
  \bibfield  {author} {\bibinfo {author} {\bibfnamefont {P.}~\bibnamefont
  {Schmidt}}, \bibinfo {author} {\bibfnamefont {I.~W.}\ \bibnamefont {Harry}},
  \ and\ \bibinfo {author} {\bibfnamefont {H.~P.}\ \bibnamefont {Pfeiffer}},\
  }\href@noop {} {\  (\bibinfo {year} {2017})},\ \Eprint
  {http://arxiv.org/abs/1703.01076} {arXiv:1703.01076 [gr-qc]} \BibitemShut
  {NoStop}%
%%CITATION = ARXIV:1703.01076;%%
\bibitem [{\citenamefont {Favata}(2014)}]{PhysRevLett.112.101101}%
  \BibitemOpen
  \bibfield  {author} {\bibinfo {author} {\bibfnamefont {M.}~\bibnamefont
  {Favata}},\ }\href {\doibase 10.1103/PhysRevLett.112.101101} {\bibfield
  {journal} {\bibinfo  {journal} {Phys. Rev. Lett.}\ }\textbf {\bibinfo
  {volume} {112}},\ \bibinfo {pages} {101101} (\bibinfo {year}
  {2014})}\BibitemShut {NoStop}%
\bibitem [{\citenamefont {Huerta}\ \emph {et~al.}(2014)\citenamefont {Huerta},
  \citenamefont {Kumar}, \citenamefont {McWilliams}, \citenamefont
  {O'Shaughnessy},\ and\ \citenamefont {Yunes}}]{PhysRevD.90.084016}%
  \BibitemOpen
  \bibfield  {author} {\bibinfo {author} {\bibfnamefont {E.~A.}\ \bibnamefont
  {Huerta}}, \bibinfo {author} {\bibfnamefont {P.}~\bibnamefont {Kumar}},
  \bibinfo {author} {\bibfnamefont {S.~T.}\ \bibnamefont {McWilliams}},
  \bibinfo {author} {\bibfnamefont {R.}~\bibnamefont {O'Shaughnessy}}, \ and\
  \bibinfo {author} {\bibfnamefont {N.}~\bibnamefont {Yunes}},\ }\href
  {\doibase 10.1103/PhysRevD.90.084016} {\bibfield  {journal} {\bibinfo
  {journal} {Phys. Rev. D}\ }\textbf {\bibinfo {volume} {90}},\ \bibinfo
  {pages} {084016} (\bibinfo {year} {2014})}\BibitemShut {NoStop}%
\bibitem [{\citenamefont {Mandel}\ \emph {et~al.}(2014)\citenamefont {Mandel},
  \citenamefont {Berry}, \citenamefont {Ohme}, \citenamefont {Fairhurst},\ and\
  \citenamefont {Farr}}]{Mandel:2014tca}%
  \BibitemOpen
  \bibfield  {author} {\bibinfo {author} {\bibfnamefont {I.}~\bibnamefont
  {Mandel}}, \bibinfo {author} {\bibfnamefont {C.~P.~L.}\ \bibnamefont
  {Berry}}, \bibinfo {author} {\bibfnamefont {F.}~\bibnamefont {Ohme}},
  \bibinfo {author} {\bibfnamefont {S.}~\bibnamefont {Fairhurst}}, \ and\
  \bibinfo {author} {\bibfnamefont {W.~M.}\ \bibnamefont {Farr}},\ }\href
  {\doibase 10.1088/0264-9381/31/15/155005} {\bibfield  {journal} {\bibinfo
  {journal} {Class. Quant. Grav.}\ }\textbf {\bibinfo {volume} {31}},\ \bibinfo
  {pages} {155005} (\bibinfo {year} {2014})},\ \Eprint
  {http://arxiv.org/abs/1404.2382} {arXiv:1404.2382 [gr-qc]} \BibitemShut
  {NoStop}%
%%CITATION = ARXIV:1404.2382;%%
\bibitem [{\citenamefont {Lower}\ \emph {et~al.}(2018)\citenamefont {Lower},
  \citenamefont {Thrane}, \citenamefont {Lasky},\ and\ \citenamefont
  {Smith}}]{PhysRevD.98.083028}%
  \BibitemOpen
  \bibfield  {author} {\bibinfo {author} {\bibfnamefont {M.~E.}\ \bibnamefont
  {Lower}}, \bibinfo {author} {\bibfnamefont {E.}~\bibnamefont {Thrane}},
  \bibinfo {author} {\bibfnamefont {P.~D.}\ \bibnamefont {Lasky}}, \ and\
  \bibinfo {author} {\bibfnamefont {R.}~\bibnamefont {Smith}},\ }\href
  {\doibase 10.1103/PhysRevD.98.083028} {\bibfield  {journal} {\bibinfo
  {journal} {Phys. Rev. D}\ }\textbf {\bibinfo {volume} {98}},\ \bibinfo
  {pages} {083028} (\bibinfo {year} {2018})}\BibitemShut {NoStop}%
\bibitem [{\citenamefont {Moore}\ and\ \citenamefont
  {Yunes}(2019{\natexlab{c}})}]{Moore:2019vjj}%
  \BibitemOpen
  \bibfield  {author} {\bibinfo {author} {\bibfnamefont {B.}~\bibnamefont
  {Moore}}\ and\ \bibinfo {author} {\bibfnamefont {N.}~\bibnamefont {Yunes}},\
  }\href@noop {} {\  (\bibinfo {year} {2019}{\natexlab{c}})},\ \Eprint
  {http://arxiv.org/abs/1910.01680} {arXiv:1910.01680 [gr-qc]} \BibitemShut
  {NoStop}%
%%CITATION = ARXIV:1910.01680;%%
\bibitem [{\citenamefont {Aasi}\ \emph {et~al.}(2015)\citenamefont {Aasi} \emph
  {et~al.}}]{TheLIGOScientific:2014jea}%
  \BibitemOpen
  \bibfield  {author} {\bibinfo {author} {\bibfnamefont {J.}~\bibnamefont
  {Aasi}} \emph {et~al.} (\bibinfo {collaboration} {LIGO Scientific}),\ }\href
  {\doibase 10.1088/0264-9381/32/7/074001} {\bibfield  {journal} {\bibinfo
  {journal} {Class. Quant. Grav.}\ }\textbf {\bibinfo {volume} {32}},\ \bibinfo
  {pages} {074001} (\bibinfo {year} {2015})}\BibitemShut {NoStop}%
%%CITATION = ARXIV:1411.4547;%%
\bibitem [{\citenamefont {Acernese}\ \emph {et~al.}(2015)\citenamefont
  {Acernese} \emph {et~al.}}]{TheVirgo:2014hva}%
  \BibitemOpen
  \bibfield  {author} {\bibinfo {author} {\bibfnamefont {F.}~\bibnamefont
  {Acernese}} \emph {et~al.} (\bibinfo {collaboration} {VIRGO}),\ }\href
  {\doibase 10.1088/0264-9381/32/2/024001} {\bibfield  {journal} {\bibinfo
  {journal} {Class. Quant. Grav.}\ }\textbf {\bibinfo {volume} {32}},\ \bibinfo
  {pages} {024001} (\bibinfo {year} {2015})},\ \Eprint
  {http://arxiv.org/abs/1408.3978} {arXiv:1408.3978 [gr-qc]} \BibitemShut
  {NoStop}%
%%CITATION = ARXIV:1408.3978;%%
\bibitem [{\citenamefont {Thrane}\ and\ \citenamefont
  {Talbot}(2019)}]{Thrane_2019}%
  \BibitemOpen
  \bibfield  {author} {\bibinfo {author} {\bibfnamefont {E.}~\bibnamefont
  {Thrane}}\ and\ \bibinfo {author} {\bibfnamefont {C.}~\bibnamefont
  {Talbot}},\ }\href {\doibase 10.1017/pasa.2019.2} {\bibfield  {journal}
  {\bibinfo  {journal} {Publications of the Astronomical Society of Australia}\
  }\textbf {\bibinfo {volume} {36}} (\bibinfo {year} {2019}),\
  10.1017/pasa.2019.2}\BibitemShut {NoStop}%
\bibitem [{\citenamefont {Skilling}\ \emph {et~al.}(2006)\citenamefont
  {Skilling} \emph {et~al.}}]{SkillingNestedsamplinggeneral2006}%
  \BibitemOpen
  \bibfield  {author} {\bibinfo {author} {\bibfnamefont {J.}~\bibnamefont
  {Skilling}} \emph {et~al.},\ }\href@noop {} {\bibfield  {journal} {\bibinfo
  {journal} {Bayesian analysis}\ }\textbf {\bibinfo {volume} {1}},\ \bibinfo
  {pages} {833} (\bibinfo {year} {2006})}\BibitemShut {NoStop}%
\bibitem [{\citenamefont {Schmidt}\ \emph {et~al.}(2015)\citenamefont
  {Schmidt}, \citenamefont {Ohme},\ and\ \citenamefont
  {Hannam}}]{PhysRevD.91.024043}%
  \BibitemOpen
  \bibfield  {author} {\bibinfo {author} {\bibfnamefont {P.}~\bibnamefont
  {Schmidt}}, \bibinfo {author} {\bibfnamefont {F.}~\bibnamefont {Ohme}}, \
  and\ \bibinfo {author} {\bibfnamefont {M.}~\bibnamefont {Hannam}},\ }\href
  {\doibase 10.1103/PhysRevD.91.024043} {\bibfield  {journal} {\bibinfo
  {journal} {Phys. Rev. D}\ }\textbf {\bibinfo {volume} {91}},\ \bibinfo
  {pages} {024043} (\bibinfo {year} {2015})}\BibitemShut {NoStop}%
\end{thebibliography}%

\end{document}